\documentstyle[12pt]{article}
\newlength{\pubnumber} 

\catcode`\@=11
\@addtoreset{equation}{section}

\def\section{\@startsection{section}{1}{\z@}{3.5ex plus 1ex minus .2ex}
 {2.3ex plus .2ex}{\large\bf}}
\def\subsection{\@startsection{subsection}{2}{\z@}{2.3ex plus .2ex}
 {2.3ex plus .2ex}{\bf}}
\newcommand\Appendix[1]{\def\thesection{Appendix \Alph{section}}
 \section{\label{#1}}\def\thesection{\Alph{section}}}
\begin{document}

\begin{titlepage}
\samepage{
\setcounter{page}{1}
\rightline{IASSNS-HEP-97/80}
\rightline{\tt hep-th/9707160}
\rightline{July 1997}
\vfill
\begin{center}
   {\Large \bf Strong/Weak Coupling Duality Relations 
    for Non-Supersymmetric String Theories\\}
\vfill
   {\large
    Julie D. Blum\footnote{
     E-mail address: julie@sns.ias.edu}
     $\,$and$\,$
      Keith R. Dienes\footnote{
     E-mail address: dienes@sns.ias.edu}
    \\}
\vspace{.12in}
 {\it  School of Natural Sciences, Institute for Advanced Study\\
  Olden Lane, Princeton, N.J.~~08540~ USA\\}
\end{center}
\vfill
\begin{abstract}
  {\rm
   Both the supersymmetric $SO(32)$ and $E_8\times E_8$ heterotic strings
   in ten dimensions have known strong-coupling duals.
    However, it has not been known whether there
   also exist strong-coupling duals for the {\it non}\/-supersymmetric
   heterotic strings in ten dimensions.
   In this paper, we construct explicit open-string duals for the
circle-compactifications
   of several of these non-supersymmetric theories,
   among them the tachyon-free $SO(16)\times SO(16)$ string.
   Our method involves the construction of heterotic and open-string
   interpolating models that continuously connect non-supersymmetric strings
    to supersymmetric strings.
   We find that our non-supersymmetric dual theories have exactly
   the same massless spectra as their heterotic counterparts within a certain
    range of our interpolations.
   We also develop a novel method for analyzing the solitons
   of non-supersymmetric open-string theories, and find that the solitons
   of our dual theories also agree with their heterotic counterparts.
   These are therefore the first known
   examples of strong/weak coupling duality relations
   between non-supersymmetric, tachyon-free string theories.
   Finally, the existence of these strong-coupling duals allows us to
   examine the non-perturbative stability of these strings,
   and we propose a phase diagram for the behavior of
   these strings as a function of coupling and radius.
   }
\end{abstract}
\vfill
\smallskip}
\end{titlepage}

\setcounter{footnote}{0}

\def\beq{\begin{equation}}
\def\eeq{\end{equation}}
\def\beqn{\begin{eqnarray}}
\def\eeqn{\end{eqnarray}}
\def\sosixteen{{$SO(16)\times SO(16)$}}
\def\V#1{{\bf V_{#1}}}
\def\half{{\textstyle{1\over 2}}}
\def\ttwo{{\vartheta_2}}
\def\tthree{{\vartheta_3}}
\def\tfour{{\vartheta_4}}
\def\ttwob{{\overline{\vartheta}_2}}
\def\tthreeb{{\overline{\vartheta}_3}}
\def\tfourb{{\overline{\vartheta}_4}}
\def\etainv{{\overline{\eta}}}
\def\Str{{{\rm Str}\,}}
\def\bone{{\bf 1}}
\def\chibar{{\overline{\chi}}}
\def\Jbar{{\overline{J}}}
\def\qbar{{\overline{q}}}
\def\calO{{\cal O}}
\def\calE{{\cal E}}
\def\calT{{\cal T}}
\def\calM{{\cal M}}
\def\calF{{\cal F}}
\def\calY{{\cal Y}}
\def\rep#1{{\bf {#1}}}
\def\ten{{(10)}}
\def\nine{{(9)}}
\hyphenation{su-per-sym-met-ric non-su-per-sym-met-ric}
\hyphenation{space-time-super-sym-met-ric}
\hyphenation{mod-u-lar mod-u-lar--in-var-i-ant}


\def\inbar{\,\vrule height1.5ex width.4pt depth0pt}

\def\IC{\relax\hbox{$\inbar\kern-.3em{\rm C}$}}
\def\IQ{\relax\hbox{$\inbar\kern-.3em{\rm Q}$}}
\def\IR{\relax{\rm I\kern-.18em R}}
 \font\cmss=cmss10 \font\cmsss=cmss10 at 7pt
\def\IZ{\relax\ifmmode\mathchoice
 {\hbox{\cmss Z\kern-.4em Z}}{\hbox{\cmss Z\kern-.4em Z}}
 {\lower.9pt\hbox{\cmsss Z\kern-.4em Z}}
 {\lower1.2pt\hbox{\cmsss Z\kern-.4em Z}}\else{\cmss Z\kern-.4em Z}\fi}

\def\NPB#1#2#3{{\it Nucl.\ Phys.}\/ {\bf B#1} (19#2) #3}
\def\PLB#1#2#3{{\it Phys.\ Lett.}\/ {\bf B#1} (19#2) #3}
\def\PRD#1#2#3{{\it Phys.\ Rev.}\/ {\bf D#1} (19#2) #3}
\def\PRL#1#2#3{{\it Phys.\ Rev.\ Lett.}\/ {\bf #1} (19#2) #3}
\def\PRT#1#2#3{{\it Phys.\ Rep.}\/ {\bf#1} (19#2) #3}
\def\CMP#1#2#3{{\it Commun.\ Math.\ Phys.}\/ {\bf#1} (19#2) #3}
\def\MODA#1#2#3{{\it Mod.\ Phys.\ Lett.}\/ {\bf A#1} (19#2) #3}
\def\IJMP#1#2#3{{\it Int.\ J.\ Mod.\ Phys.}\/ {\bf A#1} (19#2) #3}
\def\NUVC#1#2#3{{\it Nuovo Cimento}\/ {\bf #1A} (#2) #3}
\def\etal{{\it et al.\/}}

\long\def\@caption#1[#2]#3{\par\addcontentsline{\csname
  ext@#1\endcsname}{#1}{\protect\numberline{\csname
  the#1\endcsname}{\ignorespaces #2}}\begingroup
    \small
    \@parboxrestore
    \@makecaption{\csname fnum@#1\endcsname}{\ignorespaces #3}\par
  \endgroup}
\catcode`@=12

\input epsf

\section{Introduction and Overview}
\setcounter{footnote}{0}

\subsection{Why find duals of non-supersymmetric strings?}

During the past several years, significant advances have taken place
in our understanding of the strong-coupling behavior of
string theory.  Perhaps the biggest surprise was the fundamental conjecture
that the strong-coupling behavior of certain string theories can be described
as the weak-coupling behavior of corresponding ``dual'' theories which,
in many cases, are also string theories.
This observation makes it possible to address numerous non-perturbative
questions which have, until now, been beyond reach.

Among these dualities, those describing the strong-coupling behavior
of the supersymmetric ten-dimensional heterotic string theories
play a central role.
In ten dimensions, there are only two supersymmetric heterotic
string theories:  these are the $SO(32)$ theory, and the $E_8\times E_8$
theory.
As is well-known, the $SO(32)$ heterotic theory is believed to
be dual to the $SO(32)$ Type~I theory \cite{W,PW}, and the $E_8\times E_8$
theory is believed to be dual to a theory whose
low-energy limit is eleven-dimensional supergravity \cite{HW}.
Given this information, much has been learned about the strong-coupling
behavior of supersymmetric heterotic strings.

There are, however, additional heterotic string theories in ten dimensions.
These strings are non-supersymmetric, and while the
majority of them are tachyonic, one of them is tachyon-free.
The question then arises:  does this tachyon-free string have a dual theory
as well?  More generally, one can even ask whether the tachyonic heterotic
strings
might have duals.

There are numerous reasons why this is an important issue.
One fundamental reason, of course,  is to shed light on the structure  of
string
duality itself.
Is duality a property of supersymmetry or of supersymmetric string theory,
or is it, more generally, a property of string theory independently
of supersymmetry?
Knowing the answer to this question might tell us the extent to which
we expect these duality relations to survive supersymmetry breaking.

Another fundamental reason is perhaps the most obvious:  to give possible
insight into the
strong-coupling non-perturbative behavior of {\it non-supersymmetric}\/
strings.
In some sense, these strings are much harder to analyze, yet their
phenomenology, freed from the tight constraints of supersymmetry, might
be far richer.
Indeed, one might even imagine that our non-supersymmetric
world is described by a non-supersymmetric string theory
in which spacetime supersymmetry is broken by an analogue of the
Scherk-Schwarz mechanism \cite{scherkschwarz}
and in which various unexpected
stringy effects maintain finiteness,  stabilize the gauge hierarchy,
and even ensure successful gauge-coupling unification.
Various proposals in this direction can be found
in Refs.~\cite{misaligned,supertraces,review,antoniadis,others,others2}.
However, because they have non-vanishing one-loop cosmological constants,
these string theories are not believed to be stable, and are presumed
to flow to other points in moduli space at which stability is restored.
Unfortunately, an analysis of this question has been beyond reach because the
non-perturbative properties of these string theories have been, until
now, unknown.

Finally, one might even imagine that such studies could yield a new
method of supersymmetry breaking.  Indeed, if the duals of non-supersymmetric
strings were somehow found to be supersymmetric, then, viewing the duality
relation
in reverse, one would have found a supersymmetric string theory which manages
to break supersymmetry at strong coupling.

\begin{figure}[htb]
\centerline{\epsfxsize 6.0 truein \epsfbox {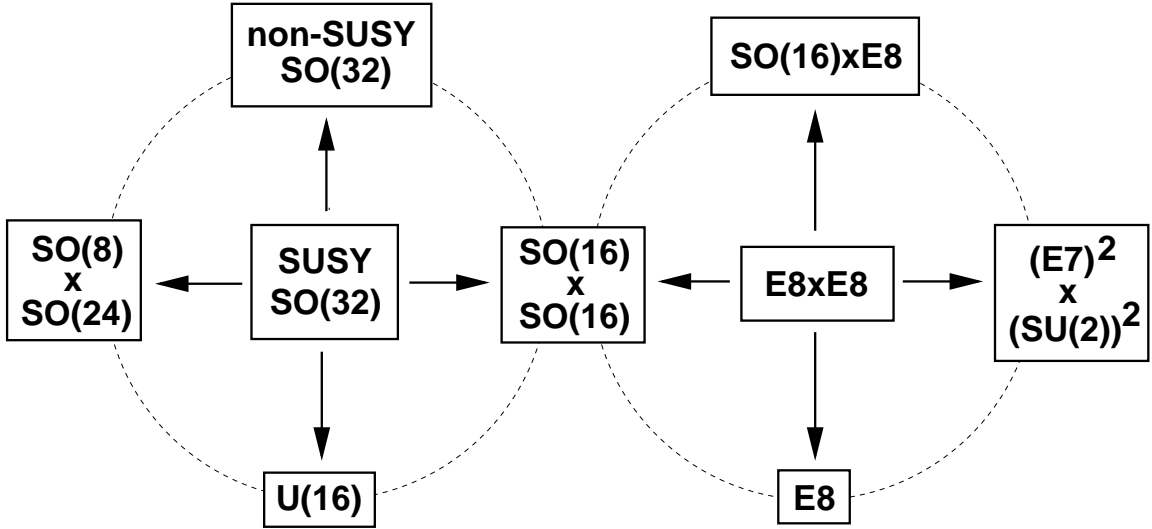}}
\caption{ The relation between the seven non-supersymmetric
    heterotic string models in ten dimensions and the
    supersymmetric $SO(32)$ and $E_8\times E_8$ string models.
    Each arrow indicates a $\IZ_2$ orbifold relation that breaks
    spacetime supersymmetry.   Only the tachyon-free $SO(16)\times SO(16)$
    string can be realized as a $\IZ_2$ orbifold of both the
    supersymmetric $SO(32)$ string and the $E_8\times E_8$ string.}
\label{wheels}
\end{figure}

What are the non-supersymmetric ten-dimensional heterotic string models?
Like their supersymmetric $SO(32)$ and $E_8\times E_8$ counterparts,
there are only a limited number of such non-supersymmetric heterotic strings.
These have been classified \cite{KLTclassification}, and are as follows:
\begin{itemize}
\item  a tachyon-free $SO(16)\times SO(16)$ model \cite{AGMV,DH};
\item  a tachyonic $SO(32)$ model \cite{DH,SW};
\item  a tachyonic $SO(8)\times SO(24)$ model \cite{DH};
\item  a tachyonic $U(16)$ model \cite{DH};
\item  a tachyonic $SO(16)\times E_8$ model \cite{DH,SW};
\item  a tachyonic $(E_7)^2 \times SU(2)^2$ model \cite{DH};  and
\item  a tachyonic $E_8$ model \cite{KLTclassification}.
\end{itemize}
In all but the last case, the gauge symmetries are realized at affine
level one.
In Fig.~\ref{wheels}, we show how these seven models are related
to the supersymmetric $SO(32)$ and $E_8\times E_8$ models through
$\IZ_2$ orbifold relations that break spacetime supersymmetry.

In this paper, we shall undertake the task of deriving strong-coupling
duals for some of these non-supersymmetric theories.
Throughout, we shall primarily concentrate on the case of the
non-supersymmetric $SO(16)\times SO(16)$ string.
Since this is the unique non-supersymmetric heterotic string
which is tachyon-free, it may be expected to have special strong-coupling
properties.
Indeed, as is evident from Fig.~\ref{wheels}, this string occupies a rather
special,
central position among the non-supersymmetric string theories, and thus
its dual theory can be expected to do so as well.

After completing our analysis for the \sosixteen\ case, we will then apply our
techniques to some of the other non-supersymmetric ten-dimensional heterotic
models listed above.  Even though these models are tachyonic, we shall
nevertheless
find that a similar analysis can be performed, and that suitable duals can be
constructed.  However, our analysis will show that these tachyonic strings
have a completely different stability behavior at strong coupling.
Our results for these strings will therefore serve to underline the unique role
of the tachyon-free \sosixteen\ string.

\subsection{Our approach}

Of course, finding the duals of non-supersymmetric strings is not a simple
undertaking, for many of the techniques that have been exploited in
finding evidence for supersymmetric duality relations no longer apply
when supersymmetry is absent.  For example, there do not {\it a priori}\/
exist any special non-supersymmetric string states (the analogues
of BPS-saturated states) whose masses are protected against strong-coupling
effects.  Likewise, upon compactification, the moduli spaces of
non-supersymmetric strings are not nearly as well understood as their
supersymmetric counterparts.

One natural idea for deriving duals of the non-supersymmetric theories
might be to start with the duals of the supersymmetric $SO(32)$ or $E_8\times
E_8$
theories, and then to duplicate the action of the appropriate $\IZ_2$ orbifolds
on these dual theories.  Unfortunately, since these orbifolds
break supersymmetry, they need not necessarily commute with strong/weak
coupling duality.
This issue has been discussed in Ref.~\cite{sen}.
Indeed, in some sense, the fundamental problem associated with this approach is
that orbifold relations are discrete:  one is dealing either with the
original theory or with the orbifolded final theory.
One cannot examine how, and where, the duality relation might begin to go wrong
in passing between the original and final theories.

Therefore, in order to derive strong-coupling duals of these non-supersymmetric
strings, our approach will be to relate the non-supersymmetric heterotic
strings
to the supersymmetric heterotic strings via {\it continuous}\/ deformations.
Since the duals for the supersymmetric theories are known,
it is then hoped that one can continuously deform both sides of such
supersymmetric
duality relations in order to obtain non-supersymmetric duality relations.
Moreover, if the duality relation were to fail to commute with such a
continuous
deformation, we would expect to see explicitly how and at what point this
failure arises as a function of the deformation parameter.

While such continuous deformations do not exist in ten dimensions,
it turns out that such deformations can indeed be performed in nine dimensions.
Specifically, we shall show that it is possible to construct
nine-dimensional {\it interpolating models}\/ by compactifying
ten-dimensional models on a circle of radius $R$ with a twist in such a way
that a given interpolating model reproduces a ten-dimensional {\it
supersymmetric}\/ model
as $R\to\infty$ and a ten-dimensional {\it non-supersymmetric}\/
model as $R\to 0$.
Such interpolating models are similar to those considered a decade ago
in Refs.~\cite{Rohm,IT,GV}.  For example, one of the nine-dimensional
interpolating models
that we shall construct reproduces the supersymmetric $SO(32)$ heterotic
string model as $R\to\infty$, but yields the
$SO(16)\times SO(16)$ model as $R\to 0$.
Thus, since the strong-coupling dual of the heterotic $SO(32)$ string model
is believed to be the $SO(32)$ Type~I model, it is natural to expect that there
will be a
corresponding nine-dimensional continuous deformation of the $SO(32)$ Type~I
model which
would produce a candidate dual for the heterotic $SO(16)\times SO(16)$ model.

\begin{figure}[htb]
\centerline{\epsfxsize 5.3 truein \epsfbox {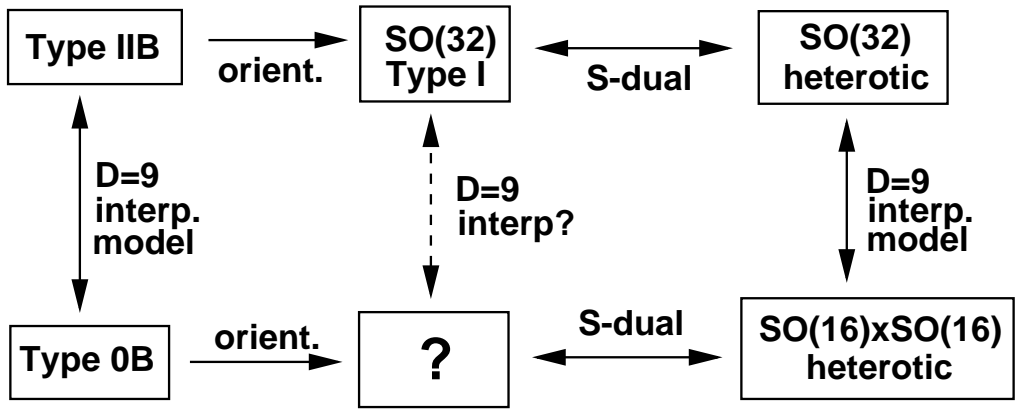}}
\caption{
      Proposed method for deriving the dual of the heterotic $SO(16)\times
SO(16)$
      string model through continuous deformations away from the supersymmetric
      $SO(32)$ string model.  Analogous deformations exist on both the
heterotic
      and the Type~II sides, from which corresponding Type~I deformations can
be
      obtained through orientifolding.}
\label{interp_approach}
\end{figure}

In order to analyze the potential continuous deformations
of Type~I models,\footnote{
    Throughout this paper, we shall use the phrase ``Type~I'' to indicate
    the general class of open-string theories, regardless of whether they
    have spacetime supersymmetry.  Similarly, we shall use the phrase
``Type~II''
    to signify closed strings with left- and right-moving worldsheet
supersymmetry,
    regardless of whether they have spacetime supersymmetry.  These terms
    will therefore be used in the same way as the general term ``heterotic''.
}
it turns out to be easier to analyze the continuous deformations of the
closed Type~II models from which they are realized as orientifolds.
Specifically, in the case of the $SO(32)$ Type~I model,
we would seek a continuous nine-dimensional deformation of the Type~IIB model.
As we shall see, there indeed exists an analogous
deformation of the Type~IIB model which breaks supersymmetry.
This deformation is described by a nine-dimensional Type~II interpolating model
which connects the Type~IIB model at $R\to \infty$ with the non-supersymmetric
so-called Type~0B model at $R\to 0$.
This situation is illustrated in Fig.~\ref{interp_approach}.

At first glance, the question would then appear to simply boil down to
determining
the orientifold of the Type~0B model that is produced at $R\to 0$.
In fact, all possible orientifolds of the Type~0B model have been derived
\cite{Sagnotti}:
the resulting non-supersymmetric Type~I models are found to have gauge
groups of total rank $32$ rather than $16$, and (except in one case) have
tachyons in their spectra.
Unfortunately, while these Type~I theories are interesting (and one of them has
recently
been conjectured \cite{BG} to be the strong-coupling dual of the {\it
bosonic}\/
string compactified to ten dimensions), their large rank precludes their
identification
as the duals of the non-supersymmetric heterotic strings.\footnote{
     Possible dualities for similar tachyonic theories have also been discussed
     in Ref.~\cite{Gates}.
     Of course, in our case
     we should actually be considering 
     the {\it $T$-dual}\/ of the Type~0B theory,
     which is the Type~0A theory.  
     However, the orientifold of the Type~0A theory
     is also a tachyonic theory.}

We shall therefore follow a slightly different course.
Rather than take the limit $R\to 0$ before orientifolding, we shall perform
our orientifold directly in nine dimensions, {\it at arbitrary radius}.
This will then yield a set of ``interpolating'' Type~I models directly in nine
dimensions,
formulated at arbitrary radius $R$.

As we shall show, these non-supersymmetric nine-dimensional interpolating
Type~I models
can be viewed as the strong-coupling duals of
our non-supersymmetric nine-dimensional interpolating heterotic models.
Specifically, for certain ranges of the radius $R$, we shall find that
\begin{itemize}
\item  both the heterotic and Type~I interpolating models are
non-supersymmetric
        and tachyon-free;
\item  their massless spectra coincide exactly;  and
\item  the D1-brane soliton of the non-supersymmetric
     Type~I theory yields the worldsheet theory of the corresponding
     non-supersymmetric heterotic string.
\end{itemize}
Moreover, because of the nature of these interpolating models,
\begin{itemize}
\item  we can smoothly take the limit as $R\to \infty$ in order
     to reproduce the known supersymmetric $SO(32)$ heterotic/Type~I duality.
\end{itemize}
Taken together, then, this provides strong evidence for the dual relation
between these non-supersymmetric string models.

As it turns out, taking the opposite limit as $R\to 0$ involves some
subtleties,
and we do not expect our duality relations to be valid in that limit.
Thus, we have not found a strong-coupling dual for the ten-dimensional
\sosixteen\ string;  rather, we have found a strong-coupling dual for the
($T$-dual of the) twisted compactification of this string on a circle
of radius $R$, valid only for $R$ taking values within a certain range.
A similar situation holds for each of the other
non-supersymmetric strings we will be discussing.

As far as we are aware, our results imply the first known duality
relations between non-supersymmetric, tachyon-free theories.
Furthermore, as we shall see, our derivation of the Type~I soliton is
rather involved.  In some respects, the standard derivation \cite{DabHull,PW}
of the Type~I soliton in the supersymmetric $SO(32)$ case
is facilitated by the fact that the heterotic $SO(32)$ theory
 {\it factorizes}\/ into separate left- and right-moving components.
This crucial property is true of all
supersymmetric string theories in ten dimensions, and makes it relatively
straightforward to realize such heterotic theories as Type~I solitons.
In the present non-supersymmetric case,
by contrast, the heterotic strings do not factorize so neatly, and consequently
a more intricate analysis is required.
Thus, we consider our derivation of this non-supersymmetric soliton --- as
well as our derivation of the techniques involved --- to be another primary
result of this paper.
Indeed, we shall see that the successful matching of the soliton to the
heterotic
string is due in large part to certain ``miracles'' that are involved
in these interpolating models.

Given these non-supersymmetric duality relations, we then
address several of the questions raised earlier.  Specifically, with
our dual theories in hand, we then consider the non-perturbative
stability for each of our candidate dual theories.
In this way, we are able to make some conjectures concerning
the ultimate fate of each of the ten-dimensional
non-supersymmetric heterotic strings.

\subsection{Outline of this paper}

This paper is organized as follows.  In Sect.~2, we provide a brief review
of some of the non-supersymmetric heterotic string models that we shall be
considering.
Then, in Sect.~3, we shall present our interpolating models, and discuss how
they are constructed.  In Sect.~4, we shall focus on the interpolation of the
non-supersymmetric \sosixteen\ string, and construct its Type~I dual.
We shall then proceed to discuss the soliton of this Type~I theory in Sect.~5.
In Sect.~6, we then apply our techniques to some of the other
non-supersymmetric
heterotic string models,  and in Sect.~7 we use our duality relations in order
to
study the perturbative and non-perturbative stability of these interpolating
models.
Sect.~8 then contains our conclusions, as well as a proposed phase diagram
for our non-supersymmetric string models and speculations about
lower-dimensional
theories.
Finally, in an Appendix we present the explicit free-fermionic realizations
of many of the ten- and nine-dimensional models we shall be considering in this
paper.
Note that a brief summary of some of the results
of this paper can be found in Ref.~\cite{short}.

\section{Review of Ten-Dimensional Models}
\setcounter{footnote}{0}

We begin by reviewing some of the ten-dimensional heterotic and Type~II string
models
that will be relevant for our analysis.
For notational convenience, in Sects.~2--5 we shall limit our attention to the
five heterotic models in Fig.~\ref{wheels} whose gauge groups contain
$SO(16)\times SO(16)$ as a subgroup.
In addition to the supersymmetric $SO(32)$ and $E_8\times E_8$ models,
these include the non-supersymmetric $SO(32)$, $SO(16)\times SO(16)$,
and $SO(16)\times E_8$ models.

Since much of our analysis of these models will proceed
via their partition functions, we begin by establishing some general
conventions and notation.

\subsection{Conventions}

Throughout this paper, we shall express partition functions in terms of the
characters of the corresponding level-one affine Lie algebras.
It turns out that we shall only need to consider the $SO(2n)$ gauge groups,
which
have central charge $c=n$ at level one.
For such groups, the only level-one
unitary representations are the identity (I), the vector (V),
and the spinor (S) and conjugate spinor (C).   This is equivalent
to the statement that $SO(2n)$ has four conjugacy classes.
These representations have
multiplicities $\lbrace 1,2n,2^{n-1},2^{n-1}\rbrace$ respectively,
and have conformal dimensions $\lbrace h_I,h_V,h_S,h_C\rbrace=
\lbrace 0,1/2,n/8,n/8\rbrace$.
Their characters can then be expressed in terms of the
Jacobi $\vartheta$ functions as follows:
\beqn
 \chi_I &=&  \half\,(\tthree^n + \tfour^n)/\eta^n ~=~ q^{h_I-c/24}
        \,(1 + n(2n-1)\,q + ...)\nonumber\\
 \chi_V &=&  \half\,(\tthree^n - \tfour^n)/\eta^n ~=~ q^{h_V-c/24} \,(2n +
...)\nonumber\\
 \chi_S &=&  \half\,(\ttwo^n + {\vartheta_1}^n)/\eta^n ~=~ q^{h_S-c/24}
\,(2^{n-1} + ...)\nonumber\\
 \chi_C &=&  \half\,(\ttwo^n - {\vartheta_1}^n)/\eta^n ~=~ q^{h_C-c/24}
\,(2^{n-1} + ...)~
\eeqn
where $q\equiv e^{2\pi i\tau}$.
Note that $n(2n-1)$ is the dimension of the adjoint representation of $SO(2n)$.
This reflects the general fact that
the adjoint is the first descendant
of the identity field in such affine Lie algebra conformal field theories.
Even though the spinor and conjugate spinor representations are distinct,
we find that $\chi_S=\chi_C$ (due to the fact that $\vartheta_1=0$).
For the ten-dimensional little Lorentz group $SO(8)$,
the distinction between $S$ and $C$ is equivalent to relative
spacetime chirality.

We can now express the partition functions of our string models
in terms of these characters.  This will also be useful for determining
the massless spectra of these models.
For convenience, we shall henceforth adopt the convention
that all left-moving (unbarred) $\vartheta$-functions will be written in
terms of the characters of $SO(16)$, while those that are right-moving
(barred) will be written in terms of those of the Lorentz group $SO(8)$.
Occasionally we shall also use the notation $\tilde\chi$
to signify left-moving $SO(32)$ characters.
Furthermore, we shall define
\beq
         Z^{(n)}_{\rm boson} ~\equiv ~ {\tau_2}^{-n/2}\, (\etainv\eta)^{-n}~,
\eeq
which represents
the contribution of $n$ transverse bosons to the total partition function.
Finally, we shall also define the Jacobi factor
\beq
     \overline{J}
       ~\equiv~ \half\,\overline{\eta}^{\,-4}\,
\left(\tthreeb^4-\tfourb^4-\ttwob^4 - {\overline{\vartheta}_1}^4\right)
     ~=~ \overline{\chi}_V - \overline{\chi}_S ~=~0~.
\eeq
This factor $J$ vanishes as the result of
the ``abstruse'' identity of Jacobi, or equivalently as a result of
$SO(8)$ triality, under which the $SO(8)$ vector and spinor representations are
indistinguishable.
Partition functions that are proportional to $J$ therefore correspond to string
models
with spacetime supersymmetry.

\subsection{Supersymmetric $SO(32)$ string model}

Given this notation, we begin by examining the supersymmetric $SO(32)$ string
model.
In terms of the characters $\tilde X$ of $SO(32)$ for the left-movers
and the characters $\overline{\chi}$ $SO(8)$ for right-movers,
its partition function takes the form
\beq
         Z_{SO(32)} ~=~ Z^{(8)}_{\rm boson}~
         (\overline{\chi}_V-\overline{\chi}_S)
         \,(\tilde \chi_I+ \tilde \chi_S)~.
\label{SO32partfunctso32}
\eeq
In terms of the characters of $SO(16)\times SO(16)$ for the left-movers,
this partition function is equivalent to
\beq
         Z_{SO(32)} ~=~ Z^{(8)}_{\rm boson}~
         (\overline{\chi}_V-\overline{\chi}_S)
         \,(\chi_I^2 + \chi_V^2 + \chi_S^2 + \chi_C^2)~.
\label{SO32partfunct}
\eeq
Note that it makes no physical difference whether $\chibar_S$ or $\chibar_C$ is
written in this partition function since all spinors in this model
have the same spacetime chirality.

It is clear that this model is supersymmetric.
It is also straightforward to determine from this partition function that
the gauge group is $SO(32)$, since
the massless gauge bosons in this model can only contribute to
terms of the form $\overline{\chi}_V \times \lbrace h=1\rbrace$.
In this partition function, the only such terms are
$\chibar_V\tilde\chi_I$ or  $ \chibar_V\, (\chi_I^2 + \chi_V^2)$
where we must consider the first excited state within $\tilde \chi_I$
(in order to produce $h=1$)
or equivalently the first excited state within $\chi_I^2$
and the ground state of
$\chi_V^2$ (which already has $h=1$).
However, the first descendant representation within $\tilde\chi_I$
is the adjoint of $SO(32)$.
Equivalently, the first excited state within $\chi_I^2$ transforms as
$({\bf adj,1})\oplus ({\bf 1,adj})$ under $SO(16)\times SO(16)$, while
the ground state of $\chi_V^2$ transforms as $({\bf vec,vec})$.
Thus, under $SO(16)\times SO(16)$, the
the gauge bosons of this model
\beq
        ({\bf adj,1})\oplus ({\bf 1,adj}) \oplus ({\bf vec,vec}) ~,
\eeq
which fill out the 496 states of the adjoint of $SO(32)$.

It is also clear from this partition function that
in addition to the gravity supermultiplet,
the gauge bosons and their superpartners
are the only states in the massless spectrum.

\subsection{$E_8\times E_8$ string model}

The partition function of the $E_8\times E_8$ model can likewise
be written in terms of the $SO(16)$ characters for the left-movers, yielding
\beq
        Z_{E_8\times E_8} ~=~ Z^{(8)}_{\rm boson} ~
(\chibar_V-\chibar_S)~(\chi_I + \chi_S)^2~.
\eeq
Once again, supersymmetry is manifest.
Moreover, we immediately recognize that the level-one $E_8$ character is
\beq
 \chi_{E_8}~\equiv~ \half \,\eta^{-8}\,\left( {\vartheta_1}^8 + \ttwo^8 +
\tthree^8 + \tfour^8 \right)
           ~=~ \chi_I+\chi_S~,
\label{chie8}
\eeq
which verifies that the gauge group of this model
is indeed $E_8\times E_8$.
For $h=1$, the statement (\ref{chie8}) reflects the fact that
the adjoint plus spinor representations of $SO(16)$ combine to fill out
the adjoint representation of $E_8$.
(Note that there is only one character of $E_8$ because the identity
is only unitary representation of $E_8$ at affine level one, or equivalently
because $E_8$ has only one conjugacy class.)
As with the $SO(32)$ model, we see that the gauge bosons, their superpartners,
and the gravity supermultiplet
are the only states in the massless spectrum.

\subsection{$SO(16)\times SO(16)$ model}

Next, we turn to the $SO(16)\times SO(16)$ model which will be, in large
part, the main focus of this paper.
In terms of the characters of $SO(16)$ for the left-movers
(and those of $SO(8)$ for the right-movers), the
partition function of this model takes the form
\beqn
       Z ~=~ Z^{(8)}_{\rm boson}~\times &\biggl\lbrace&
    \chibar_I  \,(\chi_V\chi_C+\chi_C\chi_V) ~+~ \chibar_V \,(\chi_I^2 +
\chi_S^2) \nonumber\\
      && -~\chibar_S\, (\chi_V^2 + \chi_C^2) ~- ~\chibar_C \, (
\chi_I\chi_S+\chi_S\chi_I)
          ~\biggr\rbrace~.
\label{sosixteenpartfunct}
\eeqn
It is immediately clear that supersymmetry is broken,
and that the gauge group of this model is simply $SO(16)\times SO(16)$.
In addition to the gravity multiplet, the complete massless spectrum
of this model consists of the following $SO(16)\times SO(16)$ representations:
\beqn
          {\rm vectors}:&~~& ({\bf 120}, {\bf 1}) \oplus ({\bf 1},{\bf
120})\nonumber\\
          {\rm spinors}:&~~& ({\bf 16}, {\bf 16})_+ \oplus
           ({\bf 128},{\bf 1})_- \oplus ({\bf 1}, {\bf 128})_-
\eeqn
Here the spinor subscripts $\pm$ indicate $SO(8)$ spinors and conjugate spinors
respectively;
only the relative chirality between these two groups of spinors is physically
relevant.
Thus cancellation of the irreducible gravitational anomaly is manifest,
even though supersymmetry is absent.
(The other irreducible anomalies can be shown to cancel as well.)
Furthermore, it is also clear that no physical tachyonic states are present in
this model.
In order to see this from the partition function, note that such physical
tachyonic
states would have to contribute to terms of the form
$\chibar_I \times \lbrace h=1/2\rbrace$.  The only possible term of this
form would be $\chibar_I\,( \chi_I \chi_V+\chi_V\chi_I)$,
yet no such term appears in the partition function.

This model is the unique tachyon-free non-supersymmetric heterotic string model
in ten dimensions.  As such, we may expect it to have a number of special
properties.
For example,
we have already seen in Fig.~\ref{wheels} that it is the only
heterotic non-supersymmetric model in ten dimensions which
can be realized as a $\IZ_2$ orbifold of either the supersymmetric $SO(32)$
model or the $E_8\times E_8$ model.
Likewise, because it is tachyon-free, it is the only non-supersymmetric
ten-dimensional
string model to have a finite but non-vanishing one-loop cosmological constant.

\begin{figure}[thb]
\centerline{\epsfxsize 3.5 truein \epsfbox {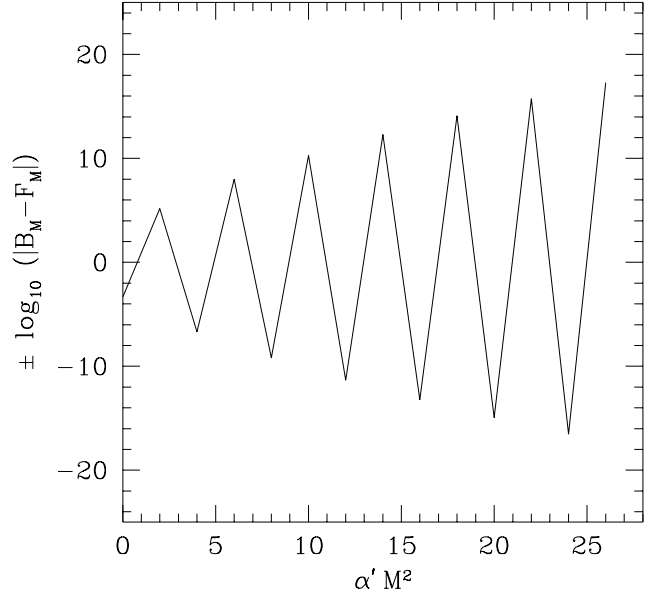}}
\caption{ Alternating boson/fermion surpluses
    in the $D=10$ non-supersymmetric tachyon-free
    $SO(16)\times SO(16)$ heterotic string.
   For each mass level $\alpha' M_2\in 2\IZ$ in this model,
   we plot $\pm \log_{10}(|B_M-F_M|)$ where $B_M$ and $F_M$ are respectively
  the numbers of spacetime bosonic and fermionic states at that level.
  The overall sign is chosen positive if $B_M>F_M$, and negative otherwise.
   The points are connected in order to stress the alternating, oscillatory
   behavior of the boson and fermion surpluses throughout the string
   spectrum.  These oscillations
   insure that $ \Str M^0= \Str M^2= \Str M^4= \Str M^6=0$
   in this model, even though there is no spacetime supersymmetry.  }
\label{missusyplot}
\end{figure}

Another remarkable property of the \sosixteen\ string
concerns its distribution of bosonic and fermionic states at all mass levels.
In this string model, one has a surplus of 2112 fermionic states over bosonic
states
at the massless level.  This reflects the broken supersymmetry.
However, this surplus of fermionic states at the massless level is
balanced by a surplus of 147,456 bosonic states at the
first excited level (for which $L_0=\overline{L_0}=1/2$), and this is balanced
in turn
by a surplus of 4,713,984 fermionic states
at the next excited level (for which $L_0=\overline{L_0}=1$).
This pattern then continues all the way through the infinite towers of
massive string states, and is illustrated in Fig.~\ref{missusyplot} where
we plot the sizes of these surpluses as a function of the spacetime
mass $M^2= 2(L_0+\overline{L_0})/\alpha'$.

These oscillations are the signature of a hidden so-called
``misaligned supersymmetry'' \cite{misaligned} in the string spectrum.
Misaligned supersymmetry is a general feature of non-supersymmetric string
models,
and serves as the way in which string theory
manages to maintain finiteness {\it even without spacetime supersymmetry}.
Moreover, for {\it tachyon-free}\/ string models,
there is an added bonus:  these oscillating surpluses are even sufficient
to cause certain mass supertraces to vanish when evaluated over the
entire string spectrum.
Specifically, if one defines a regulated string supertrace via
\beq
    \Str M^{2\beta} ~\equiv ~
        \lim_{y\to 0} \,  \sum_{\rm states}\,(-1)^F \,M^{2\beta}\,
               e^{-y \alpha' M^2}~,
\eeq
then in ten dimensions it can be shown \cite{supertraces} that
\beq
        \Str M^0~=~ \Str M^2 ~=~ \Str M^4 ~=~ \Str M^6 ~=~ 0~.
\label{supertracerelations}
\eeq
Similar results also exist for non-supersymmetric tachyon-free string theories
in lower dimensions \cite{supertraces}.

It is remarkable that such constraints can be satisfied
in string theory, especially given the fact that string theory gives rise
to infinite towers of states
whose degeneracies grow exponentially as a function of mass.
Such supertrace relations (\ref{supertracerelations}) are similar to those
that hold in field theories with spontaneously broken supersymmetry, and
suggest that even though the \sosixteen\ string is non-supersymmetric,
there may yet be a degree of finiteness in this theory which might allow
a consistent strong-coupling dual to be constructed.

\subsection{Non-supersymmetric $SO(32)$ model}

Next, we examine the tachyonic, non-supersymmetric $SO(32)$ model.
In terms of the $SO(32)$ characters $\tilde \chi$,
the partition function of this model is given by:
\beq
   Z ~=~ Z^{(8)}_{\rm boson}~ (\chibar_I\,\tilde \chi_V + \chibar_V\,\tilde
\chi_I
     - \chibar_S \,\tilde \chi_S - \chibar_C \,\tilde \chi_C)~,
\eeq
which decomposes into the characters of $SO(16)\times SO(16)$, yielding
\beqn
       Z ~=~ Z^{(8)}_{\rm boson}~\times &\biggl\lbrace &
    \chibar_I  \,(\chi_I\chi_V+\chi_V\chi_I) ~+~ \chibar_V \,(\chi_I^2 +
\chi_V^2) \nonumber\\
      && -~\chibar_S\, (\chi_S^2 + \chi_C^2) ~- ~\chibar_C \, (
\chi_S\chi_C+\chi_C\chi_S)
          ~\biggr\rbrace~.
\label{nonsusyso32}
\eeqn
We see that once again supersymmetry is broken, with gauge group $SO(32)$
and with bosonic scalar tachyons of right- and left-moving masses
$M_R^2=M_L^2= -1/2$ transforming in the vector representation of $SO(32)$.
Aside from the gravity multiplet and the adjoint gauge bosons, this model
contains
no other massless states.

\subsection{$SO(16)\times E_8$ model}

Finally, we examine the
tachyonic, ten-dimensional, non-supersymmetric $SO(16)\times E_8$ model.
Its partition function is given by:
\beq
   Z ~=~Z^{(8)}_{\rm boson}~
        (\chibar_I \,\chi_V ~+~ \chibar_V \,\chi_I ~-~ \chibar_S \,\chi_S
         ~-~\chibar_C \,\chi_C )
       ~(\chi_I+\chi_S)~,
\eeq
and its complete massless spectrum (in addition to the gravity multiplet)
consists of the following representations of $SO(16)\times E_8$:
\beqn
          {\rm vectors}:&~~& ({\bf 120}, {\bf 1}) \oplus ({\bf 1},{\bf
248})\nonumber\\
          {\rm spinors}:&~~& ({\bf 128}, {\bf 1})_+ \oplus
          ({\bf 128}, {\bf 1})_- ~.
\eeqn
Once again, cancellation of the irreducible gravitational anomaly
is manifest, even without supersymmetry.
Note that unlike the $SO(16)\times SO(16)$ case, this string model contains
tachyonic states.  These are bosonic scalars with tachyonic right- and
left-moving
masses $M_R^2=M_L^2= -1/2$ transforming
in the $({\bf 16},{\bf 1})$ representation of $SO(16)\times E_8$.

\subsection{Other non-supersymmetric heterotic string models}

A similar analysis can be given for each of the other non-supersymmetric
ten-dimensional heterotic string models, but for notational convenience we
shall
defer a discussion of these models until Sect.~6.  Since these models do
not have gauge groups which contain $SO(16)\times SO(16)$ as a subgroup,
their partition functions cannot be expressed in terms of the characters
of $SO(16)$ or $SO(8)$.

\subsection{Type~II models in ten dimensions}

We now turn to the four Type~II models that exist in ten dimensions.
These are the supersymmetric Type~IIA and Type~IIB models and the
non-supersymmetric so-called Type~A and Type~B models.
The latter two models are tachyonic, and
only the Type~IIB model is chiral.
We shall henceforth refer to the Type~A and Type~B models
as being of Types~0A and 0B respectively (to reflect their levels of
supersymmetry).

The partition functions of these four models are as follows:
\beqn
      Z_{\rm IIA} &=& Z_{\rm boson}^{(8)} ~
(\chibar_V-\chibar_S)\,(\chi_V-\chi_C) ~\nonumber\\
      Z_{\rm IIB} &=& Z_{\rm boson}^{(8)} ~
(\chibar_V-\chibar_S)\,(\chi_V-\chi_S) ~\nonumber\\
      Z_{\rm 0A} &=& Z_{\rm boson}^{(8)} ~
         (\chibar_I\chi_I + \chibar_V\chi_V + \chibar_S\chi_C +
\chibar_C\chi_S) ~\nonumber\\
      Z_{\rm 0B} &=& Z_{\rm boson}^{(8)} ~
         (\chibar_I\chi_I + \chibar_V\chi_V + \chibar_S\chi_S +
\chibar_C\chi_C) ~.
\eeqn

Note that for these Type~II partition functions,
both the left- and right-moving characters
are the characters of the $SO(8)$ transverse Lorentz group.
Also note that there is no physical distinction between a given partition
function and one in which all occurrences of $\chi_S$ and $\chibar_S$ are
respectively replaced by $\chi_C$ and $\chibar_C$ and vice versa.

\section{Nine-Dimensional Models and Continuous Deformations}
\setcounter{footnote}{0}

\subsection{Interpolating Models:  General Procedure}

As discussed in the Introduction,
our approach will be to consider nine-dimensional models
which interpolate smoothly between different ten-dimensional models.
The most straightforward procedure will be to
compactify ten-dimensional models on a circle of radius $R$.
As $R\to \infty$, we expect to reproduce our original ten-dimensional
string model which we can denote $M_1$;  likewise,
as $R\to 0$, by $T$-duality arguments we also expect to reproduce
a ten-dimensional string model which we can denote $M_2$.
Specifically, this means that as $R\to 0$, we will produce a degenerate
nine-dimensional model $\tilde M_2$ which can be identified as the
$T$-dual of the ten-dimensional model $M_2$.
Under these conditions we shall then say that such
a nine-dimensional model interpolates between $M_1$ and $M_2$.
Models~$M_1$ and $M_2$
need not be the same ten-dimensional model
if the nine-dimensional interpolating string model is
not self-dual (in the sense of $R\to 1/R$ duality).
This will generally occur if, in addition to the circle compactification,
there are some twists introduced in the compactification.

Our goal will therefore be to construct ``twisted'' nine-dimensional
string models which interpolate between the various ten-dimensional
string models.
We seek, in particular, a nine-dimensional string model which
interpolates between the non-supersymmetric ten-dimensional $SO(16)\times
SO(16)$
string model in one limit, and a {\it supersymmetric}\/ ten-dimensional
string model in the other.  As we shall see, such nine-dimensional
models can indeed be constructed.

There are two procedures that one can follow in order
to construct such nine-dimensional interpolating models.
The first procedure is to employ
the nine-dimensional free-fermionic construction \cite{KLT,ABK,KLST}.
This procedure results in models formulated at the fixed
radius  $R=\sqrt{2\alpha'}$, and our use of the
free-fermionic construction, with all of its built-in rules and constraints,
guarantees their internal consistency.
Within these models, we then identify the radius modulus corresponding to
the tenth dimension, and extrapolate to arbitrary radius.
This procedure is unambiguous, and leads to
nine-dimensional string models formulated at arbitrary radius.
By examining the two ten-dimensional limits of such models, it is simple
to identify those that interpolate between different ten-dimensional models.

The second (ultimately equivalent) procedure is to employ an orbifold
construction.  This has the advantage of exposing, from the start,
the geometric radius-dependence of the resulting nine-dimensional models.
Furthermore, we shall also be able to easily determine the
different ten-dimensional limits.
Our procedure for constructing the desired nine-dimensional models
rests upon a simple but crucial observation which we shall now explain.

Let us assume that
we wish to construct a nine-dimensional model interpolating between
two ten-dimensional models $M_1$ and $M_2$.
As stated above, this means that we wish to construct a nine-dimensional
model whose $R\to \infty$ limit produces $M_1$, and whose $R\to 0$ limit
produces the nine-dimensional degenerate model $\tilde M_2$
which is the $T$-dual of the ten-dimensional model $M_2$.
(Note that for ten-dimensional heterotic strings, $\tilde M=M$ in
all cases.)
Our procedure to construct such an interpolating model is as follows.
Let us assume that $M_1$ and $\tilde M_2$ are
related to each other in such a way that $\tilde M_2$ is the
$Q$-orbifold of $M_1$, where $Q$ is any $\IZ_2$ action.
Given this relation, let us then
compactify $M_1$ directly on a circle of radius $R$, and
denote the resulting model as $M_1^{(9)}$.
Since there are no twists involved in this compactification, $M_1^{(9)}$
contains only states with integer momentum- and winding-mode quantum numbers,
and reproduces the ten-dimensional model $M_1$ both as $R\to \infty$ and as
$R\to 0$.
By definition, the degenerate $R\to 0$ limit of $M_1$ is equivalent to the
$R\to \infty$ limit of $\tilde M_1$, where $\tilde M_1$ is the
$T$-dual of $M_1$.  Thus, Model~$M_1^{(9)}$ is a nine-dimensional
model that trivially interpolates between the ten-dimensional models
$M_1$ and $\tilde M_1$.

This much is fairly straightforward.
However, let us now consider orbifolding the model $M_1^{(9)}$ by $Q'\equiv
\calT Q$
where $Q$ is the above ten-dimensional orbifold relating $M_1$ to $\tilde M_2$
and where
\beq
                \calT:~~~~ X_1 ~\to~  X_1 + \pi R~.
\label{Tdef}
\eeq
Here $X_1$ represents the coordinate of the compactified direction,
and $R$ is the radius of compactification.
Since $\calT$ represents only half of a complete translation around the
compactified direction, the $\calT$-invariant states of $M_1^{(9)}$
are those with {\it even}\/ momentum quantum numbers (and arbitrary winding
numbers),
while the twisted sectors of this orbifold re-introduce
the odd momentum quantum numbers along with {\it half}\/-integer
winding numbers.
Thus, orbifolding the model $M_1^{(9)}$
by the combined $\IZ_2$-action $\calT Q$
has the net effect of mixing the ten-dimensional orbifold $Q$ with
the compactification orbifold $\calT$ in a non-trivial way.
However,
if we denote the resulting model as $M^{(9)}$,
it turns out that $M^{(9)}$ reproduces
Model~$M_1$ as $R\to\infty$, but reproduces
Model~$\tilde M_2$ as $R\to 0$.
The latter is equivalent to the $R\to \infty$ limit of Model~$M_2$,
whereupon we can identify Model~$M_2$ as the
ten-dimensional model that is produced in the $R\to 0$ limit.
Thus, given this construction, we see that $M^{(9)}$ is the desired
nine-dimensional model that interpolates between the ten-dimensional
models $M_1$ and $M_2$ in its $R\to \infty$ and $R\to 0$ limits.

It is simple to see intuitively why this procedure works in yielding
the desired nine-dimensional interpolating model.
Roughly speaking, as $R\to \infty$,
the odd momentum states are
degenerate with the even momentum states.
This causes $\calT$ to act as zero,
which merely amounts to a rescaling of the effective volume normalization of
the partition function of Model~$M_1$.
Thus, we are simply left with the ten-dimensional $M_1$ theory.
By contrast, as $R\to 0$, the $\calT$-twisted and $\calT$-untwisted sectors
contribute
equally.  This thereby reproduces the $Q$-orbifold of $M_1$, which is $\tilde
M_2$.
Of course, this is merely a heuristic explanation of why this procedure
works.  A detailed
proof will be presented in Sect.~3.4, when we explicitly construct
our desired nine-dimensional models.

\subsection{Ten-dimensional orbifold relations}

The first step in the above procedure is to
determine the orbifolds that relate
the ten-dimensional models in Sect.~2 to each other.
Indeed, as is well-known, many of these ten-dimensional
string models can be realized as $\IZ_2$ orbifolds of each other.
We shall therefore now give several of the orbifold relations between
the ten-dimensional heterotic and Type~II string models.

We begin with a simple example.
Let us seek to realize the $E_8\times E_8$ heterotic model as an
orbifold of the supersymmetric $SO(32)$ heterotic string model.
This orbifold can be explicitly described as follows.
Starting from the $SO(32)$ model, we can decompose
our $SO(32)$ representations into representations of $SO(16)\times SO(16)$
and then modding out by an action
which changes the signs of the vector and conjugate spinor representations of
the first $SO(16)$ factor but which leaves the identity and spinor
representations of this $SO(16)$ factor (as well as all of the representations
of the second $SO(16)$ factor) invariant.
This action, which we shall denote $R^{(1)}_{VC}$,
thus has the following effect:
\beq
\begin{tabular}{c||c|c}
        ~~~~~~ & ~~first $SO(16)$~~    & ~~second $SO(16)$~~    \\
\hline
     $I$ & $+$ & $+$ \\
     $V$ & $-$ & $+$ \\
     $S$ & $+$ & $+$ \\
     $C$ & $-$ & $+$ \\
\end{tabular}
\label{Ydef}
\eeq
Note that this is indeed a $\IZ_2$ action, in that it squares to the identity.
Given this action $R^{(1)}_{VC}$, it is then straightforward to determine the
effect of
orbifolding the $SO(32)$ theory by $R^{(1)}_{VC}$.
Since the original $SO(32)$ theory is the untwisted sector of this orbifold,
with partition function $Z_+^+\equiv Z_{SO(32)}$
as given in (\ref{SO32partfunct}),
the first step is to mod out by $R^{(1)}_{VC}$.  Given the
definition in (\ref{Ydef}), we see that this modding can be achieved
by adding to $Z_+^+$ the contribution from the projection sector
\beq
     Z_+^-~=~
         Z^{(8)}_{\rm boson}~
         (\overline{\chi}_V-\overline{\chi}_S)
         \,(\chi_I^2 - \chi_V^2 + \chi_S^2 - \chi_C^2)~.
\eeq
This projection sector is determined from the unprojected sector by acting with
$R^{(1)}_{VC}$.
To restore modular invariance, however, we must then also include the
twisted sector
\beq
     Z_-^+(\tau)~\equiv~ Z_+^-(-1/\tau)
    ~=~ Z^{(8)}_{\rm boson}~
         (\overline{\chi}_V-\overline{\chi}_S)
         \,(\chi_I\chi_S + \chi_S\chi_I + \chi_C\chi_V + \chi_V\chi_C)~
\eeq
along with its corresponding projection sector
\beq
     Z_-^-~=~
         Z^{(8)}_{\rm boson}~
         (\overline{\chi}_V-\overline{\chi}_S)
         \,(\chi_I\chi_S + \chi_S\chi_I - \chi_C\chi_V - \chi_V\chi_C)~.
\eeq
The result of this orbifolding procedure then yields a model with the
total partition function
\beq
   Z_{\rm total}~=~ {1\over 2}\,
   \bigl\lbrack Z_+^+ + Z_+^- + Z_-^+ + Z_-^- \bigr\rbrack~,
\eeq
and we find that $Z_{\rm total}=Z_{E_8\times E_8}$.
Thus, the $E_8\times E_8$ model can be realized as the
$R^{(1)}_{VC}$-orbifold of the $SO(32)$ model.

\begin{figure}[thb]
\centerline{\epsfxsize 5.0 truein \epsfbox {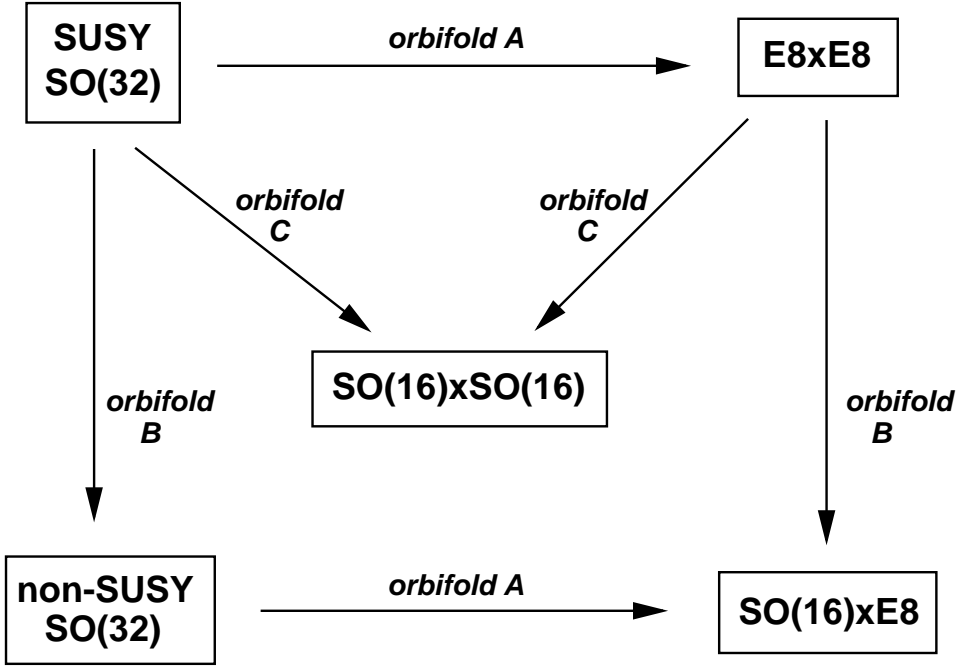}}
\caption{
   Several $\protect\IZ_2$ orbifold connections between
   ten-dimensional heterotic string models, as discussed in the text.
    Note the highly symmetric position of the tachyon-free
    $SO(16)\times SO(16)$ model.  }
\label{orbrelations}
\end{figure}

It turns out that all of the ten-dimensional heterotic string models
that we have considered can be related in this way as different $\IZ_2$
orbifolds of each other.
Some of these orbifold relations,
each of which is analogous to the above example,
are indicated in Fig.~\ref{orbrelations}.
In this figure, the orbifolds $A$ through $C$ are defined as follows:
\beqn
          A &\equiv& R^{(1)}_{VC} \nonumber\\
          B &\equiv& \tilde R_{SC} \, R^{(1)}_{SC} \nonumber\\
          C &\equiv& \tilde R_{SC} \, R^{(1)}_{SC} \, R^{(2)}_{VS} ~.
\label{hetorbs}
\eeqn
Here $\tilde R$ refers to the right-moving Lorentz $SO(8)$ representations;
$R^{(1)}$ and $R^{(2)}$ refer to the two left-moving internal $SO(16)$
representations;
and the subscripts in each case indicate which representations are odd under
the corresponding action.
Note that $\tilde R_{SC}$ is simply $(-1)^F$ where $F$ is the
spacetime fermion number.
As is evident from Fig.~\ref{orbrelations}, orbifolds involving
$(-1)^F$ are responsible for breaking supersymmetry.
We also remark that although the same orbifolds
$A$, $B$, and $C$ are responsible for the different mappings as indicated
on opposite sides of Fig.~\ref{orbrelations},
the inverse orbifolds in each case will generally be different.
For example, it turns out that the supersymmetric $SO(32)$ model can be
realized
as the $\tilde R_{IS}$-orbifold of the non-supersymmetric $SO(32)$ model,
while the $E_8\times E_8$ model turns out to be the $A$-orbifold of the
$SO(16)\times E_8$ model.
Note that $\tilde R_{IS}$ is equivalent to $-(-1)^F$
after an $SO(8)$ triality rotation.

It is striking that the \sosixteen\ theory
occupies such a central and symmetric position in Fig.~\ref{orbrelations}.
However, this is not entirely unexpected.
Given the gauge groups of these different string models,
we see that \sosixteen\ is in some sense the unique
``greatest common denominator'', for \sosixteen\ is largest common subgroup
that all of these models share.
Note from Fig.~\ref{wheels} that this central position for the \sosixteen\
theory
persists even if we include the other non-supersymmetric heterotic string
models in ten dimensions.
Furthermore, the \sosixteen\ theory is the unique ten-dimensional
heterotic string model which is non-supersymmetric yet
simultaneously tachyon-free.

The orbifold relations between Type~II strings are somewhat simpler.
Indeed, the Type~0B model can be simply realized
as the
$\tilde R_{SC} R_{SC}$
orbifold of the Type~IIB model, where
now the $R$-factors refer to the right- and left-moving
Lorentz $SO(8)$ groups.
Thus $\tilde R_{SC} R_{SC} = (-1)^F$
where $F=F_L+F_R$ is the total spacetime fermion number.
Inverting this, we likewise find that the Type~IIB model can
be realized as the $\tilde R_{IC}$
or $R_{IC}$
orbifold of the Type~0B model.

\subsection{Circle compactifications}

As with the ten-dimensional string models, our nine-dimensional
models will be most easily studied by analyzing their partition functions.
Therefore, let us first establish some notation appropriate for
partition functions at an arbitrary radius $R$.

As is traditional,
we first define the {\it dimensionless}\/ inverse radius $a\equiv
\sqrt{\alpha'}/R$.
In terms of $a$,
the right- and left-moving momenta resulting from compactification
can then be written as
\beq
   p_R = {1\over{\sqrt{2\alpha'}}}\, (ma-n/a)~,~~~~~
   p_L = {1\over{\sqrt{2\alpha'}}}\, (ma+n/a)~
\label{compmomenta}
\eeq
and their corresponding partition function contribution is
\beq
    (\etainv\eta)^{-1}\, \sum_{m,n}\,
     \qbar^{\alpha' p_R^2/2}\,q^{\alpha' p_L^2/2} ~=~
    (\etainv\eta)^{-1}\, \sum_{m,n}\, \exp\left\lbrack
        2\pi i mn\tau_1 ~-~
     \pi \tau_2 (m^2a^2 + n^2/a^2) \right\rbrack~.
\label{expre}
\eeq
Here $m$ and $n$ are the momentum- and winding-mode
excitation numbers.  {\it A priori}\/, these numbers are only restricted to be
integers.
However, as we discussed in Sect.~3.1, we will be interested in orbifolds
which restrict the momentum quantum numbers to either even or odd integers,
and likewise will also be interested in the corresponding twisted sectors
for which the winding-mode quantum numbers can also be half-integer.
Thus, starting from the expression (\ref{expre}),
we follow Ref.~\cite{Rohm} in defining
four functions ${\cal E}_{0,1/2}$ and ${\cal O}_{0,1/2}$
depending on the types of values $m$ and $n$ may have in (\ref{expre}):
\beqn
      {\cal E}_0 &\equiv& \lbrace  m ~{\rm even},~ n\in\IZ\rbrace
\nonumber\\
      {\cal E}_{1/2} &\equiv& \lbrace  m ~{\rm even},~ n\in\IZ+\half\rbrace
\nonumber\\
      {\cal O}_{0} &\equiv& \lbrace  m ~{\rm odd},~ n\in\IZ\rbrace
\nonumber\\
      {\cal O}_{1/2} &\equiv& \lbrace  m ~{\rm odd},~ n\in\IZ+\half\rbrace~.
\eeqn
Note that under $T:\tau\to \tau+1$, $\calE_0\pm \calE_{1/2}$ are each invariant
while $\calO_0\pm \calO_{1/2}$ are exchanged;
likewise, under $S:\tau\to -1/\tau$, $\calE_0+ \calE_{1/2}$ and
$\calO_0-\calO_{1/2}$ are each invariant
while $\calE_0- \calE_{1/2}$ and
$\calO_0+ \calO_{1/2}$ are exchanged.
Also note that
at the free-fermion radius $a=1/\sqrt{2}$, these four functions respectively
become
\beqn
      {\cal E}_0 &\longrightarrow& \half (\tthreeb \tthree  + \tfourb
\tfour)/\etainv\eta \nonumber\\
      {\cal O}_{1/2} &\longrightarrow& \half (\tthreeb \tthree  - \tfourb
\tfour)/\etainv\eta \nonumber\\
      {\cal O}_{0} &\longrightarrow& \half (\ttwob \ttwo  +
\overline{\vartheta}_1 \vartheta_1)/\etainv\eta
              \nonumber\\
      {\cal E}_{1/2} &\longrightarrow& \half (\ttwob \ttwo  -
\overline{\vartheta}_1 \vartheta_1)/\etainv\eta ~.
\label{fflimit}
\eeqn
This explicitly demonstrates the equivalence between a boson and a Dirac
fermion at this radius.
Finally, note that while the combination $\calE_0 + \calO_0$ is invariant
under $a\to 1/a$ (since this is the original combination that includes the
contributions from
all integer momentum and winding modes),
the other combinations of these functions break this symmetry explicitly.
Thus only those string models whose partition functions contain this
combination alone
are self-dual under $a\to 1/a$.

It is straightforward to take the limits of these functions as $a\to 0$ and
$a\to \infty$, for in each limit the spectrum of momentum modes or winding
modes
become dense and the summation over these modes can be replaced by an integral.
We then find the limiting behaviors
\beqn
     a\to 0:&~~~~~~~~& \calE_0, \calO_0\to
(2a\sqrt{\tau_2}\,\etainv\eta)^{-1}\,;
                  ~~ \calE_{1/2}, \calO_{1/2}\to 0 \nonumber\\
     a\to\infty:&~~~~~~~~& \calE_0, \calE_{1/2}\to
a(\sqrt{\tau_2}\,\etainv\eta)^{-1}\,;
                  ~~ \calO_0, \calO_{1/2}\to 0 ~.
\label{limits}
\eeqn
Since $(\sqrt{\tau_2}\etainv\eta)^{-1} = Z_{\rm boson}^{(1)}$ is the
partition function of a single uncompactified boson, we see that the relations
(\ref{limits}) permit us to obtain the partition
functions of ten-dimensional string models as the limits of those in nine
dimensions.
In this connection,
note that
the partition functions of ten-dimensional theories
are generally obtained as the limits
of those of lower-dimensional theories
via a relation of the form
\beq
            Z^{(10)} ~\equiv~ \lim_{V\to \infty} \,{ 1\over \calM^D  V }
\,Z^{(10-D)}
\label{partvol}
\eeq
where $V$ represents the effective volume of the $D$-dimensional
compactification
and where $\calM$ is the compensating mass scale
\beq
       \calM~\equiv~(4\pi^2 \alpha')^{-1/2}~.
\label{Mdef}
\eeq
The volume factor in (\ref{partvol}) then absorbs the divergent factors of $a$
in (\ref{limits}) as $a\to 0$ or $a\to \infty$.

\subsection{Nine-dimensional heterotic interpolating models:  General
construction}

Given these definitions,
we can now construct our nine-dimensional
interpolating models.

As discussed in Sect.~3.1, for every pair of
ten-dimensional models $M_1$ and $\tilde M_2$ that are related to
each other via $\IZ_2$ orbifold,
there exists a corresponding
nine-dimensional interpolating model $M^{(9)}$
which reproduces $M_1$ as $R\to \infty$ and $\tilde M_2$ as $R\to 0$.
Thus, we shall say that $M^{(9)}$ interpolates between the ten-dimensional
models $M_1$ and $M_2$.
We shall assume, for simplicity, that both $M_1$ and $M_2$ are $T$-selfdual,
so that $\tilde M_1=M_1$ and $\tilde M_2=M_2$.
This is true for all ten-dimensional heterotic string models.
We shall discuss the Type~II case in the next section.

Following the procedure outlined in Sect.~3.1,
we begin with the untwisted compactification of $M_1$ on a circle
of radius $R$.
If Model~$M_1$ has partition function
\beq
             Z_{M_1} ~=~ Z_{\rm boson}^{(8)}  ~Z^+_+~,
\label{origpart}
\eeq
then this compactification results in a nine-dimensional model
with partition function
\beq
        Z^{(9)+}_+ ~=~ Z_{\rm boson}^{(7)}~(\calE_0 + \calO_0)~ Z^+_+~.
\label{z9one}
\eeq
As a check,
note that this
expression indeed reproduces (\ref{origpart}) in the limits
$a\to 0$ and $a\to \infty$.
Specifically, for $a\to 0$ (or $R\to \infty$),
the volume of compactification is $V= 2\pi R = 2\pi a^{-1} \sqrt{\alpha'}$,
while for $a\to \infty$,
the effective volume of compactification is $V=  2\pi a \sqrt{\alpha'}$.

Let us now orbifold this theory by $\calT Q$ where
$\calT$ is defined in (\ref{Tdef}) and $Q$ is the action such that $M_2$
is the $Q$-orbifold of $M_1$.
Clearly, while $\calT$ acts on the compactification sums $\calE$ and $\calO$,
$Q$ acts on the purely internal part $Z^+_+$.
Specifically, since $\calT$ represents only half of a complete translation
around the
compactified direction, the states which are invariant under $\calT$ are
those with even integer momentum quantum numbers.
Thus, at the level of the partition
function, the states that contribute to $\calE_{0}$ are even under $\calT$
while those that contribute to $\calO_{0}$ are odd.
Thus, in order to project onto the states invariant under $\calT Q$, we
add to (\ref{z9one})
the contributions from the projection sector
\beq
        Z^{(9)-}_+ ~=~ Z_{\rm boson}^{(7)}~(\calE_0 - \calO_0)~ Z^-_+~.
\eeq
Here $Z^-_+$ is the $Q$-projection sector of the internal contribution $Z^+_+$.
In the usual fashion, modular invariance then requires us
to add the contribution from the twisted sector
\beq
        Z^{(9)+}_- ~=~ Z_{\rm boson}^{(7)}~(\calE_{1/2} + \calO_{1/2})~ Z^+_-
\eeq
as well as its corresponding projection sector
\beq
        Z^{(9)-}_- ~=~ Z_{\rm boson}^{(7)}~(\calE_{1/2} - \calO_{1/2})~ Z^-_-~.
\eeq
The net result, then, is a nine-dimensional
model $M^{(9)}$ with total partition function
\beq
      Z^{(9)} ~=~  {1\over 2}\, Z_{\rm boson}^{(7)} \,
   \biggl\lbrace
      \calE_0\,(Z^+_+ + Z^-_+) ~+~
      \calE_{1/2}\,(Z^+_- + Z^-_-) ~+~
      \calO_0\,(Z^+_+ - Z^-_+) ~+~
      \calO_{1/2}\,(Z^+_- - Z^-_-) \biggr\rbrace ~.
\label{mainresult}
\eeq

Using the relations (\ref{limits}),
we can now determine what ten-dimensional models
are reproduced in the $R\to\infty$ and $R\to 0$ limits.
In the $R\to \infty$ limit ({\it i.e.}, as $a\to 0$),
we find that we immediately obtain the original partition function
(\ref{origpart})
upon recognizing that the effective volume of compactification
in this case is given by $V= 2\pi(R/2) = \pi \sqrt{\alpha'}/a$.
This factor of two in the effective radius
simply reflects the fact that our original orbifold projection
onto states with even momentum quantum numbers is equivalent
to halving the effective radius of compactification.\footnote{
    In general, the effective volume of compactification can be
    easily determined by demanding that the resulting ten-dimensional
    partition function have the correct overall normalization.  Such
    normalizations are unique, since they essentially rescale the numbers of
    states in the string spectrum.  Since we know that there can be only
    one graviton in a consistent string model,
    the term $Z_{\rm boson}^{(8)} \chibar_V \chi_I^2$
    must always appear with coefficient one in the ten-dimensional
    partition function.}
Thus, we see that the $R\to \infty$ limit of Model~$M^{(9)}$
reproduces Model~$M_1$.
Likewise, in the $R\to 0$ limit ({\it i.e.}, the limit $a\to \infty$),
we find that we reproduce the ten-dimensional partition function
\beq
     {1\over 2}\, Z_{\rm boson}^{(8)}\,
       (Z^+_+ + Z_+^- + Z_-^+ + Z_-^-)
\label{ss}
\eeq
where in this case we have identified the effective volume
as $V=2\pi a\sqrt{\alpha'}$.
However, the expression (\ref{ss}) is simply the partition
function of the $Q$-orbifold of Model~$M_1$.
This is therefore
the partition function of Model~$M_2$.

Thus we conclude that Model~$M^{(9)}$, with partition function
given in (\ref{mainresult}), is the desired nine-dimensional
model that interpolates
between $M_1$ as $R\to \infty$ and $M_2$ as $R\to 0$.

\subsection{Nine-dimensional heterotic interpolating models}

Having established the validity of our general procedure,
it is now straightforward to construct our
nine-dimensional interpolating models.
Indeed, since all of the orbifolds in Fig.~\ref{orbrelations}
are $\IZ_2$ orbifolds, for each such relation there exists
a corresponding nine-dimensional interpolating model.
Note that an alternate derivation of each of these
nine-dimensional interpolating models using the free-fermionic
construction appears in the Appendix.

\begin{figure}[thb]
\centerline{\epsfxsize 5.0 truein \epsfbox {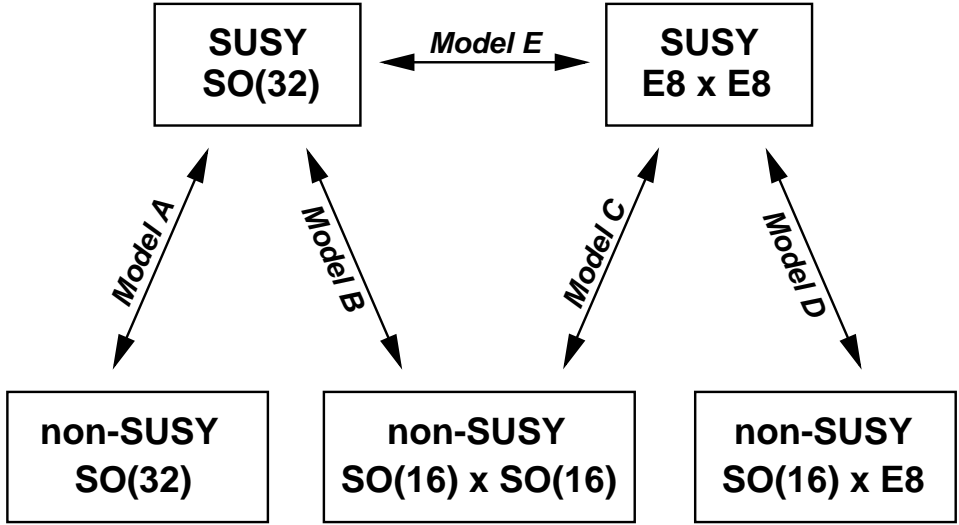}}
\caption{Nine-dimensional Models~A through D interpolate
   between different supersymmetric and non-supersymmetric
   ten-dimensional string models.  In each case, supersymmetry
   is restored in one limit only.  Model~E interpolates between
   the supersymmetric $SO(32)$ and $E_8\times E_8$ models, and
   is therefore supersymmetric at all radii.}
\label{interpfig}
\end{figure}

For the purposes of this paper, the models that will most interest
us are those that interpolate between
a ten-dimensional supersymmetric string model and a ten-dimensional
non-supersymmetric string model.
There are indeed four nine-dimensional models of this type,
and their interpolations are sketched in Fig.~\ref{interpfig}.
We shall refer to these as Models~A through D.
There is also a fifth model which will interest us, but which interpolates
between the supersymmetric $SO(32)$ and $E_8\times E_8$ models.
We shall refer to this as Model~E.

At arbitrary radii, these partition functions of our
nine-dimensional models take the following
forms:
\beqn
    Z_A ~=~  Z^{(7)}_{\rm boson} \,\times \,\bigl\lbrace ~
    \phantom{+}&\calE_0 &  \lbrack
          \chibar_V \,(\chi_I^2 + \chi_V^2)  ~-~  \chibar_S \,(\chi_S^2 +
\chi_C^2)
          \rbrack\nonumber\\
   +&\calE_{1/2}  & \lbrack
           \chibar_I \,(\chi_I\chi_V+\chi_V\chi_I) ~-~ \chibar_C
\,(\chi_S\chi_C+\chi_C\chi_S)
          \rbrack\nonumber\\
   +&\calO_0 & \lbrack
          \chibar_V \,(\chi_S^2+\chi_C^2)  ~-~  \chibar_S \,(\chi_I^2+\chi_V^2)
          \rbrack\nonumber\\
   +&\calO_{1/2} & \lbrack
           \chibar_I \,(\chi_S\chi_C+\chi_C\chi_S) ~-~ \chibar_C
\,(\chi_I\chi_V+\chi_V\chi_I)
          \rbrack ~~\bigr\rbrace~\nonumber\\
     ~&& \nonumber\\
    Z_B ~=~  Z^{(7)}_{\rm boson} \,\times \,\bigl\lbrace ~
    \phantom{+}&\calE_0 &  \lbrack
          \chibar_V \,(\chi_I^2 + \chi_S^2)  ~-~  \chibar_S
\,(\chi_V^2+\chi_C^2)\rbrack\nonumber\\
   +&\calE_{1/2}  & \lbrack
           \chibar_I \,(\chi_V\chi_C+\chi_C\chi_V) ~-~ \chibar_C \,(\chi_I
\chi_S+\chi_S\chi_I)\rbrack\nonumber\\
   +&\calO_0 & \lbrack
          \chibar_V \,(\chi_V^2 + \chi_C^2)  ~-~  \chibar_S
\,(\chi_I^2+\chi_S^2)\rbrack\nonumber\\
   +&\calO_{1/2} & \lbrack
           \chibar_I \,(\chi_I\chi_S+\chi_S\chi_I) ~-~ \chibar_C
\,(\chi_V\chi_C+\chi_C\chi_V)\rbrack
       ~~\bigr\rbrace~\nonumber\\
     ~&& \nonumber\\
    Z_C ~=~  Z^{(7)}_{\rm boson} \,\times \,\bigl\lbrace ~
    \phantom{+}&\calE_0 &  \lbrack
          \chibar_V \,(\chi_I^2 + \chi_S^2)  ~-~  \chibar_S \,(\chi_I\chi_S +
\chi_S\chi_I)
          \rbrack\nonumber\\
   +&\calE_{1/2}  & \lbrack
           \chibar_I \,(\chi_V\chi_C+\chi_C\chi_V) ~-~ \chibar_C \,(\chi_V^2
+\chi_C^2)
          \rbrack\nonumber\\
   +&\calO_0 & \lbrack
          \chibar_V \,(\chi_I\chi_S+\chi_S\chi_I)  ~-~  \chibar_S
\,(\chi_I^2+\chi_S^2)
          \rbrack\nonumber\\
   +&\calO_{1/2} & \lbrack
           \chibar_I \,(\chi_V^2+\chi_C^2) ~-~ \chibar_C
\,(\chi_V\chi_C+\chi_C\chi_V)
          \rbrack ~~\bigr\rbrace~\nonumber\\
     ~&& \nonumber\\
    Z_D ~=~  Z^{(7)}_{\rm boson} \,\times
   \,\bigl\lbrace ~
    \phantom{+}&\calE_0 &  \lbrack
     \chibar_V\,\chi_I ~-~ \chibar_S\,\chi_S
     \rbrack\nonumber\\
   +&\calE_{1/2}  & \lbrack
     \chibar_I\,\chi_V ~-~ \chibar_C\,\chi_C
     \rbrack\nonumber\\
   +&\calO_0 & \lbrack
       \chibar_V\,\chi_S ~-~ \chibar_S\,\chi_I
       \rbrack\nonumber\\
   +&\calO_{1/2} & \lbrack
      \chibar_I\,\chi_C  ~-~ \chibar_C\,\chi_V
  \rbrack ~~\bigr\rbrace~ \times~ (\chi_I+\chi_S)~\nonumber\\
     ~&& \nonumber\\
    Z_E ~=~  Z^{(7)}_{\rm boson}  \, \times \,\bigl\lbrace ~
    \phantom{+}&\calE_0 &  (\chi_I^2 + \chi_S^2)
        \nonumber\\
   +&\calE_{1/2}  & (\chi_I \chi_S+\chi_S\chi_I)
        \nonumber\\
   +&\calO_0 & (\chi_V^2 + \chi_C^2)
        \nonumber\\
   +&\calO_{1/2} & (\chi_V \chi_C+\chi_C\chi_V)
             ~~\bigr\rbrace~\times ~ (\chibar_V-\chibar_S)~. \nonumber\\
   ~&& ~
\label{9dparts}
\eeqn
For each of these partition functions, the partition function
of the corresponding interpolating model
with opposite endpoints can be obtained by exchanging $a\leftrightarrow
a^{-1}$.
A similar effect can also be achieved by exchanging
$a\leftrightarrow (2a)^{-1}$, which is equivalent to
exchanging $\calE_{1/2}\leftrightarrow\calO_0$.

There is an important comment we must make concerning
these nine-dimensional partition functions.
When writing these partition functions, we have continued to use the
characters of the transverse $SO(8)$ Lorentz group for the right-movers.
Strictly speaking, of course, it is improper to use such $SO(8)$ characters,
and instead we should be expressing our partition functions in terms of
level-one $SO(7)$ characters and the characters of a residual Ising model.
Thus, in these partition functions, the $SO(8)$ characters should really be
understood as a shorthand for combinations of $SO(7)$ and Ising-model
characters,
and only in the $R\to \infty$ and $R\to 0$ limits
should they be interpreted as the corresponding $SO(8)$ characters.
There is, however, an important subtlety connected with this
interpretation in the case of $SO(8)$ spinors.
While $SO(8)$ has two distinct
spinor representations (namely the spinor $S$ and conjugate spinor $C$),
$SO(7)$ has only one spinor representation.
Thus, in passing from ten dimensions to nine
dimensions, all chirality information is lost, and one cannot
look to these partition functions in order to determine which
$SO(8)$ spinors emerge in the $R\to \infty$ and $R\to 0$ limits.
Fortunately, in the present heterotic case,
this distinction is immaterial and
does not affect our identification
of the $R\to \infty$ and $R\to 0$ limiting theories.
Moreover, since we have constructed the actual string models and
not merely their partition functions,
we can also directly examine the resulting states and
their representations under the ten-dimensional Lorentz group.
The results of both approaches confirm the identifications shown in
Fig.~\ref{interpfig}.

Note that this issue is also related to the
behavior of the heterotic string models under a (one-dimensional)
$T$-duality transformation from $R=\infty$ to $R=0$.
As explained in Refs.~\cite{DHS,DLP}, this transformation
generally flips the chirality of the spinor ground state,
and can be realized in the partition function by exchanging
$\chibar_S$ with $\chibar_C$ for the right-movers only.
In the case of heterotic strings, however,
this overall chirality is simply a matter of convention,
for there is no physically significant relative $SO(8)$ chirality
between right-movers and left-movers such as there is for Type~II strings.
Thus heterotic string models
are self-dual in the sense that any nine-dimensional heterotic string model
is physically the same whether formulated at $R\to \infty$ or at $R\to 0$.
Consequently, in the heterotic case, it is easy not only to
identify the $R\to 0$ limits of these interpolating models,
but also to interpret these limiting theories as equivalent
ten-dimensional ({\it i.e.}, $R\to \infty$) theories.
We shall see that this issue becomes slightly more subtle for
the Type~II case.

Given the heterotic partition functions given in (\ref{9dparts}),
it is easy to deduce the physical properties of the corresponding models.
For example, it is immediately evident that except for Model~E, each of these
models
is supersymmetric only at one limiting point (corresponding to $R\to \infty$,
or $a\to 0$).
Moreover, Models~B and C are tachyon-free for all values of their radii.
It is also clear that Models~B and C have gauge
symmetry $SO(16)\times SO(16)\times U(1)^2$
for all generic values of their radii, and that there are no
enhanced gauge symmetry points for finite values of $R$.

The massless spectrum of Model~B will be of particular interest to us.
At generic radii $0<R<\infty$, the massless spectrum of this model consists
of
\begin{itemize}
\item  the nine-dimensional gravity multiplet:  graviton, anti-symmetric
tensor, and dilaton;
\item  gauge bosons (vectors) transforming in the adjoint representation
         of $SO(16)\times SO(16)\times U(1)^2$; and
\item  a spinor transforming in the  $(\rep{16},\rep{16})$ representation of
$SO(16)\times SO(16)$,
          with no $U(1)$ charges.
\end{itemize}
There are also extra states that appear in the massless spectrum at certain
 {\it discrete}\/
finite, non-zero radii.  These are
\begin{itemize}
\item  at $R=\sqrt{\alpha'}$:  two spinors with $U(1)$ charges transforming as
singlets under
      $SO(16)\times SO(16)$ ;  and
\item  at $R=\sqrt{2\alpha'}$:  two scalars with $U(1)$ charges transforming in
the
     $(\rep{128},\rep{1})\oplus (\rep{1},\rep{128})$ representation of
\sosixteen.
\end{itemize}
As $R\to 0$, a massive opposite-chirality spinor in the
$(\rep{128},\rep{1})\oplus (\rep{1},\rep{128})$ representation of \sosixteen\
with no $U(1)$ charges becomes massless, and the gauge bosons of the
Kaluza-Klein
$U(1)$ gauge factors disappear (augmenting the gravity multiplet from
nine-dimensional
to ten-dimensional).
This then reproduces the spectrum of the ten-dimensional \sosixteen\ string.

The spectrum of Model~C is similar.
At generic radii $0<R<\infty$, the massless spectrum of this model consists of
\begin{itemize}
\item  the nine-dimensional gravity multiplet:  graviton, anti-symmetric
tensor, and dilaton;
\item  gauge bosons (vectors) transforming in the adjoint representation
         of $SO(16)\times SO(16)\times U(1)^2$; and
\item  a spinor transforming in the
     $(\rep{128},\rep{1})\oplus (\rep{1},\rep{128})$
     representation of $SO(16)\times SO(16)$,
          with no $U(1)$ charges.
\end{itemize}
Extra states appearing at discrete radii are
\begin{itemize}
\item  at $R=\sqrt{\alpha'}$:  two spinors with $U(1)$ charges transforming as
singlets under
        $SO(16)\times SO(16)$;  and
\item  at $R=\sqrt{2\alpha'}$:  two scalars with $U(1)$ charges transforming in
the
        $(\rep{16},\rep{16})$ representation of \sosixteen.
\end{itemize}

\begin{figure}
\centerline{\epsfxsize 3.0 truein \epsfbox {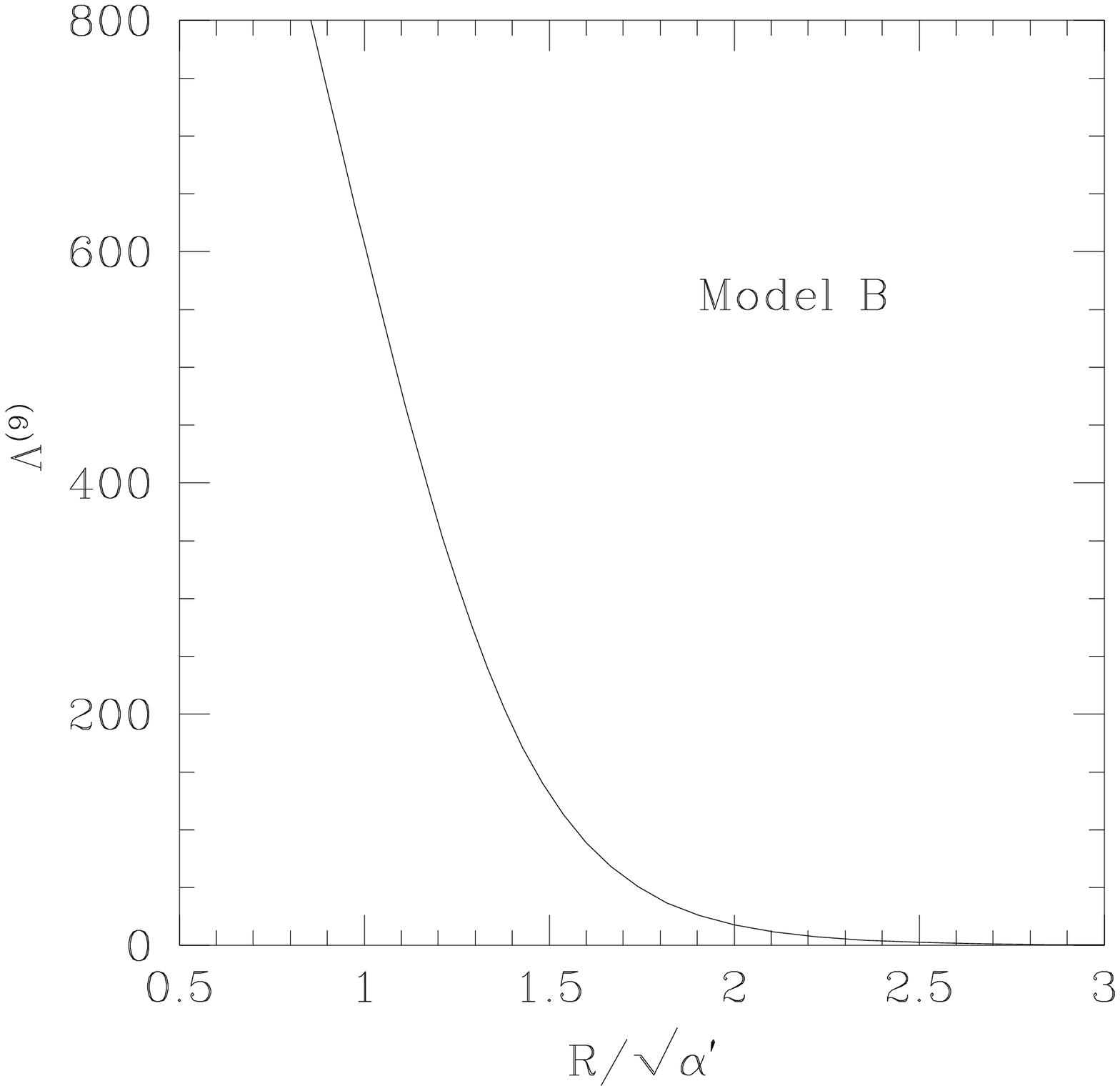}}
\caption{
  The one-loop cosmological constant
    $\Lambda^{(9)}$
   of Model~B, plotted in units of $\half\calM^{9}$, as a function of the
radius
   $R$ of the compactified dimension.
   This model reproduces the supersymmetric $SO(32)$ heterotic string
    as $R\to \infty$ and the non-supersymmetric \sosixteen\ heterotic
    string as $R\to 0$.  The divergence in $\Lambda^{(9)}$ as $R\to 0$
    reflects the effective decompactification of the tenth dimension.}
\label{firstlambdaplot}
\vskip 0.22 truein
\centerline{\epsfxsize 3.0 truein \epsfbox {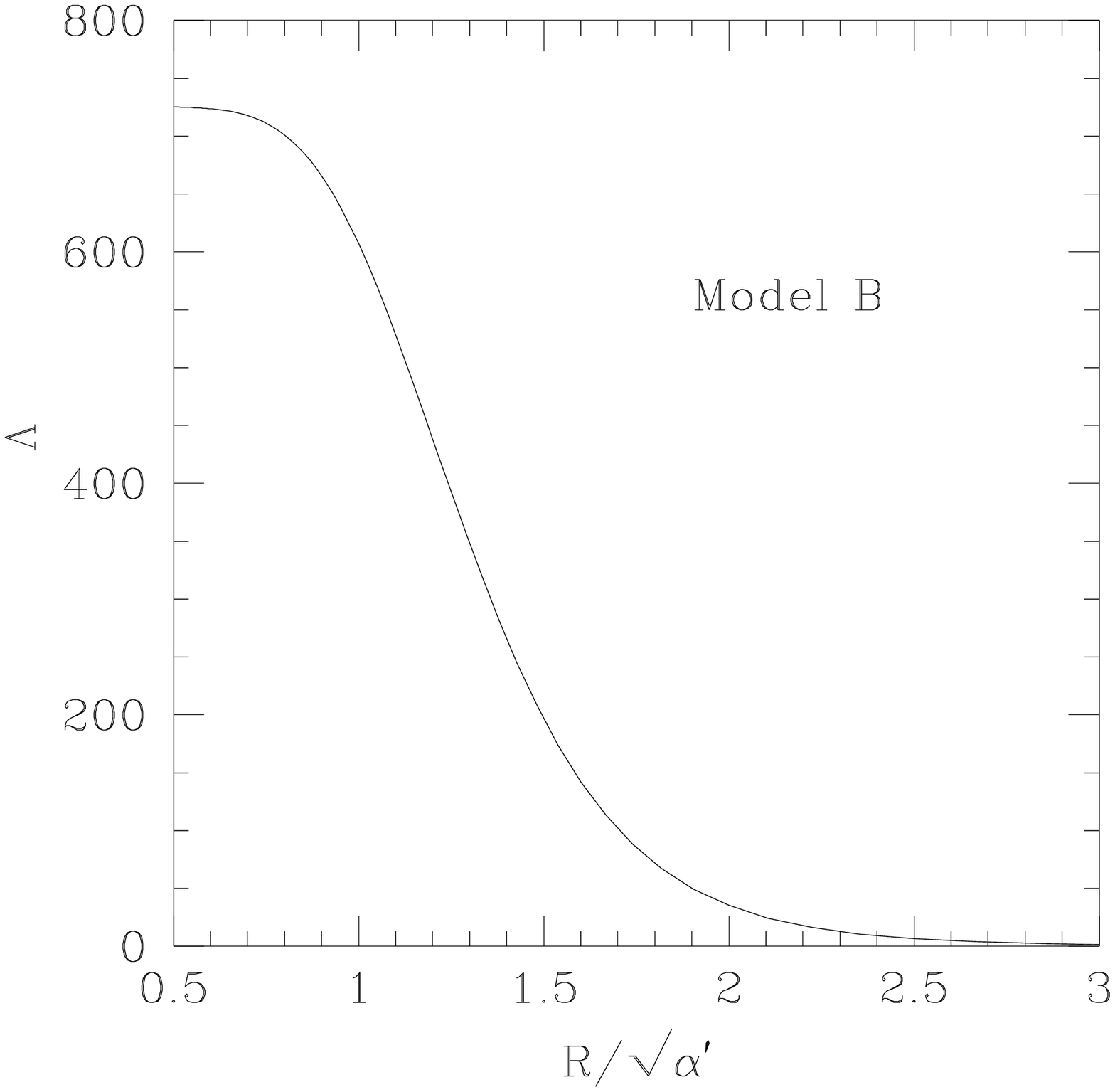}}
\caption{
  The one-loop cosmological constant
    $\tilde \Lambda$
   of Model~B, plotted in units of $\half\calM^{10}$, as a function of the
radius
   $R$ of the compactified dimension.
   This model reproduces the supersymmetric $SO(32)$ heterotic string
    as $R\to \infty$ and the non-supersymmetric \sosixteen\ heterotic
    string as $R\to 0$.  The $R\to 0$ limiting value
$\tilde\Lambda/(\half\calM^{10})
    \approx 725$ agrees with the ten-dimensional cosmological constant
   $(2\pi \alpha')^5 \Lambda^{(10)} \approx 0.0371 $
    originally calculated for the \sosixteen\ string in Ref.~\cite{DH}.}
\label{interplambda}
\end{figure}

We can dramatically illustrate the interpolation properties of these
models by calculating their one-loop vacuum amplitudes
(or cosmological constants) as functions of their radii $R$.
Note that such radius-dependent cosmological constants are similar
to those calculated in Refs.~\cite{IT,GV}.
For this purpose we shall focus on Model~B, which interpolates between
the supersymmetric $SO(32)$ theory as $R\to \infty$ and the
non-supersymmetric \sosixteen\ theory as $R\to 0$.
(We shall discuss Model~A in Sect.~6.1.)
For arbitrary radius $R$, it is relatively straightforward to calculate the
one-loop vacuum amplitude of this nine-dimensional model
\beq
     \Lambda^{(9)}(R) ~=~  -\,\half \,\calM^9\,
            \int_{\cal F} {d^2 \tau \over ({\rm Im}\, \tau)^2} \,Z(\tau,R)
\eeq
where $\calF$ is the fundamental domain of the modular group and
$\calM$ is the overall scale given in (\ref{Mdef}).
The result is shown in Fig.~\ref{firstlambdaplot}.
Note that the divergence in $\Lambda^{(9)}$ as $R\to 0$
reflects the effective decompactification of the tenth dimension
in this limit.

While $\Lambda^{(9)}$ is the actual nine-dimensional cosmological constant,
this quantity is not particularly suitable for examining the
$R\to \infty$ and  $R\to 0$ limits where
the theory effectively becomes ten-dimensional.
Specifically, if a given nine-dimensional model interpolates between
ten-dimensional models $M_1$ and $M_2$ with ten-dimensional cosmological
constants $\Lambda_1$ and $\Lambda_2$ respectively, we would like to define
a new radius-dependent quantity, which we shall denote $\tilde \Lambda(R)$,
with the property that
\beq
       \lim_{R\to \infty}  \tilde\Lambda(R) ~=~ \Lambda_1
               ~~~~~~{\rm and}~~~~~~
       \lim_{R\to 0}  \tilde\Lambda(R) ~=~ \Lambda_2~.
\eeq
This would most appropriately demonstrate the interpolating nature of
the model.  In the cases for which the $R\to \infty$ limiting model $M_1$
is supersymmetric (so that $\Lambda_1=0$), the only subtlety is
the $R\to 0$ limit.  We therefore define
\beq
     \tilde \Lambda (R) ~\equiv~ \calM\, {R\over \sqrt{\alpha'}}\,
\Lambda^{(9)}(R)~.
\label{tildelambdadef}
\eeq
Note that this extra factor $\calM R/\sqrt{\alpha'}$
is simply $1/V$ where $V=2\pi(\alpha'/R)$ is the effective volume of
compactification in the $R\to 0$ limit.
This volume factor thus absorbs
the divergence that arises in the $R\to 0$ limit, as
indicated in (\ref{partvol}), and produces an extra power
of the mass scale $\calM$, as appropriate for a (dimensionful) cosmological
constant as an extra dimension develops.
In Fig.~\ref{interplambda}, we show the behavior of
$\tilde\Lambda$ as a function of $R$.
It is evident from this figure that as $R\to \infty$,
 $\tilde\Lambda$ approaches zero;
this reflects the supersymmetry that re-emerges in this limit.
Likewise, as $R\to 0$, we see that
 $\tilde\Lambda$ approaches a fixed positive value ($\tilde \Lambda/(\half
\calM^{10})\approx
725$) which can be identified as the one-loop
cosmological constant of the ten-dimensional \sosixteen\ string.
Moreover, for intermediate values of $R$, we see that
Model~B smoothly interpolates between these two limits.

The existence of Models~B and C thus shows that it is possible
to connect the non-supersymmetric
ten-dimensional $SO(16)\times SO(16)$ model to the
supersymmetric ten-dimensional $SO(32)$ and $E_8\times E_8$
models through {\it continuous}\/ deformations.
The existence of Model~B, in particular,
suggests that a natural method of finding a strong-coupling dual for the
$SO(16)\times SO(16)$ string might be to start with the known
heterotic/Type~I duality relation 
for the $SO(32)$ theory, and then to ``deform'' away from this duality relation
in a continuous manner through nine dimensions.
Specifically, one would hope to derive a dual for the $SO(16)\times SO(16)$
string
by starting with the ten-dimensional supersymmetric $SO(32)$ Type~I string,
compactifying this theory on a circle of radius $R$ along with an appropriate
Wilson line (to match the Wilson-line effects that are incorporated on the
heterotic
side within Model~B), and then taking the $R\to 0$ limit.
As we shall see, this is roughly the correct procedure.

\subsection{Type II interpolating models}

In order for these nine-dimensional heterotic interpolations to be useful
for exploring duality relationships, there must be analogous nine-dimensional
interpolations between Type~I models.  Since Type~I models can generally
be realized as orientifolds of Type~II models, we therefore first search for
nine-dimensional interpolations
between supersymmetric and non-supersymmetric ten-dimensional Type~II models.

\begin{figure}[thb]
\centerline{\epsfxsize 5.0 truein \epsfbox {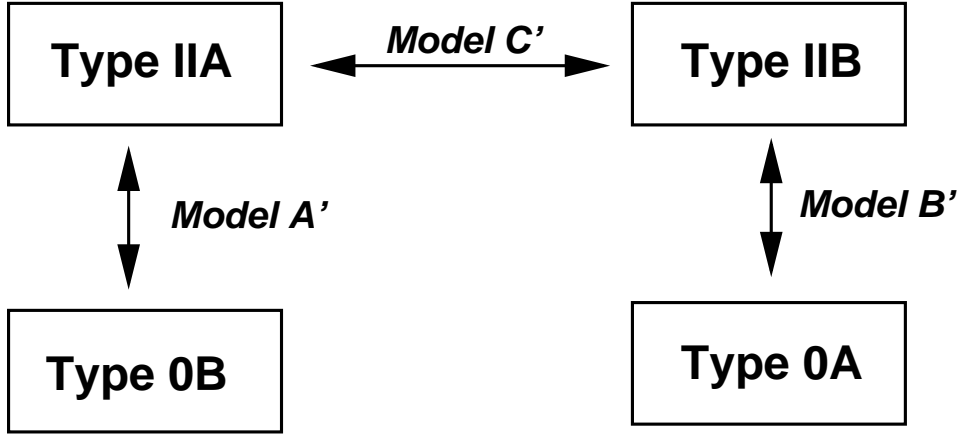}}
\caption{Nine-dimensional Type~II Models~${\rm A}^\prime$ and ${\rm B}^\prime$
   interpolate between the different supersymmetric and non-supersymmetric
   ten-dimensional Type~II string models as shown.  Model~C$^\prime$, by
contrast,
    interpolates between the supersymmetric Type~IIA and Type~IIB string
models.}
\label{interpfig2}
\end{figure}

Remarkably, we find that there exist two Type~II interpolating models of this
sort.
As we shall see, they are completely analogous to the interpolating models
on the heterotic side.
We shall refer to these nine-dimensional Type~II models as
Models~${\rm A}^\prime$ and ${\rm B}^\prime$, and their explicit free-fermionic
construction is presented in the Appendix.
They interpolate between the four ten-dimensional Type~II models
as indicated in Fig.~\ref{interpfig2}.
Likewise, in this figure we also include Model~C$^\prime$, which interpolates
between the supersymmetric Type~IIA and Type~IIB strings.

\vfill\eject

Models~${\rm A}^\prime$ and ${\rm B}^\prime$
have the following partition functions:
\beqn
    Z_{A'} ~=~  Z^{(7)}_{\rm boson} \,\times \,\bigl\lbrace ~
    \phantom{+}&\calE_0 &  \lbrack
    \chibar_V\chi_V + \chibar_C\chi_S
        \rbrack\nonumber\\
   +&\calE_{1/2}  & \lbrack
     \chibar_I\chi_I + \chibar_S\chi_C
    \rbrack\nonumber\\
   -&\calO_0 & \lbrack
     \chibar_V\chi_S + \chibar_C\chi_V
     \rbrack\nonumber\\
   -&\calO_{1/2} & \lbrack
     \chibar_I \chi_C + \chibar_S\chi_I
  \rbrack ~~\bigr\rbrace~ \nonumber\\
    && \nonumber\\
    Z_{B'} ~=~  Z^{(7)}_{\rm boson} \,\times \,\bigl\lbrace ~
    \phantom{+}&\calE_0 &  \lbrack
     \chibar_V\chi_V + \chibar_S\chi_S
        \rbrack\nonumber\\
   +&\calE_{1/2}  & \lbrack
    \chibar_I\chi_I + \chibar_C\chi_C
    \rbrack\nonumber\\
   -&\calO_0 & \lbrack
      \chibar_V\chi_S + \chibar_S\chi_V
     \rbrack\nonumber\\
   -&\calO_{1/2} & \lbrack
     \chibar_I\chi_C + \chibar_C\chi_I
  \rbrack ~~\bigr\rbrace~ ,
\label{typeIIinterps}
\eeqn
while Model~C$^\prime$ has partition function:
\beq
    Z_{C'} ~=~ Z^{(7)}_{\rm boson} \,
      (\calE_0 +\calE_{1/2})\,(\chibar_V-\chibar_S)\,(\chi_V-\chi_S)  ~.
\label{ModelCprime}
\eeq

As we discussed below (\ref{9dparts}), an important subtlety arises
when identifying the ten-dimensional theories that are produced
via these interpolations in the $R\to 0$ limit.
Specifically, it is important to carefully
determine the chirality of the $SO(8)$ spinors in the $R\to 0$ limit
in order to correctly identify the corresponding ten-dimensional
string models.  As a well-known example of this subtlety,
let us consider the untwisted nine-dimensional compactification
of the Type~IIA or IIB theory which is then orbifolded, as
described in Sect.~3.1, by only the half-rotation $\calT$.
By construction, this does not alter the internal theory, and therefore
as $R\to 0$ we obtain again the Type~IIA or IIB theory formulated at zero
radius.
However, as explained in Refs.~\cite{DHS,DLP}, the Type~IIA (or IIB) theory
at zero radius is equivalent to the Type~IIB (or IIA)
theory at infinite radius, and it is therefore the latter theory
which must be regarded as the effective ten-dimensional
model that is produced in the $R\to 0$ limit.
Indeed, as has become standard terminology,
we would say that such a nine-dimensional compactification
interpolates between the Type~IIA and Type~IIB ten-dimensional theories,
and this is precisely Model~C$^\prime$ as indicated in Fig.~\ref{interpfig2}.
Of course, this subtlety is merely a reflection of the $T$-duality
between the Type~IIA and IIB theories,
and is an issue for us only in this Type~II case
because it is only for Type~II theories that a physically significant
left/right {\it relative}\/ chirality exists which cannot be
absorbed into an overall chirality convention.

Note that this subtlety also explains
our identification of the $R\to 0$ limits
of Models~A$^\prime$ and B$^\prime$ as corresponding to the ten-dimensional
Type~0B and Type~0A theories respectively.
In each case, the original ten-dimensional
$\IZ_2$ orbifold relation exists between
ten-dimensional theories of similar
chirality ({\it i.e.}\/, between IIA and 0A,
and between IIB and 0B).  However,
the Type~0A theory at zero radius is equivalent to the
Type~0B theory at infinite radius, and vice versa.
Thus we find that Models~A$^\prime$ and B$^\prime$ interpolate
between ten-dimensional theories of opposite chiralities,\footnote{
   Note that this conclusion corrects the apparent inconsistency
   in Fig.~1 of Ref.~\cite{DHS}.}
as indicated in Fig.~\ref{interpfig2}.

Given these partition functions, we can again immediately deduce a
number of properties of the corresponding models.
Of most interest to us will be the existence of tachyons in Model~B$^\prime$.
In general, this model can only contain tachyons which contribute
to the term $\calE_{1/2} \chibar_I \chi_I$ in the partition function.
Analyzing this term in detail, however, we find that it contains tachyonic
contributions only for inverse radii $a> a^\ast$
where $a^\ast$ is the critical radius
\beq
         a^\ast ~\equiv~ {1\over 2\sqrt{2}}~.
\label{acrit}
\eeq
Thus, for $a<a^\ast$ there are no tachyons in the theory;
then as $a$ increases (or as $R$ decreases), certain massive scalar
states become lighter;
then at $a=a^\ast$ these massive states become exactly massless;
and finally for $a> a^\ast$ these states become tachyonic.
Of course, as expected,
these tachyonic states settle into the Type~II ground state
as $a\to \infty$, where they become the tachyons
of the Type~0A model.

\begin{figure}[thb]
\centerline{\epsfxsize 4.0 truein \epsfbox {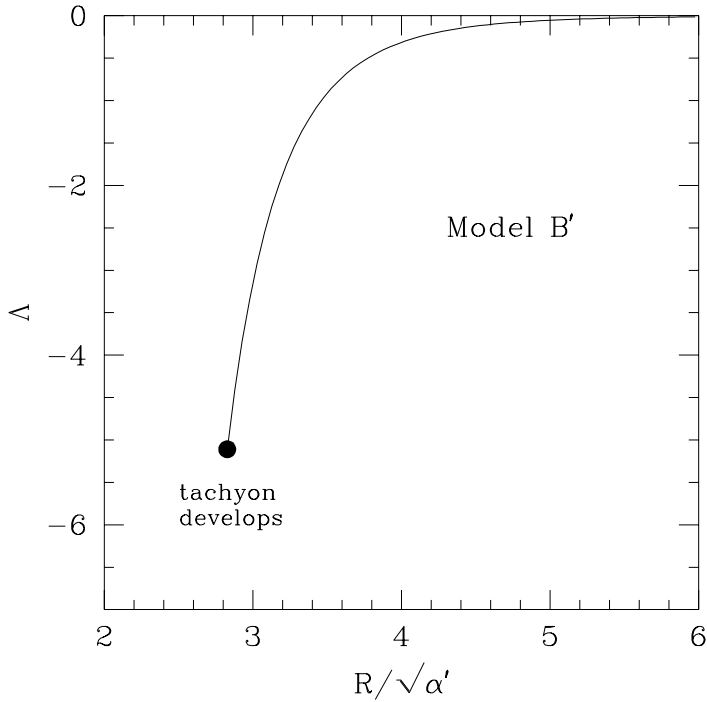}}
\caption{
  The one-loop cosmological constant
    $\tilde \Lambda$
   of Model~B$^\prime$, plotted in units of $\half \calM^{10}$,
   as a function of the radius $R$ of the compactified dimension.
   This model reproduces the supersymmetric Type~IIB string
    as $R\to \infty$ and the non-supersymmetric Type~0A string
    as $R\to 0$.  The one-loop amplitude $\tilde \Lambda$ develops
    a divergence below $R^\ast/\sqrt{\alpha'}\equiv 2\sqrt{2}\approx 2.83$,
    which reflects the appearance of a tachyon in the
    spectrum of Model~B$^\prime$ below this radius.  This discontinuity
    in $\tilde \Lambda$ is indicated with a solid dot at $R=R^\ast$.}
\label{interplambdaII}
\end{figure}

Finally, we can also examine the behavior of the one-loop
vacuum amplitude $\tilde \Lambda$ of
Model~B$^\prime$ as a
function of its radius $R$ of compactification.
As $R\to \infty$, we expect this amplitude to vanish,
reflecting the supersymmetry of the Type~IIB string.
Likewise, as $R$ decreases from this limit,
this amplitude should no longer vanish,
and in fact it should become divergent (to negative infinity)
below $R^\ast/\sqrt{\alpha'} = 2\sqrt{2}$ where Model~B$^\prime$
develops a tachyon.
In Fig.~\ref{interplambdaII}, we show the behavior
of $\tilde \Lambda$ for $R>R^\ast$.
Note that in calculating $\tilde \Lambda$, we have continued
to use the definition in (\ref{tildelambdadef}) with its overall
factor of $R$;  thus Figs.~\ref{interplambda} and \ref{interplambdaII}
may be compared directly.
Also note that even though the one-loop {\it amplitude}\/
has a discontinuity below $R^\ast$, jumping from a finite to an infinite value,
the apparent perturbative spectrum of Model~B$^\prime$ is completely continuous
though this point, with heavy states smoothly becoming massless and then
tachyonic.
This behavior is entirely expected, since
Model~B$^\prime$ represents a smooth interpolation
between the supersymmetric Type~IIB string and the tachyonic
Type~0A string.
Of course, strictly speaking, this interpolating model is not well-defined
below $R^\ast$ due to the appearance of the tachyons,
and hence the apparent perturbative spectrum is not relevant
in that range.

\section{Construction of the $SO(16)\times SO(16)$ Dual}
\setcounter{footnote}{0}

In this section we will explicitly construct the Type~I
theory which is our candidate dual for the non-supersymmetric
tachyon-free $SO(16)\times SO(16)$ heterotic interpolating model.

\subsection{General Approach}

We have seen in the previous section that the \sosixteen\
heterotic string is continuously connected to the $SO(32)$ heterotic string
via a tachyon-free interpolating model.
Since the strong-coupling dual of the $SO(32)$ string
is given by the $SO(32)$ Type~I string, it is natural to
assume that the strong-coupling dual of the \sosixteen\ string
is given by a Type~I theory which is similarly
connected to the $SO(32)$ Type~I theory.

There are then several methods that one might follow towards
constructing this theory.
One option might simply be to
deform the $SO(32)$ Type~I theory in such a way that supersymmetry
is broken and the gauge group is reduced to \sosixteen.
However, it is not necessarily clear how to perform such a deformation
in a unique way while maintaining the consistency of the Type~I theory.
If this had been a deformation of a heterotic string, modular invariance
would have served as a guide as to how any deformation of a given
sector of the theory translates into a corresponding deformation
of a corresponding twisted sector.
However, in the Type~I case, the analogous constraint is tadpole anomaly
cancellation, and there are {\it a priori}\/ many different ways of
deforming a given Type~I theory while maintaining tadpole cancellation.
Not all of these approaches will necessarily yield a consistent Type~I theory.

Indeed, if we take the point of view that a Type~I theory is consistent
if and only if it can be realized as an orientifold (or a generalized
orientifold) of a Type~II theory, then we are naturally led to consider
 {\it reversing}\/ the order of the deformation and the orientifold.
Thus, in this approach, we would first seek to perform our
deformation in the Type~II theory, and subsequently
orientifold the deformed Type~II theory to produce our Type~I model.
This realization of our Type~I model as an orientifold of a consistent Type~II
theory
then guarantees its internal consistency.

Since the $SO(32)$ Type~I theory can be realized as an orientifold of the
Type~IIB theory, we are therefore led to consider
continuous deformations of the Type~IIB theory.
However, as we showed in the last section, there is only one such deformation
of the Type~IIB theory that breaks supersymmetry:  this deformation corresponds
to the nine-dimensional model that interpolates between the Type~IIB and
Type~0A
theories.
Our procedure will therefore be to take the orientifold of this
nine-dimensional
interpolating model.
This will explicitly give us a nine-dimensional Type~I string model
formulated at arbitrary radius, and we can then use this model 
in order to examine
the behavior of the dual theory as a function of the radius.

\subsection{Orientifold procedure}

We begin by briefly reviewing the orientifold procedure.
Since the general orientifolding procedure is completely standard
\cite{orientifolds,Sagnotti},
our main purpose in this section is to establish our notation
and normalizations by presenting the appropriate formulas.
(Readers familiar with the formalism of Type~I string theory are encouraged
to skip to Sect.~4.3.)

In general, a Type~I theory can be specified through its one-loop amplitude.
This has four contributions, two from the closed-string sector
(whose one-loop geometries have the topologies of a torus and Klein bottle
respectively),
and two from the open-string sector (with the topologies of a cylinder and
M\"obius strip).
In each case, the contribution can be
obtained by evaluating a trace over relevant string states and then integrating
over all corresponding conformally inequivalent geometries.
In general, these traces are defined as:
\beqn
      T(\tau)  &\equiv&  {\rm Tr} ~{\textstyle{1\over 2}}\cdot (-1)^F \cdot
 {\rm orb} \cdot
            {\rm GSO}_{L}\cdot {\rm GSO}_{R} \cdot
          e^{2\pi i\tau L_0}
          \, e^{-2\pi i\overline{\tau} \overline{L_0}} \nonumber\\
      K(t)  &\equiv&  {\rm Tr} ~{\textstyle{\Omega\over 2}}\cdot (-1)^F \cdot
 {\rm orb} \cdot
            {\rm GSO}_{L}\cdot {\rm GSO}_{R} \cdot
               e^{-2\pi t (L_0+\overline{L_0})} \nonumber\\
      C(t)  &\equiv&  {\rm Tr} ~{\textstyle{1\over 2}}\cdot (-1)^F \cdot {\rm
orb} \cdot
            {\rm GSO}\cdot
               e^{-2\pi t L_0} \nonumber\\
      M(t)  &\equiv&  {\rm Tr} ~{\textstyle{\Omega\over 2}}\cdot (-1)^F \cdot
 {\rm orb} \cdot
            {\rm GSO}\cdot
               e^{-2\pi t L_0} ~.
\label{fourtraces}
\eeqn
Thus, with these normalizations, $T+K$ gives the trace over the closed-string
states
while $C+M$ gives the trace over the open-string states.
In these traces, $F$ is the spacetime fermion number;
GSO$_{L,R}$ are the closed-string left- and right-moving GSO projection
operators
\beq
        {\rm GSO}_L ~\equiv~ \half\,(1+(-1)^G)~,~~~~
        {\rm GSO}_R ~\equiv~ \half\,(1+(-1)^{\tilde G})~
\eeq
where $G$ and $\tilde G$ are the left- and right-moving $G$-parities (related
to worldsheet fermion number);
`GSO' without any subscripts is the open-string GSO projection operator
$\half(1+G)$
where $G$ is the open-string $G$-parity;
`orb' indicates the orbifold projection operator, which can be expressed
for a $\IZ_2$ orbifold as
\beq
           {\rm orb} ~\equiv~ \half\,(1+g)
\eeq
where $g$ is the $\IZ_2$ orbifold action;
and $\Omega$ is the worldsheet orientation-reversing ``orientifold'' action,
defined
on the individual bosonic excitation modes $\alpha_{-n}$ and fermionic
excitation modes $\psi_{-r}$ as:
\beqn
    {\rm closed~strings:}&&
         \Omega \alpha_{-n} \Omega^{-1} = \tilde \alpha_{-n}~,~
         \Omega \psi_{-r} \tilde \psi_{-s} \Omega^{-1} = \psi_{-s}
\tilde\psi_{-r}~\nonumber\\
    {\rm open~strings:}&&
           \cases{
         \Omega \alpha_{-n} \Omega^{-1} = +e^{-i\pi n} \tilde \alpha_{-n}~,~
         \Omega \psi_{-r} \Omega^{-1} = +e^{-i\pi r} \tilde \psi_{-r} &  in NN
sector \cr
         \Omega \alpha_{-n} \Omega^{-1} = -e^{-i\pi n} \tilde \alpha_{-n}~,~
         \Omega \psi_{-r} \Omega^{-1} = -e^{-i\pi r} \tilde \psi_{-r} &  in DD
sector \cr} \nonumber\\
      ~& &~
\eeqn
where in general $2r\in \IZ$.
In the torus and Klein-bottle contributions, the traces are to be taken over
both the
untwisted and twisted sectors of the $g$-orbifold,
while in the cylinder and M\"obius contributions, the traces are instead taken
over all
Chan-Paton factors.
Throughout,
the closed-string operators $L_0$ and $\overline{L_0}$ are the left- and
right-moving worldsheet energies
which are defined and normalized in the standard way such that
\beq
      L_0+\overline{L_0}~=~ N_L + \tilde N_R +
      \half \alpha' (p^2_{\rm n.c.}  + p^2_{\rm c.}) ~+~ {\rm VE}
\label{closednormalization}
\eeq
where ${\rm VE}={\rm VE}_L + {\rm VE}_R = -1$ indicates the closed-string
vacuum energy,
where $N$ are oscillator numbers, and where $p_{\rm n.c.}$ (respectively
$p_{\rm c.}$) are
momenta in non-compact (respectively compact) directions.
In the case of a single circle-compactified dimension, we have $p^2_{\rm c.} =
p_L^2 + p_R^2$
where $p_{L,R}$ are defined in (\ref{compmomenta}).
By contrast, the open-string energy operator $L_0$ is normalized as
\beq
       L_0 ~=~ N + \alpha' (p^2_{\rm n.c.}  + p^2_{\rm c.})~.
\label{opennormalization}
\eeq
Note that in the Klein-bottle trace,
the $\Omega$ projection enforces $L_0=\overline{L_0}$.

Once these traces are calculated, the corresponding one-loop amplitudes are
then determined by integrations over the appropriate modular parameters.
Because the precise relative normalizations of these integrals will be
important for
what follows, we shall now briefly review the derivation of these integrals.
Recall that the general field-theoretic expression for the one-loop amplitude
(cosmological constant) in $D$ dimensions is given by
\beqn
       \Lambda &=& \half \sum_i \,(-1)^F\, \int {d^D p \over (2\pi)^D}
\,\log(p^2+M_i^2)\nonumber\\
    &=&  -\half \sum_i \,(-1)^F\, \int {d^D p \over (2\pi)^D} \,
           \int_0^\infty {d\hat t\over \hat t} e^{-(p^2+M_i^2)\hat
t}\nonumber\\
    &=&  -{1\over 2} \,{1\over (4\pi)^{D/2}}\, \sum_i \,(-1)^F\,
       \int_0^\infty {d\hat t\over \hat t^{1+D/2}} e^{-M_i^2 \hat t}~.
\label{Lambdafield}
\eeqn
In these expressions, we have summed over the contributions from bosonic and
fermionic
states with masses $M_i$, and we have used a representation in terms of a
Schwinger proper-time parameter $\hat t$.  Also, in passing to the final line,
we have explicitly performed the integral over the $D$ uncompactified momenta.
Given this form for the field-theoretic result, it is then easy to write down
the
corresponding string-theory amplitudes.  Indeed, the only crucial issue is that
of correctly
identifying $\hat t$ with the appropriate modular parameter for the different
Type~I genus-one surfaces.
In order to do this, we must first identify the spacetime mass $M_i$
of each state in terms of $L_0$ and $\overline{L_0}$.  Fortunately, these
latter
identifications are standard, following from (\ref{closednormalization}) and
(\ref{opennormalization}), and are given as:
\beqn
                {\rm closed}:&~~~ L_0 + \overline{L_0}  &=~ \half \alpha' M_i^2
 \nonumber\\
                {\rm open}:&~~~~~~~~~~~ L_0  &=~  \alpha' M_i^2~.
\eeqn
Given the form of the traces in (\ref{fourtraces}) and comparing with
(\ref{Lambdafield}),
it is then easy to identify the one-loop Type~I modular parameters relative to
$\hat t$:
\beqn
             {\rm torus}:&&~~~  {\rm Im}\,\tau ~=~ \hat t/(\pi \alpha')
\nonumber\\
             {\rm Klein~bottle}:&&~~~  t ~=~ \hat t/(\pi \alpha') \nonumber\\
             \hbox{cylinder,~M\"obius}:&&~~~  t ~=~ \hat t/(2\pi \alpha')~.
\eeqn
Inserting these expressions for $\hat t$ into (\ref{Lambdafield}),
we then obtain the proper normalizations for our amplitudes:
\beqn
        \Lambda_T &=&   -\, \half\, \calM^D\,
                       \int_{\calF} {d^2\tau\over ({\rm Im}\,\tau)^{1+D/2}}
\,T'(\tau) \nonumber\\
        \Lambda_K &=&   -\,\half\,  \calM^D\,  \int_0^\infty \, {dt\over
t^{1+D/2}}\,K'(t) \nonumber\\
        \Lambda_C &=&   -\,\half\,  2^{-D/2}\, \calM^D\,  \int_0^\infty\,
 {dt\over t^{1+D/2}}\, C'(t) \nonumber\\
        \Lambda_M &=&   -\,\half\,  2^{-D/2}\, \calM^D\,  \int_0^\infty\,
 {dt\over t^{1+D/2}}\, M'(t)~.
\label{relativenormalizations}
\eeqn
Here $\calF$ is the fundamental domain of the modular group,
$\calM$ is the overall normalization scale defined in (\ref{Mdef}), and the
primes on the integrands
indicate that these traces {\it no longer include}\/ the integrations over
non-compact momenta.
Thus, calculating the total one-loop amplitude requires summing these four
contributions
with the relative normalizations given in (\ref{relativenormalizations}).
Note that since the torus trace $T(\tau)$ as defined in (\ref{fourtraces}) is
half the
one-loop torus partition function $Z(\tau)$ of the corresponding Type~II string
model,
we find that $\Lambda_T=\half\Lambda$ where $\Lambda$ is the cosmological
constant
of the corresponding Type~II string model.
This is of course expected, since there are only half as many Type~I states
that propagate in
the torus of a Type~I model as there are in the corresponding Type~II model
prior to orientifolding.

Thus, given these definitions,
the procedure for constructing a given Type~I model as an orientifold of a
given Type~II model is relatively simple.  We start with the Type~II model
(whose partition
function is identified as $Z=2T$) along with its associated GSO and orbifold
projections.
Given this information, we then perform the Klein-bottle, cylinder, and
M\"obius-strip traces, and take the limit as $t\to 0$ of these three traces in
order to
determine the massless Ramond-Ramond and NS-NS tadpoles.
Imposing a cancellation of these tadpoles then determines the gauge structure
of the theory, and ultimately yields a set of conditions on the Chan-Paton
factors
(or equivalently on the numbers of Dirichlet nine-branes and anti-nine-branes
in
the theory).  The spectrum of the resulting Type~I theory can then be
determined by examining
the resulting traces subject to these constraints, and its one-loop amplitudes
are calculated
as described above.

\subsection{The orientifold of Model~B$^\prime$}

Having reviewed the general orientifolding procedure,
let us now specialize to the case at hand:  the construction of the orientifold
of the nine-dimensional Type~II Model~B$^\prime$ whose partition function is
given
in (\ref{typeIIinterps}).
Recall that this model can be realized
by starting from
the untwisted circle-compactified Type~IIB theory
whose partition function is
\beq
            Z ~=~ Z^{(7)}_{\rm boson}\, (\calE_0 + \calO_0)\,
                   (\chibar_V-\chibar_S)\,(\chi_V - \chi_S)~,
\eeq
and then orbifolding by the $\IZ_2$ action
\beq
          g ~\equiv ~  \calT  (-1)^F~
\eeq
where $F\equiv F_L+F_R$ is the Type~II spacetime fermion number and where
$\calT$ is the half-rotation given in (\ref{Tdef}).

Our first step is to consider the torus contribution to the one-loop amplitude.
This torus contribution $\Lambda_T$ is given in (\ref{relativenormalizations})
where the torus trace is half the corresponding Type~II partition function
of Model~B$^\prime$ as given in (\ref{typeIIinterps}).  Thus we immediately
have
\beqn
            T'(\tau) ~=~ \half\,(\eta \overline{\eta})^{-7} \,\times
\,\bigl\lbrace ~
    \phantom{+}&\calE_0 &  \lbrack
     \chibar_V\chi_V + \chibar_S\chi_S
        \rbrack\nonumber\\
   +&\calE_{1/2}  & \lbrack
    \chibar_I\chi_I + \chibar_C\chi_C
    \rbrack\nonumber\\
   -&\calO_0 & \lbrack
      \chibar_V\chi_S + \chibar_S\chi_V
     \rbrack\nonumber\\
   -&\calO_{1/2} & \lbrack
     \chibar_I\chi_C + \chibar_C\chi_I
  \rbrack ~~\bigr\rbrace~ .
\label{torustrace}
\eeqn

We now evaluate the remaining Klein-bottle, cylinder, and M\"obius-strip
traces.
To this end, it proves convenient to define the functions \cite{PC}
\beqn
      f_1(q) \equiv&  \eta(q^2) &=~ q^{1/12}\,\prod_{n=1}^\infty\, (1-q^{2n})
\nonumber\\
      f_2(q) \equiv&  \sqrt{\vartheta_2(q^2)\over \eta(q^2)} &=~
                   \sqrt{2}\,q^{1/12}\,\prod_{n=1}^\infty\, (1+q^{2n})
\nonumber\\
      f_3(q) \equiv&  \sqrt{\vartheta_3(q^2)\over \eta(q^2)} &=~
                             q^{-1/24}\,\prod_{n=1}^\infty\, (1+q^{2n-1})
\nonumber\\
      f_4(q) \equiv&  \sqrt{\vartheta_4(q^2)\over \eta(q^2)} &=~
                             q^{-1/24}\,\prod_{n=1}^\infty\, (1-q^{2n-1})~
\label{fdefs}
\eeqn
which satisfy the so-called ``abstruse'' identity
\beq
             f_3^8(q) ~=~ f_2^8(q) ~+~ f_4^8(q)~.
\label{abstruse}
\eeq
Each of the traces $K(t)$, $C(t)$, and $M(t)$ can then be evaluated in terms of
these functions.

We begin with the Klein-bottle trace $K'(t)$, and define $q\equiv e^{-2\pi t}$.
Given the definition in (\ref{fourtraces}), we see that this trace can
naturally
be decomposed into two terms according to how the GSO projections are
performed:
\beq
          {\rm GSO}_L \,{\rm GSO}_R
         ~=~ {1\over 4}\,
        \left\lbrack 1+(-1)^G\right\rbrack \,\left\lbrack 1+(-1)^{\tilde
G}\right\rbrack
         ~=~ {1\over 2}  ~+~ {1\over 2} \,(-1)^{G}~.
\label{GSOdecomp}
\eeq
In passing to the final expression we have used the fact that the
orientation-reversing
projection $\Omega$ ensures that $G=\tilde G$ for all contributing states.
We can therefore consider the two final terms in (\ref{GSOdecomp}) separately.
Since these two projections respectively correspond \cite{PC} to NS-NS and
Ramond-Ramond
states in the corresponding tree amplitude, we shall refer to the different
contributions
to the total trace as $K'_{\rm NSNS}(t)$ and $K'_{\rm RR}(t)$, with
\beq
              K'(t) ~=~ K'_{\rm NSNS}(t) ~-~ K'_{\rm RR}(t)~.
\eeq
Calculating these traces for our model, we then find that $K'_{\rm
RR}(t)=K'_{\rm NSNS}(t)$,
where
\beq
       K'_{\rm NSNS}(t)  ~=~
       {1\over 8}\, { f_4^8(q)\over f_1^8(q) } \,
         \left\lbrace \sum_{m=-\infty}^\infty\, [1+(-1)^m]\,q^{m^2 a^2/2}
\right\rbrace~.
\label{Ktrace}
\eeq
The leading factor of $1/8$ reflects the three factors of $1/2$ in
(\ref{fourtraces}).
As an aside, note that if we define
$\tau_K \equiv 2it$ and $q_K\equiv e^{2\pi i\tau_K}$, then
this total expression for $K'(t)$ can be equivalently written in terms of our
usual $SO(8)$ characters as
\beq
         K'(t)~=~ {1\over 2}\,
             {1\over \eta^7(q_K)}\, \biggl\lbrack \chi_V(q_K) -\chi_S(q_K)
\biggr\rbrack\,
                  \hat \calE_0 (q_K)
\eeq
where
\beq
         \hat \calE_0(q_K) ~\equiv ~
           {1\over \eta(q_K)} \, \sum_{m\,{\rm even}} \, (q_K)^{m^2 a^2/4}~.
\eeq
This form for the trace, which is reminiscent of the
notation used by Sagnotti \cite{orientifolds,Sagnotti},
makes it clear how the Klein-bottle
trace explicitly implements the orientifold projection on the torus trace:
the $\Omega$ projection removes all states with non-zero
winding modes as well as all sectors with different left- and right-moving
$SO(8)$ characters,
and produces a total Klein-bottle trace which is, in some sense, the ``square
root'' of
the first line of (\ref{torustrace}).
Note that this comparison between the torus and Klein-bottle traces implicitly
implies the
relation $\tau_K= 2i ({\rm Im}\,\tau)$, so that $q_K=\overline{q_T} q_T$ where
$q_T=e^{2\pi i\tau}$.
Such a rewriting can be performed for each of the remaining traces that we will
consider, but we shall not do so here.

The cylinder and M\"obius traces can be calculated similarly.
In the cylinder case we shall again split our trace into two contributions,
with
GSO insertions of $\half$ and $\half (-1)^G$ respectively,
and in the M\"obius case we shall also split our trace into two contributions,
one from the open-string NS sector and the other from the open-string Ramond
sector.
As for the Klein-bottle trace, we decompose our traces in this way
because in each case they then correspond respectively to the NS-NS or
Ramond-Ramond
sectors of the corresponding tree amplitudes.  We then have
\beq
                C'(t)~=~ C'_{\rm NSNS}(t) - C'_{\rm RR}(t) ~,~~~~
                M'(t)~=~ M'_{\rm NSNS}(t) - M'_{\rm RR}(t) ~,
\eeq
and in terms of the functions (\ref{fdefs}), and with $q\equiv e^{-2\pi t}$,
these traces are as follows:
\beqn
    C'_{\rm NSNS}(t) &=&
     {1\over 8}\, {f_4^8(q^{1/2}) \over f_1^8(q^{1/2})} \,
        ({\rm Tr}\,\gamma_I )^2\,
              \sum_{m= -\infty}^\infty\, q^{m^2 a^2} \nonumber\\
      && ~+~
     {1\over 8}\, \left\lbrack {f_2^8(q^{1/2})+f_3^8(q^{1/2}) \over
f_1^8(q^{1/2})} \right\rbrack \,
        ({\rm Tr}\,\gamma_g )^2\,
              \sum_{m= -\infty}^\infty\, (-1)^m \, q^{m^2 a^2} \nonumber\\
    C'_{\rm RR}(t) &=&
         {1\over 8}\, {f_4^8(q^{1/2}) \over f_1^8(q^{1/2})} \,
       \left\lbrace
        ({\rm Tr}\,\gamma_I )^2\,
              \sum_{m= -\infty}^\infty\, q^{m^2 a^2} ~+~
        ({\rm Tr}\,\gamma_g )^2\,
              \sum_{m= -\infty}^\infty\, (-1)^m \, q^{m^2 a^2}
\right\rbrace\nonumber\\
    M'_{\rm NSNS}(t) &=&
         -\,{1\over 8}\,  \,{f_2^8(q) \, f_4^8(q) \over f_1^8(q) \, f_3^8(q)}
\,\times\,\nonumber\\
        && ~ \left\lbrace
         ({\rm Tr}\,\gamma_\Omega^T \gamma_\Omega^{-1})\,
         \sum_{m= -\infty}^\infty \, q^{m^2 a^2} ~+~
         ({\rm Tr}\,\gamma_{\Omega g}^T \gamma_{\Omega g}^{-1})\,
         \sum_{m= -\infty}^\infty \, (-1)^m\, q^{m^2 a^2}  \right\rbrace
\nonumber\\
    M'_{\rm RR}(t) &=&
         -\,{1\over 8} \,{f_2^8(q) \, f_4^8(q) \over f_1^8(q) \, f_3^8(q)}
\,\times\,\nonumber\\
        && ~ \left\lbrace
         ({\rm Tr}\,\gamma_\Omega^T \gamma_\Omega^{-1})\,
         \sum_{m= -\infty}^\infty \, q^{m^2 a^2} ~-~
         ({\rm Tr}\,\gamma_{\Omega g}^T \gamma_{\Omega g}^{-1})\,
         \sum_{m= -\infty}^\infty \, (-1)^m\, q^{m^2 a^2}  \right\rbrace
 {}~.\nonumber\\
      & & ~
\label{evaluatedtraces}
\eeqn
Here $\gamma_I$, $\gamma_g$, and $\gamma_{\Omega g}$ are respectively the
actions of the identity, the orbifold element $g$, and the orientifold element
$\Omega g$ on the Chan-Paton factors (or equivalently on the nine-branes in the
theory).
In each case the leading factor of $1/8$ reflects the three factors of $1/2$ in
(\ref{fourtraces}).

In order to solve for these $\gamma$ matrices, we must impose the tadpole
anomaly cancellation constraints.
In general, tadpole anomalies appear as divergences in
the amplitudes $\Lambda_{K,C,M}$ from the $t\to 0$ region
of integration, and are interpreted as arising from the
exchange of massless (and possibly tachyonic) string states
in the tree channel.
In order to extract these divergences,
it is simplest to recast our expressions from the loop variable $t$
to the tree variable $\ell$.  With our present conventions,
the tree variable $\ell$ is related to the loop variable $t$ via
\beq
          t~=~\cases{ 1/(4\ell) & Klein bottle\cr
                      1/(2\ell) & cylinder \cr
                      1/(8\ell) & M\"obius strip ~,\cr}
\label{tlconversion}
\eeq
where the different numerical factors reflect, among other things, the
changes in the conventional normalization of the string length
in passing from the open strings of the loop channel to the closed strings
of the tree channel.
The tadpole divergences can then be extracted by taking the $\ell\to \infty$
limit.

Passing to the tree formalism and
taking this limit is facilitated
by defining $\tilde q\equiv e^{-2\pi /t}$
and making use of the identities
\beq
       f_1(q^A)=\sqrt{1\over 2At}\,f_1(\tilde q^{1/4A})~,~~~
       f_2(q^A)=f_4(\tilde q^{1/4A})~,~~~
       f_3(q^A)=f_3(\tilde q^{1/4A})~
\eeq
for the $f$-functions and the Poisson resummation formula
\beq
      \sum_{m=-\infty}^\infty \, (-1)^{Bm} \,q^{Am^2} ~=~
         \sqrt{1\over 2At}\, \sum_{m=-\infty}^\infty \, \tilde q^{(m+B/2)^2/4A}
\eeq
for the circle momentum sums.
We then find that our traces
can be written as
\beqn
       K'_{\rm NSNS}(t)  &=&
        {2 \,t^{7/2}\over a}\,
       { f_2^8(\tilde q^{1/4})\over f_1^8(\tilde q^{1/4}) } \,
         \sum_{m=-\infty}^\infty\, \left(
         \tilde q^{m^2/ 2 a^2} ~+~ \tilde q^{(m+1/2)^2/ 2 a^2} \right)
\nonumber\\
       K'_{\rm RR}(t) &=& K'_{\rm NSNS}(t)   \nonumber\\
    C'_{\rm NSNS}(t) &=&
     {1\over 8}\,
        {t^{7/2}\over \sqrt{2}\, a}\,
         {f_2^8(\tilde q^{1/2}) \over f_1^8(\tilde q^{1/2})} \,
        ({\rm Tr}\,\gamma_I )^2\,
              \sum_{m= -\infty}^\infty\, \tilde q^{m^2/4 a^2} \nonumber\\
      && ~+~
     {1\over 8}\,
        {t^{7/2}\over \sqrt{2}\, a}\,
      \left\lbrack {f_3^8(\tilde q^{1/2})+f_4^8(\tilde q^{1/2}) \over
f_1^8(\tilde q^{1/2})} \right\rbrack \,
        ({\rm Tr}\,\gamma_g )^2\,
              \sum_{m= -\infty}^\infty\, \tilde q^{(m+1/2)^2/ 4a^2} \nonumber\\
    C'_{\rm RR}(t) &=&
         {1\over 8}\,
        {t^{7/2}\over \sqrt{2}\, a}\,
       {f_2^8(\tilde q^{1/2}) \over f_1^8(\tilde q^{1/2})} \,
       \left\lbrace
        ({\rm Tr}\,\gamma_I )^2\,
              \sum_{m= -\infty}^\infty\, \tilde q^{m^2/4 a^2} ~+~
        ({\rm Tr}\,\gamma_g )^2\,
              \sum_{m= -\infty}^\infty\, \tilde q^{(m+1/2)^2/4 a^2}
\right\rbrace\nonumber\\
    M'_{\rm NSNS}(t) &=&
         -\,{1\over 8}\,
        {2^{7/2} \,t^{7/2}\over a}\,
       {f_2^8(\tilde q^{1/4}) \, f_4^8(\tilde q^{1/4}) \over f_1^8(\tilde
q^{1/4}) \, f_3^8(\tilde q^{1/4})}
             \,\times\,\nonumber\\
        && ~ \left\lbrace
         ({\rm Tr}\,\gamma_\Omega^T \gamma_\Omega^{-1})\,
         \sum_{m= -\infty}^\infty \, \tilde q^{m^2/4 a^2} ~+~
         ({\rm Tr}\,\gamma_{\Omega g}^T \gamma_{\Omega g}^{-1})\,
         \sum_{m= -\infty}^\infty \, \tilde q^{(m+1/2)^2/4 a^2}  \right\rbrace
\nonumber\\
    M'_{\rm RR}(t) &=&
         -\,{1\over 8}\,
        {2^{7/2} \,t^{7/2}\over a}\,
       {f_2^8(\tilde q^{1/4}) \, f_4^8(\tilde q^{1/4}) \over f_1^8(\tilde
q^{1/4}) \, f_3^8(\tilde q^{1/4})}
             \,\times\,\nonumber\\
        && ~ \left\lbrace
         ({\rm Tr}\,\gamma_\Omega^T \gamma_\Omega^{-1})\,
         \sum_{m= -\infty}^\infty \, \tilde q^{m^2/4 a^2} ~-~
         ({\rm Tr}\,\gamma_{\Omega g}^T \gamma_{\Omega g}^{-1})\,
         \sum_{m= -\infty}^\infty \, \tilde q^{(m+1/2)^2/4 a^2}  \right\rbrace
 {}~,
\eeqn
and in this form it is straightforward to take the $t\to 0$ (or
$\ell\to\infty$) limit.
Using (\ref{tlconversion}), and assuming $a<\infty$, we find that the leading
behavior of these traces as $\ell\to \infty$ is given by
\beqn
     K'_{\rm NSNS}(\ell) &\sim& \ell^{-7/2}\, a^{-1} \,[1/4 ~+~... ]
\nonumber\\
     K'_{\rm RR }(\ell) &\sim& \ell^{-7/2}\,a^{-1}\,  [1/4 ~+~...] \nonumber\\
     C'_{\rm NSNS}(\ell) &\sim& \ell^{-7/2}\,a^{-1}\, \biggl\lbrack
           {1\over 8}\,({\rm Tr}\,\gamma_I)^2 \nonumber\\
       && ~~~~+~
           {1\over 64}\,({\rm Tr}\,\gamma_g)^2 \,
           \sum_{m=-\infty}^\infty
          \exp\left\lbrace \pi \ell \left( 2- {(m+1/2)^2 \over a^2}  \right)
\right\rbrace
          ~+~...\biggr\rbrack \nonumber\\
     C'_{\rm RR}(\ell) &\sim& \ell^{-7/2}\, a^{-1}\, [ ({\rm Tr}\,\gamma_I)^2 /
8 ~+~...]\nonumber\\
     M'_{\rm NSNS}(\ell) &\sim&
      \ell^{-7/2}\,a^{-1}\,
       [ - ({\rm Tr}\,\gamma_\Omega^T \gamma_\Omega^{-1})/64 ~+~...
]\nonumber\\
     M'_{\rm RR}(\ell) &\sim&
      \ell^{-7/2}\,a^{-1}\,
       [ - ({\rm Tr}\,\gamma_\Omega^T \gamma_\Omega^{-1})/64 ~+~... ]~.
\eeqn
Thus, since our integrals (\ref{relativenormalizations}) become
\beqn
        \Lambda_K &=&   -\,\half\,  2^9\, \calM^9\,
             \int_0^\infty \, d\ell ~ \ell^{7/2} ~ K'(\ell) \nonumber\\
        \Lambda_C &=&   -\,\half\,  \calM^9\,
             \int_0^\infty\, d\ell ~ \ell^{7/2} ~  C'(\ell) \nonumber\\
        \Lambda_M &=&   -\,\half\,  2^9\, \calM^9\,
             \int_0^\infty\, d\ell ~ \ell^{7/2} ~  M'(\ell)~,
\label{amplitudestree}
\eeqn
we see that the total NS-NS tadpole
divergence is proportional to
\beqn
    && 64 ~+~ {1\over 16}\,({\rm Tr}\,\gamma_I)^2 ~-~ 4\,({\rm
Tr}\,\gamma_\Omega^T \gamma_\Omega^{-1})
      \nonumber\\
    && ~~~~~~~~~~~~~+ {1\over 128}\,({\rm Tr}\,\gamma_g)^2 \,
           \sum_{m=-\infty}^\infty
       \exp\left\lbrace \pi \ell \left( 2- {(m+1/2)^2 \over a^2}  \right)
\right\rbrace
\label{NSNStadpole}
\eeqn
and that the total Ramond-Ramond tadpole divergence is proportional to
\beq
    64 ~+~ {1\over 16}\,({\rm Tr}\,\gamma_I)^2 ~-~ 4\,({\rm
Tr}\,\gamma_\Omega^T \gamma_\Omega^{-1})~.
\label{RRtadpole}
\eeq
The second line of (\ref{NSNStadpole}) represents potentially tachyonic
contributions,
while the remaining contributions in (\ref{NSNStadpole}) and (\ref{RRtadpole})
are all
due to massless states.

Thus, in order to cancel both tadpole divergences for
general values of the radius $R/\sqrt{\alpha'}\equiv a^{-1}$,
we see that there is only one solution for the Chan-Paton factors:  we must
choose ${\rm Tr}\,\gamma_g=0$.   Taking $\gamma_\Omega$ symmetric, and letting
the dimensionality
of these matrices be $N$, we then find that $N=32$.
Thus, the only symmetric choice for $\gamma_g$ is
\beq
          \gamma_g ~=~ \pmatrix{ {\bf 1}_{16} & 0 \cr
           0 & -{\bf 1}_{16} \cr}~
\label{gammaX}
\eeq
which corresponds to the gauge group \sosixteen.
Note that the anti-symmetric possibility for $\gamma_g$ would yield
the gauge group $U(16)$;  this case will be discussed in Sect.~6.3.

It is quite remarkable that the case for which all tree-channel tachyons
cancel,
and for which the massless tadpole divergences cancel for general radii $a$,
corresponds to the gauge group \sosixteen.
Indeed, these observations precisely mirror the situation on the heterotic
side,
where it is likewise only for the gauge group $SO(16)\times SO(16)$ that our
nine-dimensional heterotic interpolating Model~B is consistent and
tachyon-free.
Moreover, we see from the above traces that with the choice (\ref{gammaX}), our
massless
open-string states consist of a vector in the adjoint representation
of $SO(16)\times SO(16)$ as well as a spinor
in the $(\rep{16},\rep{16})$ representation.
Once we include the gravity multiplet and $U(1)^2$ gauge bosons from the
closed-string
sector, we see that
this exactly matches the massless spectrum of our nine-dimensional
interpolating Model~B.
Note that at the discrete radii for which extra massless states appear on the
heterotic
side, the ten-dimensional Type~I coupling is non-perturbative so that we have
no
contradiction.  This is similar to the situation discussed in Ref.~\cite{PW}.
Finally, we see that in the $a\to 0$ (or $R\to \infty$) limit, this Type~I
orientifold model
precisely reproduces the ten-dimensional supersymmetric Type~I $SO(32)$ theory
which is
known to be the strong-coupling dual of the ten-dimensional $SO(32)$ heterotic
string.
Thus, we see that it is natural and consistent to interpret this
nine-dimensional Type~I model
with the choice (\ref{gammaX}) as the strong-coupling dual of the heterotic
interpolating
Model~B presented in Sect.~3.
If this interpretation is correct, this would be the first known
example of a heterotic/Type~I strong/weak coupling duality
between non-supersymmetric, tachyon-free string models.

\subsection{Cosmological constant}

Given these results, we now seek to evaluate the one-loop amplitude or
cosmological
constant $\tilde \Lambda(R)$
for our nine-dimensional Type~I model as a function of its radius $R$, just as
we did for our interpolating models B and B$^\prime$ in Sect.~3.
This will enable us to analyze its stability properties.

It is apparent from the above results that the net contribution to the
cosmological
constant from the Klein-bottle trace vanishes in this model, and therefore the
only
non-vanishing closed-string contribution comes from the torus.
This torus contribution $\Lambda_T$ is given in (\ref{relativenormalizations})
where the torus trace $T^\prime(\tau)$ is given in (\ref{torustrace}).
As we remarked above, $\Lambda_T$ is therefore exactly half the cosmological
constant of Model~B$^\prime$, and is therefore half of what
is shown in Fig.~\ref{interplambdaII}.
If this were the sole contribution to the Type~I cosmological constant, our
Type~I model
would be as unstable as the Type~II model from which it is derived, and would
always
flow in the direction of decreasing radius, ultimately developing a tachyon.
Fortunately, however, this is not the case,
for our Type~I theory also contains open-string sectors whose contributions
must
also be included.

Let us now consider the open-string contributions.
It is immediately apparent from the results of Sect.~4.3 that the only
open-string sectors that make non-vanishing contributions to the total one-loop
amplitude
are those from the M\"obius trace.
In terms of the M\"obius loop variable $t$, these contributions
give rise to the nine-dimensional open-string cosmological constant
\beq
 \Lambda_M^{(9)} ~\equiv~ {\calM^9\over 4\sqrt{2}}\,\int_0^\infty  {dt \over
t^{11/2}}~
       {f_2^8 (q)\,f_4^8(q) \over
        f_1^8 (q)\,f_3^8(q) } \,
    \sum_{m=-\infty}^{\infty} \, (-1)^m \,q^{m^2 a^2} ~
\label{expressionloop}
\eeq
where $\calM$ is the overall scale defined in (\ref{Mdef}) and where $q\equiv
e^{-2\pi t}$.
Recall that the leading numerical factor of $(4\sqrt{2})^{-1}$ comprises
the following contributions:  a factor of $2^{-11/2}$ from
(\ref{relativenormalizations}),
a factor of $32$ from the Chan-Paton trace factor, a factor of two for
equal contributions from the NS-NS and Ramond-Ramond states in the M\"obius
strip, and a factor of $1/8$ from the three factors of two in the denominators
of the trace (\ref{fourtraces}).
We thus seek to evaluate (\ref{expressionloop})
as a function of $R/\sqrt{\alpha'}\equiv a^{-1}$.

Unfortunately, evaluating this expression is not straightforward.
While the absence of tachyonic states in this loop amplitude ensures that the
integrand is suitably
convergent as $t\to\infty$,
there is, however, an apparent divergence in this integral in the $t\to 0$
limit.
This is, of course, simply the apparent tadpole divergence,
but its cancellation is not manifest in this expression in terms of the loop
variable $t$.
One possible way to cure this $t\to 0$ ``divergence'' is to
rewrite\footnote{
    Note, in particular, that we are only {\it algebraically rewriting}\/ the
loop amplitude,
    not constructing to the tree amplitude, and consequently we may simply
define $t'=1/t$ as
    a convenient variable and do not require the tree variable $\ell=1/8t$.}
this expression in terms of the new variable $t'=1/t$, yielding
\beq
   \Lambda_M^{(9)} ~=~ {\calM^9\over 4\sqrt{2}} \,{2^{7/2}\over a}\,
\int_0^\infty dt' ~
       {f_2^8 (\tilde q^{1/4})\,f_4^8(\tilde q^{1/4}) \over
       f_1^8 (\tilde q^{1/4})\,f_3^8(\tilde q^{1/4}) } \,
    \sum_{m=-\infty}^{\infty} \,  \tilde q^{(m+1/2)^2/4a^2}
\label{expressiontree}
\eeq
where $\tilde q\equiv e^{-2\pi t'}$.
Written in this form, the amplitude is then manifestly finite
in the corresponding $t'\to \infty$ limit.
Moreover, unlike the previous closed-string cases, the expression in
(\ref{expressiontree})
can be evaluated exactly, without need for numerical integration, for if
we expand the integrand in the form
\beq
       {f_2^8 (\tilde q^{1/4})\,f_4^8(\tilde q^{1/4}) \over
       f_1^8 (\tilde q^{1/4})\,f_3^8(\tilde q^{1/4}) } \,
    \sum_{m=-\infty}^{\infty} \,  \tilde q^{(m+1/2)^2/4a^2}
          ~=~ \sum_{i=0}^\infty \, d_i \,e^{-m_i t'}
\eeq
where $d_i$ and $m_i$ are respectively the degeneracies and masses of states at
the
$i^{\rm th}$ excited string level, we find
\beq
           \Lambda_M^{(9)} ~=~
   {\calM^9\over 4\sqrt{2}}\, {2^{7/2}\over a}\,  \sum_{i=0}^\infty \,
d_i/m_i~.
\label{sum}
\eeq
Unfortunately, while this formulation cures the apparent $t\to 0$ (or $t'\to
\infty$)
divergence, there now arises an apparent divergence as $t'\to 0$ (or $t\to
\infty$).
This is apparent from (\ref{expressiontree}), as well as from the fact that the
degeneracies $d_i$ in (\ref{sum}) typically grow in magnitude as $|d_i|\sim
e^{C\sqrt{m_i}}$ for
some constant $C>0$, rendering the summation in (\ref{sum}) meaningless.

To evaluate this amplitude, therefore, we shall split the original integral
(\ref{expressionloop})
into two pieces, with ranges of integration $1\leq t\leq \infty$ and $0\leq
t\leq 1$ respectively.
We then transform only the second piece into the variable $t'\equiv 1/t$,
so that the range of integration in (\ref{expressiontree}) for this piece is
$1\leq t'\leq \infty$.
This in turn modifies the result in (\ref{sum}) for our second term, inserting
an extra
factor of $e^{-m_i}$ into the sum (and thereby rendering it manifestly finite).
We thus have
\beq
 \Lambda_M^{(9)} ~=~  {\calM^9\over 4\sqrt{2}} \,\int_1^\infty  {dt \over
t^{11/2}}~
       \left\lbrace {f_2^8 (q)\,f_4^8(q) \over
        f_1^8 (q)\,f_3^8(q) } \,
    \sum_{m=-\infty}^{\infty} \, (-1)^m \,q^{m^2 a^2}  \right\rbrace
    ~+~ {\calM^9\over 4\sqrt{2}}\, {2^{7/2}\over a}\,  \sum_{i=0}^\infty \,
 {d_i\over m_i}\,e^{-m_i}~.
\label{expressionmixed}
\eeq
Written in this form, this amplitude is now manifestly finite.

\begin{figure}[thb]
\centerline{\epsfxsize 4.0 truein \epsfbox {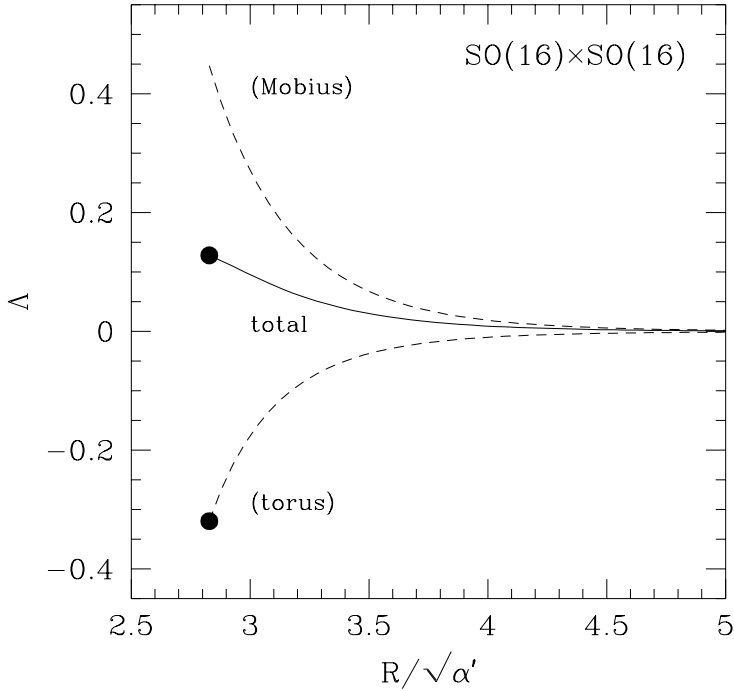}}
\caption{
  The total one-loop cosmological constant
    $\tilde \Lambda$
   for the Type~I interpolating model, plotted in units of $\half \calM^{10}$,
   as a function of the radius $R$ of the compactified dimension.
   Also shown (dashed lines) are the separate torus and M\"obius-strip
contributions;
   note that the Klein-bottle and cylinder contributions vanish.
   This Type~I model reproduces the supersymmetric $SO(32)$ Type~I string
    as $R\to \infty$, and has gauge group $SO(16)\times SO(16)$ for all finite
    radii.  The torus amplitude develops
    a divergence below $R^\ast/\sqrt{\alpha'}\equiv 2\sqrt{2}\approx 2.83$,
    which reflects the appearance of a tachyon in the
    torus amplitude below this radius.  All other contributions are
     tachyon-free for all radii.}
\label{interplambdaI}
\end{figure}

Given this result,
it is then straightforward to evaluate this expression as a function of $R$
and add it to the closed-string result from the torus amplitude.
Our results are shown in Fig.~\ref{interplambdaI}.  Note that in this figure,
we again plot our results in terms of the effective ten-dimensional
cosmological constant $\tilde \Lambda(R)$ which we now define as
\beq
     \tilde \Lambda (R) ~\equiv~ \calM\, {\sqrt{\alpha'}\over R}\,
\Lambda^{(9)}(R)~.
\label{tildelambdadefI}
\eeq
This definition is similar to that in (\ref{tildelambdadef}), except that
we have replaced $R$ with $\alpha'/R$, as suitable for the present case
in which we are only concerned with the $R\to \infty$ limit.
Indeed, the extra factor $\calM \sqrt{\alpha'}/R$
in (\ref{tildelambdadefI}) is simply $1/V$ where $V=2\pi R$
is the effective volume of compactification in the $R\to \infty$ limit.
This then ensures the formal result
\beq
       \lim_{R\to \infty}  \tilde\Lambda(R) ~=~ \Lambda^{(10)}_{SO(32)}~.
\eeq

It is clear from Fig.~\ref{interplambdaI} that although the torus amplitude
alone has a shape which would seem to render our Type~I string unstable (so
that it would always seem to flow in the direction of developing a tachyon),
there is another contribution from the M\"obius strip which succeeds in
cancelling
this behavior and which causes our Type~I model to flow to a
stable supersymmetric point.
This is a fortunate fact, and rests crucially on the relative normalization
(specifically the factor of $2^{-9/2}$)
between these two amplitudes that we derived
in (\ref{relativenormalizations}).\footnote{
   As an aside, we note the interesting fact that if
   this relative normalization factor had been $1/32$
   (as would have been appropriate in ten dimensions),
   then the above torus and M\"obius-strip contributions would have actually
   come very close to {\it cancelling}.  Indeed, at the critical radius
   $R^\ast=\sqrt{8\alpha^\prime}$, we would have found
   the total value $\tilde \Lambda(R^\ast) \approx -2.87 \times 10^{-3}$
   when plotted in the same units.
   However, this value is significantly larger than the expected numerical
error of our
   calculations, and thus is not consistent with zero.
   Furthermore, as $R$ increases from $R^\ast$, this value increases slightly
   and becomes positive before asymptotically approaching zero. \hfil\break
   \indent\indent As a separate but related matter, we also note that
   the possibility of cancellation between the torus amplitude and the
   open-string amplitudes provides a new mechanism for cancelling the
   cosmological constant.  This mechanism does not exist for heterotic strings.
   Indeed, in heterotic string theory, one must cancel the torus amplitude
   by itself, and despite many proposals \cite{cancellambda,lowerdim}, no
non-supersymmetric
   heterotic string models with vanishing cosmological constants have been
   constructed.  Type~I string theory may thus provide a new,
phenomenologically
   interesting way of addressing this paramount issue.}
Thus, not only is our Type~I string model
tachyon-free for all radii $R\geq R^\ast$, but it also consistently flows
in the direction of increasing radius towards the supersymmetric point
at $R=\infty$.


\section{The \sosixteen\ Soliton}
\setcounter{footnote}{0}

In this section, we shall discuss the soliton of the Type~I theory we
constructed in the previous section.
We shall begin by recalling how the $SO(32)$ heterotic string emerges as
a soliton of the $SO(32)$ Type~I theory, and then discuss the modifications
that lead to the \sosixteen\ theory.

\subsection{The $SO(32)$ soliton}

 The $SO(32)$ Type~I theory can be realized as an orientifold of the
Type~IIB theory, and as such it contains a variety of perturbative
sectors associated with closed strings (torus and Klein
bottle sectors) and open strings (cylinder and M\"obius strip sectors).
These sectors give rise to the perturbative states of the $SO(32)$ Type~I
theory, namely the supergravity multiplet (from the closed-string sectors)
and the $SO(32)$ gauge bosons and their superpartners (from the Neveu-Schwarz
and Ramond open-string sectors respectively).  In the language of D-branes,
these latter states can
be viewed as the excitations of an open string (the Type~I string) whose
endpoints
each end on a Dirichlet nine-brane and therefore each satisfy Neuman boundary
conditions.
These are the so-called NN states.

In order to discuss the {\it soliton}\/ states in the theory,
let us now consider one of the other D-branes in the theory,
namely the Dirichlet one-brane.
This D1-brane soliton has been described as a classical
solution in Ref.~\cite{DabHull},
 and it is this solitonic object which is eventually interpreted as
the $SO(32)$ heterotic string.  Specifically, the zero modes of this
D1-brane have been quantized \cite{PW}, and have been found
to yield the same worldsheet fields and GSO projections as the $SO(32)$
heterotic string.  It is this result that we shall now review.

Let $\mu$, $0\leq \mu\leq 9$,
index the spacetime dimensions (with $\mu=0$ identified as the time direction),
and let us assume that the D1-brane lies along the $\mu=1$ direction.
Thus $i$, $2\leq i\leq 9$, indexes the spatial dimensions transverse
to the D1-brane.

 In order to determine the perturbative spectrum of this soliton, we
consider the perturbative Type~I strings which have at least one endpoint on
this
one-brane.
There are then two possibilities for the remaining endpoint:  either it can
end on this one-brane as well, or
it can end on one of the Dirichlet nine-branes in the theory.
If both endpoints are on the D1-brane, this string must satisfy
Dirichlet boundary conditions in the directions transverse to the D1-brane.
These are the so-called DD states.  By contrast, if the remaining endpoint
lies on one of the D9-branes, then this string will satisfy
mixed Neumann-Dirichlet boundary conditions.  These are the so-called ND
states.
In either case, since at least one end of the string
is always fixed on the D1-brane, the boundary conditions imply that
the massless bosons $X^i$ have no zero modes.  Thus the massless states in
these
two sectors can depend only on $x_0$ and $x_1$, the coordinates of the
D1-brane worldsheet theory.
In the heterotic description, these coordinates will be identified as the
heterotic worldsheet coordinates $(\sigma,\tau)$.

We now consider the massless excitations that arise in each case.

In the DD case, if both endpoints of the Type~I string lie on the one-brane,
then massless states can be formed if both endpoints
end at the {\it same}\/ point on the one-brane --- {\it i.e.}\/, if the
Type~I string is closed.
There are then two options for quantizing this string, depending on whether we
assign Neveu-Schwarz or Ramond boundary conditions to the Type~I worldsheet
fermions.
In the Neveu-Schwarz sector, the usual Type~I GSO projection produces a
ten-dimensional vector which we shall denote $V_{10}$.
As we have stated above, this vector depends on only the coordinates
$(x_0,x_1)$.
Under the Lorentz decomposition
\beq
       SO(9,1)~\supset~ SO(8)~\times~ SO(1,1)~,
\label{lorentzdecomp}
\eeq
this representation decomposes as
\beq
        V_{10} ~\longrightarrow~ V_{8} ~\oplus ~ V_{1,1}~.
\eeq
However, note that the orientifold action $\Omega$ projects out the modes
parallel to the D1-brane and retains the modes transverse to the D1-brane.
We are thus left with the remaining $V_8(x_0,x_1)$ fields,
which can be interpreted as the worldsheet coordinate bosons of the heterotic
string.  We therefore identify
\beq
          V_8(x_0,x_1) ~\Longleftrightarrow~ X^i(\sigma,\tau)~.
\eeq
A $\IZ_2$ subgroup of the $U(1)$ generated by modes parallel to the D1-brane
commutes with the $\Omega$ projection, and generates the GSO projection as
a holonomy around closed cycles on the soliton worldsheet.

Let us now consider the Ramond sector.  Recall that the Type~I GSO
projection yields a ten-dimensional spinor which we can denote $S_{10}$.
Under the decomposition (\ref{lorentzdecomp}), this spinor decomposes as
\beq
           S_{10} ~\longrightarrow~ S_8^+ ~\oplus~ C_8^-
\eeq
where the superscripts indicate the corresponding $SO(1,1)$ charge.
A positive $SO(1,1)$ charge is interpreted as a right-moving degree of freedom
along the D1-brane, and a negative charge is interpreted as left-moving.
Because the orientifold projection $\Omega$ acts with an extra minus sign on
Ramond
fermions transverse to the D1-brane in the DD sector, the orientifolding action
preserves $S_8^+$ and projects out $C_8^-$.
We therefore interpret these fermions as the right-moving Green-Schwarz
worldsheet fermions $S_R^a$ of the heterotic string:
\beq
            S_{8}^+(x_0,x_1)  ~\Longleftrightarrow~ S_R^a(\sigma,\tau)~.
\eeq
Here $a$ is an index transforming in the spinor representation of the
Lorentz group $SO(8)$.
Moreover, the spacetime supersymmetry of the Type~I theory, which
related $V_{10}$ and $S_{10}$ to each other as superpartners,
now implies a {\it worldsheet}\/ supersymmetry
for the soliton, so that $X^i_R$ and $S_R^a$
are also superpartners.
Thus, we see that the excitations of the DD sector of the
D1-brane correspond directly to the coordinate bosons of the $SO(32)$ heterotic
string along with its right-moving superpartners.

Let us now turn to the contributions from the ND sector.  Recall that this
sector consists of strings with a single endpoint lying on one of the 32
different
D9-branes in the theory.
Thus all excitations in this sector carry an $SO(32)$ vector index
$A=1,...,32$.
Because of the ND boundary conditions in this sector, the excitations of the
Type~I string must be half-integrally moded in directions perpendicular to the
D1-brane,
whereas they are integrally moded in directions parallel to the D1-brane.
Thus only the excitations in the parallel directions will have zero modes,
which again implies that the fields in this sector depend on only $(x_0,x_1)$.
 Just as for the DD sector, we have both a Neveu-Schwarz sector and a Ramond
sector
depending on the boundary conditions that we assign to the Type~I worldsheet
fermions.
However, due to the half-integral moding of the Type~I worldsheet boson fields,
the vacuum energies of these sectors are altered:  while the Ramond sector
continues to have zero vacuum energy (as required by the worldsheet
supersymmetry
of the Type~I string), the Neveu-Schwarz sector
turns out to have a positive vacuum energy.
Thus only the Ramond sector contains massless excitations, which in this case
correspond to the zero-mode excitations that produce the
degenerate Ramond spinor ground state.
However, the GSO projection in the $SO(32)$ Type~I theory preserves
only one of these ground states --- the state with
negative $SO(1,1)$ charge, corresponding to a left-moving field.
Thus, the states that arise from the ND sector are
Majorana-Weyl fermions $S_1^{A-}$ which are left-moving, which are singlets
under
the $SO(8)$ Lorentz group, and which transform as vectors under $SO(32)$.
These are then naturally identified as the 32 left-moving Majorana-Weyl
worldsheet fermions $\psi^A_L$ of the $SO(32)$ heterotic string:
\beq
          S_1^{A-}(x_0,x_1) ~\Longleftrightarrow~
          \psi^A_L(\sigma,\tau)~.
\eeq
Thus, we see that the ND sector provides the
remaining left-moving worldsheet fields of the heterotic string, thereby
completing the worldsheet fields of the $SO(32)$ heterotic theory.
The previously mentioned $\IZ_2$ holonomy from the DD sector generates the GSO
projection on the worldsheet fermions from this sector.

Note that the tension of the soliton is $T \sim
T_F /\lambda$ where $T_F=(2\pi \alpha')^{-1}$ and where $\lambda$
is the ten-dimensional Type~I coupling.

\subsection{The \sosixteen\ soliton}

Having reviewed the construction of the $SO(32)$ soliton in ten dimensions,
let us now consider how these results are altered in the present case of
the \sosixteen\ string.

First, we remark that in comparison to the construction of
the supersymmetric $SO(32)$ soliton,
the construction of the soliton corresponding to the non-supersymmetric
\sosixteen\ string string is substantially more subtle and involves
new complications.  The primary reason for this concerns the nature of
the non-supersymmetric \sosixteen\ theory.  In the case of the
$SO(32)$ theory, a major simplification occurs because
the theory essentially {\it factorizes}\/ into separate left- and right-moving
components.
This factorization is reflected, for example, in the $SO(32)$ partition
function (\ref{SO32partfunct}).
Indeed, such a factorization is a general property
of all supersymmetric theories in ten dimensions.
This factorization implies that a corresponding solitonic realization
of the theory is relatively straightforward --- one simply realizes
separate sets of left- and right-moving worldsheet fields on the D1-brane
(along with their GSO projections), and then tensors the two theories together
to fill out the complete spectrum of states.
This is essentially the procedure outlined above for the $SO(32)$ soliton.
Unfortunately, the non-supersymmetric \sosixteen\ theory is substantially more
complicated, for its partition function given in (\ref{sosixteenpartfunct})
does {\it not}\/ factorize.
This essentially represents the effect of the additional twist incorporated
in this theory relative to the supersymmetric $SO(32)$ case.
The fact that the \sosixteen\ theory fails to factorize neatly into
left- and right-moving components implies that the realization of this
theory as a D1-brane soliton is far more subtle, and involves
several further assumptions regarding the ways in which the different
pieces of the soliton theory are adjoined together.  We shall discuss
these assumptions as they arise below.  We nevertheless find it remarkable
that, even with these mild assumptions, the \sosixteen\ theory can indeed
be realized as a D1-brane soliton.  In fact, as we shall see,
there are several crucial and rather surprising features which
enable this realization to occur.

We shall now present the construction of the \sosixteen\ soliton.
As in our construction of the corresponding Type~I theory discussed in
Sect.~4, we begin with the $SO(32)$ Type~I theory and its corresponding soliton
as discussed above.
We then seek to determine how this $SO(32)$ soliton is
modified when we compactify the $x_1$ direction on a circle of radius $R$
and project by the element
\beq
         \calY~\equiv~ \calT \,(-1)^F\, \gamma_g~.
\label{Ydefagain}
\eeq
Recall that $\calT$ is
the half-rotation defined in (\ref{Tdef}), while $F$
is the Type~I spacetime fermion number.  As in the preceding orientifold
calculation, $\gamma_g$ is defined in (\ref{gammaX}) where
we separate the set of 32 heterotic left-moving Majorana-Weyl fermions
(or equivalently the set of 32 D-branes in the Type~I picture) into two
subsets with indices $A_1=1,...,16$ and $A_2=17,...,32$ respectively.

Note that
although it may seem that we should derive the soliton directly in terms
of our nine-dimensional Type~I interpolating model, it is
sufficient to start with the $SO(32)$ soliton, compactify it on a circle,
and then apply the $\calY$ projection.
This is because our Type~I interpolating model
is essentially a  $\IZ_2\times \IZ_2$ orientifold of the supersymmetric
Type~IIB
theory compactified on a circle of radius $R$.
The first $\IZ_2$ factor corresponds to the worldsheet parity operator
$\Omega$, and the second to the orbifold operator $g$.
Because these two operators commute, we can consider our interpolating
open-string model either as an orientifold of the Type~II interpolation,
or as an orbifold of the $SO(32)$ Type~I theory.
In this section, we are considering our Type~I model as
an orbifold of the $SO(32)$ Type~I theory, and are therefore analyzing
the soliton accordingly.

It is also important to note that we will be compactifying the $x_1$ direction
---
 {\it i.e.}\/, the D1-brane of the $SO(32)$ Type~I theory
will itself be wrapped on the circle of radius $R$.
This is not our only choice, however, and one could have
instead chosen to compactify one of the transverse directions $x_2,...,x_9$.
This would leave the soliton unwrapped in nine dimensions.
However, it is only by choosing to wrap the $SO(32)$ soliton
that we can hope to obtain a soliton which behaves as a fundamental
({\it i.e.}\/, nearly massless) string at a perturbative value of the Type~I
coupling $\lambda_I$.  In this connection, recall that since the tension
of the D1-brane goes as $T\sim T_F/\lambda_I$ where $T_F=(2\pi\alpha')^{-1}$
is the fundamental string tension,
the total mass of the soliton goes as
\beq
             M ~\sim~ R\, T ~ \sim~ R\,T_F /\lambda_I~.
\eeq
Hence, as $R\to 0$, the mass of the soliton can vanish even for $\lambda_I \ll
1$.
By contrast, if we had compactified any of the other spatial directions,
the effective mass of the unwrapped soliton would have been infinite for
perturbative
Type~I couplings, and would only have become zero at strong coupling.
We shall therefore focus on the wrapped case.

The first step, then, is to wrap the soliton on a circle of radius $R$.
Since each of the $SO(32)$ soliton fields $X^i$, $S^a$, and $\psi^A$
is a function of $(x_0,x_1)$, this compactification
leads to a quantization of their corresponding momenta $p_1$:
\beq
        p_1 ~=~  {m\over R}~,~~~~~~~m\in\IZ~.
\label{momentump1}
\eeq
This compactification then produces the soliton corresponding to the
nine-dimensional untwisted $SO(32)$ Type~I theory,
and this quantization determines the oscillator moding of worldsheet
soliton fields.

As in our orientifold calculation, the next step is to implement
the projection by $\calY\equiv \calT (-1)^F \gamma_g$.
Let us first recall the effects of each of these operators.
The operator $\calT$ is the half-rotation, and gives an overall sign $(-1)^m$
where $m$ is the momentum quantum number in (\ref{momentump1}).
Likewise, since $F$ is the Type~I fermion number,
the operator $(-1)^F$ gives a minus sign when acting on the Ramond sector
of the Type~I theory and a plus sign when acting on the Neveu-Schwarz.
Finally, $\gamma_g$ gives a minus sign on the fundamental (vector)
representation
of the second $SO(16)$ factor under the
decomposition $SO(32)\supset SO(16)\times SO(16)$.  Since this minus sign
is naturally associated with the fields $\psi^{A_2}$, $A_2=17,...,32$,
that produce this representation, this in turn implies
that there will also be a minus sign for one of the spinor representations
of this $SO(16)$ factor.  Without loss of generality, we can choose
this to be the spinor (rather than conjugate spinor) representation
of the second $SO(16)$ factor.

Remarkably, the latter two operators have a natural interpretation in
terms of the {\it heterotic}\/ orbifold that produced
the heterotic \sosixteen\ theory from the heterotic $SO(32)$ theory
as in Fig.~\ref{orbrelations}.
It is, of course, immediately evident from the action of $\gamma_g$
that we can equivalently describe this operator in the notation of
Sect.~3 as $R_{VS}^{(2)}$.
Somewhat more surprising, however,
is the interpretation of $(-1)^F$.  Since this gives a minus sign
to the entire Type~I Ramond sector,
this operator gives a minus sign for the
Ramond sectors arising from both $S^a$ and $\lambda^A$.
In the case of the right-moving field $S^a$,
this minus sign is equivalent to the heterotic operator $\tilde R_{SC}$.
Likewise, if $S_{32}$ and $C_{32}$ denote the two spinor ground states
of the left-moving field $\lambda^A$,
then under the $SO(32)\supset SO(16)\times SO(16)$ decomposition,
we have
\beqn
    S_{32} &=& S^{(1)}_{16} S^{(2)}_{16} ~\oplus~ C^{(1)}_{16}
C^{(2)}_{16}\nonumber\\
    C_{32} &=& S^{(1)}_{16} C^{(2)}_{16} ~\oplus~ C^{(1)}_{16} S^{(2)}_{16}~.
\label{spinordecomps}
\eeqn
A minus sign for both $S_{32}$ and $C_{32}$ is therefore equivalent to
a minus sign for just
$S_{16}^{(1)}$ and $C_{16}^{(1)}$.
We therefore find that in heterotic language,
the effect of the Type~I operator $(-1)^F$ is
\beq
             (-1)^F ~\Longleftrightarrow~
        \tilde R_{SC} R_{SC}^{(1)} ~=~ (-1)^{\hat F} R_{SC}^{(1)}
\label{Ftype1het}
\eeq
where $\hat F$ is the {\it heterotic}\/ spacetime fermion number.
Thus we find that
\beq
        (-1)^{F} \,\gamma_g ~\Longleftrightarrow~
       \tilde R_{SC}\, R_{SC}^{(1)}\, R_{VS}^{(2)}~,
\label{type1hetprojections}
\eeq
which, as indicated in (\ref{hetorbs}) and
Fig.~\ref{orbrelations}, is exactly the orbifold
the produces the non-supersymmetric heterotic \sosixteen\
string from the supersymmetric $SO(32)$ heterotic string.
We find this correspondence remarkable.  In particular, note that
this correspondence rests upon the rather non-trivial relation
(\ref{Ftype1het}) between the Type~I spacetime fermion number $F$ and the
heterotic spacetime fermion number $\hat F$.

At this stage, given the correspondence (\ref{type1hetprojections}),
it might appear that our construction of the \sosixteen\ soliton
is complete.
After all, we know that orbifolding
the supersymmetric heterotic $SO(32)$ theory by the action in
(\ref{type1hetprojections})
produces the non-supersymmetric heterotic \sosixteen\ theory.
However, since we are seeking to construct this theory as a D1-soliton,
we are really working in the Type~I theory, not the heterotic,
and consequently we are not performing an {\it orbifold}\/ by $(-1)^F \gamma_g$
but rather a {\it projection}\/ by $\calT (-1)^F \gamma_g$.
Indeed,
in the Type~I theory we do not have the option (at least at weak coupling)
of adding any of the extra twisted sectors that might na\"\i vely seem to be
required.
Instead, as we shall see, these twisted sectors arise
in a completely different and surprising way.

Let us now continue our derivation by performing the projection
by $\calY$.  As a first step,
we begin by considering the effect of $\calY$ on the bosonic fields $X^i$.
It is clear that these fields carry neither gauge quantum numbers nor
spacetime fermion number;  hence a projection by $\calY$ is equivalent
to a projection by $\calT$, which in turn preserves only that subset of
momentum modes in (\ref{momentump1}) for which $m$ is even.
However, these modes can
equivalently be interpreted as {\it all}\/ of the momentum modes of
the boson fields $X^i$ where these fields are viewed as being compactified on
a circle of radius $R/2$:
\beq
        p~=~ {m\over R} ~=~ {m'\over R/2}  ~~~~\Longrightarrow~~~~
          m' ~=~ m/2~.
\eeq
This observation simply reflects the fact that, as discussed in previous
sections,
a projection by the half-rotation (\ref{Tdef}) essentially puts the theory
onto a circle of half the original radius.
We shall therefore find it convenient in the following to describe
all of the fields on the soliton in terms of the momentum modes $m'=m/2$
relative to the ``half-circle'' of radius $R/2$.

Let us now consider the effect that this radius reduction has
on the fermionic fields $S^a$ and $\psi^A$.

We begin by considering the effects of this radius reduction on
the eight Green-Schwarz fermions $S^a$.
Ordinarily, on the full circle, we would expect all of these fermions
to be integrally moded.
However, in terms of the momentum modes of the circle with radius $R/2$,
each fermion $S^a$ will now have two distinct sectors, one with momentum modes
$m'\in \IZ$ and one with $m'\in \IZ+1/2$.
Thus, in principle there are $2^8$ moding combinations that can arise
(modulo permutations and $SO(8)$ rotations in the fermion space).
However, among these combinations there are only two that are invariant under
$SO(8)$ Lorentz symmetry:  these are the combinations for which all
fermions have modes $m'\in\IZ$, and for which all
fermions have modes $m'\in\IZ+1/2$.
These sectors will have eigenvalues $\pm 1$ respectively under $\calT$,
and, upon quantization, will have integer and half-integer
excitation modes respectively.

Clearly the integer excitation modes
of $S^a$ are the ``usual'' modings from the perspective of the smaller circle
of radius $R/2$, and we shall refer to this sector
as belonging to ``Class~A''.  By contrast, the sector with modings $m'\in
\IZ+1/2$
is the new feature arising from this decomposition onto the half-radius
circle, and we shall refer to this sector as belonging to ``Class~B''.
While the Class~B sector may seem to be a ``twisted'' version of the
Class~A sector, the important point is that {\it both sectors emerge
automatically from the reduction to the half-circle of the $S^a$ field
with integer modings on the original circle.}  Specifically,
no {\it ad hoc}\/ twist has been introduced in order to produce the
sectors in Class~B.

A similar (but slightly more complicated) situation exists for the fermions
$\psi^A$.
In the Ramond sector, each of these 32 fermions would have integer modings on
the full circle, and such modings reduce to both integer and half-integer
modings on the half-radius circle.
Thus, we see that in principle, there are $2^{32}$ different modings
that are possible for these $32$ fermions (modulo permutations and rotations
in fermion space).
However, only {\it four}\/ of these combinations are invariant under
the \sosixteen\ symmetry (or equivalently, only four of these combinations
will be eigenstates of the operator $\calY$).
These are
\beqn
    {\rm Ramond~Class~A}:&~~& \psi^{A_1}_{-n}\, \psi^{A_2}_{-n} \nonumber\\
    {\rm Ramond~Class~B}:&~~&\cases{
           \psi^{A_1}_{-r}\, \psi^{A_2}_{-r} \cr
           \psi^{A_1}_{-n}\, \psi^{A_2}_{-r} \cr
           \psi^{A_1}_{-r}\, \psi^{A_2}_{-n} \cr}
\label{fermioncombosR}
\eeqn
where $n\in\IZ$ signifies integer modings, $r\in \IZ+1/2$ signifies
half-integer modings, $A_1\in\lbrace 1,...,16\rbrace$,  and
$A_2\in\lbrace 17,...,34\rbrace$.
In the Ramond sector, only the first of these combinations
is the correct moding from the perspective of the half-radius circle.
As above, we shall therefore designate this sector as belonging to Class~A,
and the remaining possibilities will be designated as belonging to Class~B.
Of course, along with the Ramond sector for the fermions $\psi^A$,
there is also a Neveu-Schwarz sector that we expect to arise
non-perturbatively,
and in this sector we will have the opposite modings
for the fermions $\psi^A$.   We will therefore identify those with all
half-integer
modings as being in Class~A, and all other combinations as being in Class~B:
\beqn
    {\rm NS~Class~A}:&~~& \psi^{A_1}_{-r}\, \psi^{A_2}_{-r} \nonumber\\
    {\rm NS~Class~B}:&~~&\cases{
           \psi^{A_1}_{-n}\, \psi^{A_2}_{-n} \cr
           \psi^{A_1}_{-r}\, \psi^{A_2}_{-n} \cr
           \psi^{A_1}_{-n}\, \psi^{A_2}_{-r} ~.\cr}
\label{fermioncombosNS}
\eeqn

Let us now consider the effects of $\calY$ on these different fermion
sectors.  For reasons that will become clear shortly, we shall begin by
merely determining the signs that each of these sectors have under $\calY$.
Once this is done, we will then discuss how the projection is to be performed.

We start by considering the Class~A sectors, {\it i.e.}, those in
which the right-moving Green-Schwarz fermion $S^a$ is integrally moded on
the half-circle, and in which all of the left-moving fermions $\psi^A$ have
integral modes in the Ramond sector and half-integral
modes in the Neveu-Schwarz sector.
As in the case of the supersymmetric $SO(32)$ soliton,
quantizing the zero modes of $S^a$ gives both a vector and spinor
representation of the transverse $SO(8)$ Lorentz group:
\beq
                    S^a_{-n} ~\Longrightarrow~
                  V_8 ~\oplus~ S_8~.
\eeq
Under $\calY$, these vector and spinor sectors
respectively have eigenvalues $\pm 1$.
This follows from the observation that each of these sectors has eigenvalue
$+1$ under $\calT$ and $\gamma_g$, and are distinguished only by $(-1)^F$;
equivalently, this also follows from (\ref{type1hetprojections}).

Let us now consider the sectors that arise from the fermions $\psi^A$.
In the Ramond sector, the vacuum is the $SO(32)$ spinor $S_{32}$
(since the $SO(32)$ conjugate spinor representation $C_{32}$ is GSO-projected
out of the spectrum).
As indicated in (\ref{spinordecomps}),
this spinor $S_{32}$ decomposes
into the $S_{16}^{(1)} S_{16}^{(2)}$
and $C_{16}^{(1)} C_{16}^{(2)}$
representations of \sosixteen.
Under $\calY$, these two representations have eigenvalues $\pm 1$ respectively.
Likewise, in the Neveu-Schwarz sector, we obtain the \sosixteen\
representations
$I_{16}^{(1)} I_{16}^{(2)}$ and $V_{16}^{(1)} V_{16}^{(2)}$.
Under $\calY$, these also accrue eigenvalues
$\pm 1$ respectively.  This once again follows from (\ref{spinordecomps}),
along with the observation that both of these sectors have eigenvalue $+1$
under
$\calT$.

Thus the Class~A sectors, along with their eigenvalues under $\calY$,
can be summarized as follows:
\beq
\begin{tabular}{||c|c||c|c||c|c||}
\hline
  \multicolumn{2}{||c||}{right-movers} & \multicolumn{4}{c||}{left-movers}
\\
\hline
\hline
  \multicolumn{2}{||c||}{~} & \multicolumn{2}{c||}{Ramond} &
\multicolumn{2}{c||}{NS} \\
\hline
   $\calY$ & sector & $\calY$ & sector &  $\calY$ & sector \\
\hline
 $+1$ & $V_8$ & $+1$ & $S_{16}^{(1)}  S_{16}^{(2)}$
    &  $+1$ & $I_{16}^{(1)}  I_{16}^{(2)}$ \\
 $-1$ & $S_8$ & $-1$ & $C_{16}^{(1)}  C_{16}^{(2)}$
    & $-1$ & $V_{16}^{(1)}  V_{16}^{(2)}$ \\
\hline
\end{tabular}
\label{classAtable}
\eeq

We now turn to the Class~B sectors.
Recall that these are the additional sectors that
were automatically generated from the reduction of our
soliton fields from the circle of radius $R$ to the circle of radius $R/2$,
and consist of the extra combinations of modings that would not
na\"\i vely seem to
be allowed from the perspective of compactification on the circle of radius
$R/2$.
As we saw above, in these sectors $S^a$ is
half-integrally moded ({\it i.e.}, has $m'\in \IZ+1/2$),
while $\psi^A$ has the complicated
moding pattern that was discussed above.

Let us first consider $S^A$.
Since the bosonic fields $X^i$ are integrally moded,
giving the Green-Schwarz fermions $S^A$ half-integer modings changes
their vacuum energy from 0 to $-1/2$.  This means that the two subsectors
which concern us now are the sector built upon the tachyonic vacuum
$|0\rangle$ (which transforms as the identity representation
$I$ under the $SO(8)$ transverse Lorentz group),
and the sector built upon the massless vacuum state
$S^a_{-1/2}|0\rangle$ (which, given our previous conventions,
transforms as the conjugate spinor representation $C$).
Thus, since the net momentum modings of these sectors are respectively
half-integer and integer, they have eigenvalues $\mp 1$ under $\calT$.
Furthermore, they are each invariant under $\gamma_g$, and,
thanks to the extra fermion number of the vacuum that accrues
for such half-integrally moded $S^A$ fields, they respectively
have eigenvalues $\mp 1$ under $(-1)^F$.
Thus, under $\calY$, both of these sectors have eigenvalue $+1$.
We shall denote these two $SO(8)$ sectors as
$I_8$ and $C_8$ respectively.

Turning to the left-moving fermions $\psi^A$,
we now wish to consider the Class~B moding combinations
given in (\ref{fermioncombosR}) and (\ref{fermioncombosNS}).
Let us consider the Ramond sector first.
The moding pattern $\psi_{-r}^{A_1} \psi_{-r}^{A_2}$
has vacuum energy $-1$, and
gives rise to the sectors
$V_{16}^{(1)} I_{16}^{(2)}$
and
$I_{16}^{(1)} V_{16}^{(2)}$.
These sectors respectively have eigenvalues  $\pm 1$ under $\calY$.
(The remaining possibilities, namely
$I_{16}^{(1)} I_{16}^{(2)}$ and
$V_{16}^{(1)} V_{16}^{(2)}$,
are GSO-projected out of the spectrum.)
Similarly, the moding pattern
$\psi_{-r}^{A_1} \psi_{-n}^{A_2}$
gives rise to sectors
$I_{16}^{(1)} S_{16}^{(2)}$ and
$V_{16}^{(1)} C_{16}^{(2)}$, each with
eigenvalue $+1$ under $\calY$, while
the moding pattern
$\psi_{-n}^{A_1} \psi_{-r}^{A_2}$
gives rise to sectors
$C_{16}^{(1)} I_{16}^{(2)}$ and
$S_{16}^{(1)} V_{16}^{(2)}$, each with
eigenvalue $-1$ under $\calY$.
The the Neveu-Schwarz sector is similar.
The moding pattern
$\psi_{-n}^{A_1} \psi_{-n}^{A_2}$
gives rise to sectors
$S_{16}^{(1)} C_{16}^{(2)}$ and
$C_{16}^{(1)} S_{16}^{(2)}$, with
eigenvalues $\pm 1$ respectively under $\calY$, while
the moding pattern
$\psi_{-n}^{A_1} \psi_{-r}^{A_2}$
gives rise to sectors
$S_{16}^{(1)} I_{16}^{(2)}$ and
$C_{16}^{(1)} V_{16}^{(2)}$, each with
eigenvalue $+1$.
Finally, the Neveu-Schwarz sector with moding
$\psi_{-r}^{A_1} \psi_{-n}^{A_2}$
gives rise to sectors
$I_{16}^{(1)} C_{16}^{(2)}$ and
$V_{16}^{(1)} S_{16}^{(2)}$,
with eigenvalues $+1$ under $\calY$.

Thus the Class~B sectors, along with their eigenvalues under $\calY$, can
be summarized as follows:
\beq
\begin{tabular}{||c|c||c|c||c|c||}
\hline
  \multicolumn{2}{||c||}{right-movers} & \multicolumn{4}{c||}{left-movers}
\\
\hline
\hline
  \multicolumn{2}{||c||}{~} &
  \multicolumn{2}{c||}{Ramond} &
  \multicolumn{2}{c||}{NS}  \\
\hline
   $\calY$ & sector & $\calY$ & sector & $\calY$ & sector \\
\hline
 $+1$ & $I_8$
      & $+1$ & $V_{16}^{(1)}  I_{16}^{(2)}$
           & $+1$ & $S_{16}^{(1)}  C_{16}^{(2)}$ \\
 $+1$ & $C_8$
      & $-1$ & $I_{16}^{(1)}  V_{16}^{(2)}$
           & $-1$ & $C_{16}^{(1)}  S_{16}^{(2)}$ \\
    ~ & ~
      & $+1$ & $I_{16}^{(1)}  S_{16}^{(2)}$
           & $+1$ & $S_{16}^{(1)}  I_{16}^{(2)}$ \\
    ~ & ~
      & $+1$ & $V_{16}^{(1)}  C_{16}^{(2)}$
           & $+1$ & $C_{16}^{(1)}  V_{16}^{(2)}$ \\
    ~ & ~
      & $-1$ & $C_{16}^{(1)}  I_{16}^{(2)}$
           & $+1$ & $I_{16}^{(1)}  C_{16}^{(2)}$ \\
    ~ & ~
      & $-1$ & $S_{16}^{(1)}  V_{16}^{(2)}$
           & $+1$ & $V_{16}^{(1)}  S_{16}^{(2)}$ \\
\hline
\end{tabular}
\label{classBtable}
\eeq

Now that we have constructed all of the sectors of the soliton and determined
their eigenvalues under $\calY$, the next step is to do the projection onto
sectors with $\calY=+1$.
Strictly speaking, of course,
we would na\"\i vely expect to
perform this projection on each sector individually, without regard
to whether the sector is left-moving or right-moving, since this is how such
a projection would ordinarily be implemented from the Type~I perspective.
Clearly, this would then simply remove all sectors with $\calY= -1$ from the
tables
(\ref{classAtable}) and (\ref{classBtable}).
Indeed, this is the correct procedure that one should follow when constructing
a
 {\it supersymmetric}\/ soliton (such as the supersymmetric $SO(32)$ Type~I
soliton)
for which the corresponding heterotic theory factorizes.
However, in the present non-supersymmetric case we no longer expect such a
factorization to occur, and consequently we expect interactions to arise
between the left- and right-moving components of the theory.
Such interactions might be interpreted as arising due to the non-perturbative
dynamics of the soliton.
We shall therefore make the mild assumption that when performing our $\calY$
projection,
we will consider those left- and right-moving combinations for which the {\it
combined}\/
sector is invariant under $\calY$.  Indeed, this will be our method for
performing
the $\calY$ projection in non-supersymmetric cases.

Given this understanding, we then perform our $\calY$ projection
by considering all possible left- and right-moving {\it combinations}\/
which meet the following qualifications:
\begin{itemize}
\item  The combined sector should be invariant under $\calY$ ({\it i.e.}\/,
         the combined $\calY$-charge should be $+1$).
\item  The left- and right-moving components should satisfy
$L_0=\overline{L_0}$.
        This is of course nothing but the level-matching condition for closed
strings,
        but it is justified for our Type~I soliton because it is equivalent
             to the Type~I momentum conservation condition $p_1^L=p_1^R$.
\item  The resulting set of combined sectors should be invariant under
        exchange of the two left-moving $SO(16)$ gauge factors.  This
restriction
       arises due to the fact that our fundamental Type~I string
       itself preserves this exchange symmetry, and that the dual heterotic
        \sosixteen\ string also preserves this exchange symmetry.
       While it may seem that this symmetry has been broken when writing
       our orbifolds in the form (\ref{type1hetprojections}), this form
       is merely one convention, and there would be no physical distinction
       between this orbifold and one in which the two $SO(16)$ gauge factors
are
       exchanged.
\end{itemize}
Finally, there is one further condition that we will need to impose.
Specifically, at sufficiently large Type~I coupling, we shall assume that
\begin{itemize}
\item  We may only choose left-right sector combinations
        from the same class.  In other words, we may choose the left- and
        right-moving sectors from {\it either}\/ Class~A {\it or}\/ Class~B,
        but we will not choose one sector from Class~A and the
        other from Class~B.
\end{itemize}
While the origin of this last restriction will become apparent from
the heterotic interpretation of the soliton,
we do not have an interpretation of this restriction based on our Type~I
construction.  We shall nevertheless assume that this restriction holds,
possibly as the result of some non-perturbative selection rule.

Given these restrictions, it is then straightforward to
collect the surviving left-right combination sectors from
tables (\ref{classAtable}) and (\ref{classBtable}).
Specifically, we obtain the following combination sectors:
\beqn
   {\rm Class~A:}&&~~~
     V_8 \,  I_{16}^{(1)} I_{16}^{(2)} ~,~~
     V_8 \, S_{16}^{(1)} S_{16}^{(2)} ~,~~
     S_8 \, V_{16}^{(1)} V_{16}^{(2)} ~,~~
     S_8 \, C_{16}^{(1)} C_{16}^{(2)}  \nonumber\\
   {\rm Class~B:}&&~~~
     I_8 \,  V_{16}^{(1)} C_{16}^{(2)} ~,~~
     I_8 \,  C_{16}^{(1)} V_{16}^{(2)} ~,~~
     C_8 \,  I_{16}^{(1)} S_{16}^{(2)} ~,~~
     C_8 \,  S_{16}^{(1)} I_{16}^{(2)} ~.
\label{resultingsectors}
\eeqn
Remarkably, this is precisely the set of sectors that comprise the
non-supersymmetric \sosixteen\ heterotic string!
(This is most easily apparent via the heterotic partition function given in
(\ref{sosixteenpartfunct}).)
Thus, under these assumptions, we see that we have succeeded in realizing the
non-supersymmetric \sosixteen\ heterotic string
as a soliton of our Type~I string model.

There are several remarkable things to note about our derivation.
First, we see that via the above assumptions, we have been able
to overcome the problems caused by the lack of factorization that is
typical of non-supersymmetric heterotic string theories.
Specifically, we have made only relatively mild assumptions regarding
the interactions between left- and right-moving sectors of the theory.
Second, we see by comparing the sectors that arise in
Type~I soliton
with their heterotic string equivalents that we have the natural
Type~I/heterotic identification
\beqn
              {\rm Class~A} &~\Longleftrightarrow~& {\rm untwisted~sectors}
\nonumber\\
              {\rm Class~B} &~\Longleftrightarrow~& {\rm twisted~sectors~.}
\eeqn
This identification is especially remarkable  because
on the heterotic side, the twisted sectors are interpreted as extra
sectors that are added to the theory
as a result of modular invariance.
By contrast, on the Type~I side, modular invariance is not a symmetry,
and we do not have the ability to add in
any twisted sectors.  Instead, we can only perform projections.
Thus, it is quite astonishing that the Class~B sectors ``magically'' manage to
give rise to precisely the same sectors that would have
been required by modular invariance on the heterotic side.
Indeed, the Class~B sectors in the Type~I theory
somehow manage to reconstruct modular invariance on the soliton,
even in this unfactorized case for which modular invariance is highly
non-trivial.

In this connection,
we emphasize again that no twists {\it of any kind}\/
were performed in the Type~I theory in order to generate the
Class~B sectors.
Rather, these sectors were generated naturally upon the reduction
of our Type~I theory from the circle of radius $R$ to
the circle of radius $R/2$.   This reduction was generated
by the action of the half-rotation operator $\calT$, which in turn
was precisely the operator that was responsible for allowing us
to construct our interpolations on the heterotic and Type~II sides.
Thus, the ``miracle'' that produces the required Class~B sectors
on the Type~I side can ultimately be traced all the way back to our heterotic
models and their nine-dimensional interpolations.

This surprising set of results and interconnections can therefore
be taken not only as compelling evidence for the conjectured duality
between our Type~I model and the \sosixteen\ heterotic string,
but also as compelling evidence for the correctness of our general procedure
(and its associated assumptions) for the construction of non-supersymmetric
Type~I solitons.  Indeed, as we shall see in the next section, we shall find
even further evidence for this picture when we consider the Type~I solitons
corresponding to other non-supersymmetric theories.

\section{Duals for Other Non-Supersymmetric Theories}
\setcounter{footnote}{0}

It is clear that the procedure we have developed for constructing the
duals of non-supersymmetric string theories and analyzing their solitons
is quite general, and should have validity beyond the case of the \sosixteen\
string.  In this section, we shall apply our methods to several other cases
of non-supersymmetric ten-dimensional heterotic strings, including
the non-supersymmetric $SO(32)$ string discussed in Sect.~2.5, the
non-supersymmetric $SO(8)\times SO(24)$ string, and the non-supersymmetric
$U(16)$ string.  All of these ten-dimensional string theories have tachyons,
and thus we expect their behavior to differ significantly from that of the
\sosixteen\ string.  Indeed, this comparison will in some sense serve to
emphasize the unique and rather special properties of the \sosixteen\ string.
However, despite the tachyons that appear in these ten-dimensional models, 
it is nevertheless remarkable that their associated nine-dimensional heterotic 
interpolations are tachyon-free in the range $R>R^\ast$, 
that these interpolations continue to have strong-coupling duals, 
and that their massless states and 
solitons continue to match exactly in this range.

\subsection{The non-supersymmetric $SO(32)$ theory}

We begin by constructing the dual theory 
for the case of the non-supersymmetric $SO(32)$ heterotic 
theory discussed in Sect.~2.5.

\begin{figure}[thb]
\centerline{\epsfxsize 4.0 truein \epsfbox {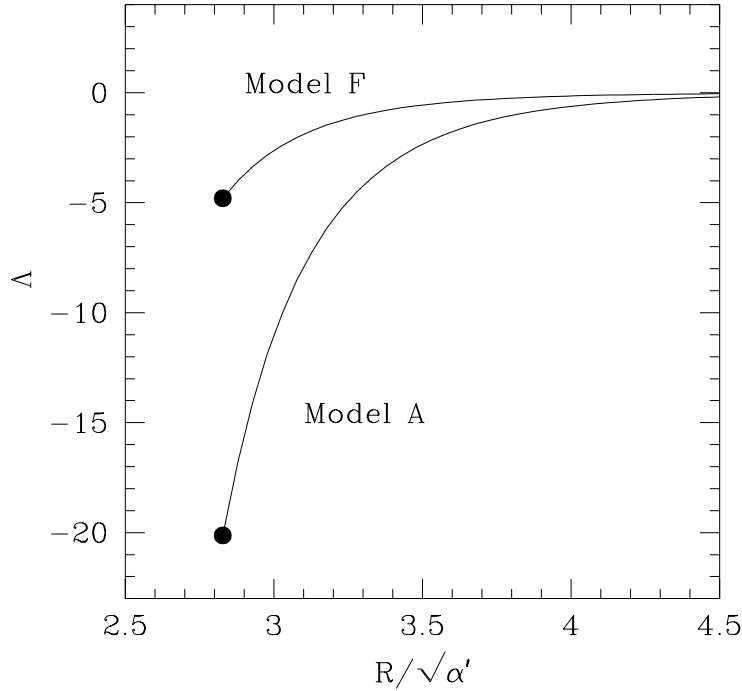}}
\caption{
  The one-loop cosmological constants
    $\tilde \Lambda$
   of Models~A and F, plotted in units of
    $\half \calM^{10}$, as functions of the radius $R$ of the compactified
dimension.
    Model~A is given in Sect.~3.5, while Model~F will be discussed
     in Sect.~6.2.
   Both of these models reproduce the supersymmetric $SO(32)$ heterotic string
    as $R\to \infty$,
    and their one-loop amplitudes $\tilde \Lambda$
    develop divergences below $R^\ast/\sqrt{\alpha'}\equiv 2\sqrt{2}\approx
2.83$.
    This reflects the appearance of tachyons in these
    models below this radius.}
\label{otherhetplots}
\end{figure}

Like the \sosixteen\ model,
it is evident from Fig.~\ref{interpfig} that the non-supersymmetric $SO(32)$
heterotic
theory can also be continuously connected to the supersymmetric
$SO(32)$ theory.
The relevant interpolating model in this case is Model~A, whose partition
function is given in (\ref{9dparts}).  It is apparent from this partition
function that Model~A is tachyon-free for
all compactification radii $R\geq R^\ast$ where the critical radius turns out
to be the
same as that for Model~B$^\prime$, namely $R^\ast=2\sqrt{2\alpha^\prime}$.
In Fig.~\ref{otherhetplots}, we have plotted
the one-loop cosmological constant
$\tilde \Lambda$ of Model~A as a function of the compactification radius $R$
in the range $R\geq R^\ast$.
Just as for Fig.~\ref{interplambdaI}, we taken our definition of $\tilde
\Lambda(R)$
as in (\ref{tildelambdadefI}) in order to ensure the correct formal limiting
behavior\footnote{
          Comparing Fig.~\ref{otherhetplots} with Fig.~\ref{interplambdaII},
the reader may
          suspect that the cosmological constant of Model~A is simply a
rescaling
          of the cosmological constant of Model~B$^\prime$ by some constant
factor $r_{A/B'}$.
          Likewise, the reader
          may suspect that the cosmological constant of Model~F (which is also
          plotted in Fig.~\ref{otherhetplots} and which will be discussed in
Sect.~6.2)
          represents another rescaling of the same function, with a different
rescaling factor
          $r_{A/F}$.  To remarkable degree of numerical accuracy, this is
indeed the case:
          if we take $r_{A/B'}\approx 31.5116$ and $r_{A/F}\approx 4.1951$,
then
          these respective cosmological constants do not differ by more than
$0.01$
          in units of $\half \calM^{10}$ at any point over the entire range
$R\geq R^\ast$.
          However, given the numerical accuracy to which these calculations
were performed,
          this difference is significant, and is not consistent with zero.
Thus, these cosmological
          constants are not simple rescalings of each other.  Of course, we do
not
          expect these cosmological constants to be related to each other (and
certainly not
          through non-integer rescaling factors), for they are derived from
different underlying string
          models with entirely different spectra.}
as $R\to \infty$.

We remind the reader that the appearance of tachyons in Model~A below a
critical
radius does not imply that the model itself is discontinuous.  Rather,
only the cosmological constant is discontinuous, and the spectrum of Model~A
passes smoothly from supersymmetric at infinite radius, to non-supersymmetric
but tachyon-free for radii $R\geq R^\ast$, and then finally to
non-supersymmetric
and tachyonic for $R<R^\ast$.
(Of course, the interpretation of such a model changes discontinuously, since
the appearance of tachyons implies that the theory can exist only at strong
coupling.)
We also wish to emphasize that the behavior of the cosmological constant
for this (tachyonic) $SO(32)$ interpolating model is significantly different
from
the behavior of the cosmological constant of the corresponding (non-tachyonic)
\sosixteen\
interpolating model sketched in Fig.~\ref{interplambda}.  Whereas the
\sosixteen\ interpolating
model had a {\it positive}\/ cosmological constant for all radii, causing the
model to flow
in the direction of increasing radius towards a stable, supersymmetric limit,
the analogous $SO(32)$
interpolating model has a {\it negative}\/ cosmological constant which causes
the model
to flow in the direction of {\it decreasing}\/ radius, {\it away}\/ from the
supersymmetric
limit.  We can attribute this difference in behavior to the appearance of the
tachyon
in the interpolating model.

Despite these observations,
the fact that the non-supersymmetric heterotic $SO(32)$ theory can
be continuously connected to the supersymmetric heterotic $SO(32)$ theory
implies that
it will be possible to realize a dual for the non-supersymmetric $SO(32)$
theory
in much the same way that we constructed a dual for the \sosixteen\ theory ---
 {\it i.e.}\/, by starting from the Type~I $SO(32)$ theory, and considering
its compactification to nine dimensions with a suitable Wilson line.
Following the logic in Sect.~4, this procedure would be rigorously implemented
by starting with an appropriate nine-dimensional Type~II interpolating model,
and then constructing its orientifold.  However, we have seen in Sect.~3.6 that
there is {\it only one}\/ nine-dimensional non-supersymmetric Type~II
interpolating model
that is continuously connected to the Type~IIB theory:  this is
Model~B$^\prime$.
Thus, the Type~I dual of the non-supersymmetric $SO(32)$ theory must be related
to the same orientifold that we have already constructed for the \sosixteen\
theory.
Indeed, tracing through the previous calculation, we see that the only
difference is that we shall choose $\gamma'_g=\bone_{32}$ (the 32-dimensional
identity matrix) instead of (\ref{gammaX}).

Choosing $\gamma'_g=\bone$ induces a number of changes in the resulting Type~I
theory relative to the \sosixteen\ case.
Of course, this is to be expected, since the corresponding heterotic
$SO(32)$ theory is itself plagued with tachyons.
This situation only serves to emphasize the uniqueness of the properties
of the \sosixteen\ theory, both on the heterotic and Type~I sides.

For radii $R\geq R^\ast$, this Type~I model continues to be tachyon-free.
We can therefore calculate the cosmological constant $\tilde\Lambda(R)$ of this
theory.
Since this theory is derived from the same interpolating Type~II model as was
the \sosixteen\ Type~I theory, its torus contribution is unchanged and its
Klein-bottle contribution continues to cancel.
Likewise, the M\"obius-strip contribution is unchanged, and the only difference
is that we now have a contribution from the cylinder.
This contribution is given by
\beq
 \Lambda_C^{(9)} ~\equiv~
     -{\calM^9\over 128\sqrt{2}}\,\left( {\rm Tr}\,\gamma_g\right)^2\,
     \int_0^\infty  {dt \over t^{11/2}}~
           {f_2^8 (q^{1/2})\over f_1^8 (q^{1/2}) } \,
         \sum_{m=-\infty}^{\infty} \, (-1)^m \,q^{m^2 a^2}~
\label{expressionloopC}
\eeq
where $\calM$ is the overall scale defined in (\ref{Mdef}) and $q\equiv
e^{-2\pi t}$.
The leading numerical factor of $(128\sqrt{2})^{-1}$ comprises
the following contributions:  a factor of $2^{-11/2}$ from
(\ref{relativenormalizations}),
a factor of two for equal contributions from the NS-NS and Ramond-Ramond states
in the cylinder,
and a factor of $1/8$ from the three factors of two in the denominators of the
trace
(\ref{fourtraces}).  For this model, we also have
$( {\rm Tr}\,\gamma_g)^2 = (32)^2 = 1024$, but for later convenience we shall
keep
$( {\rm Tr}\,\gamma_g)$ general.
We thus seek to evaluate (\ref{expressionloop})
as a function of $R/\sqrt{\alpha'}\equiv a^{-1}$.

\begin{figure}[thb]
\centerline{\epsfxsize 4.0 truein \epsfbox {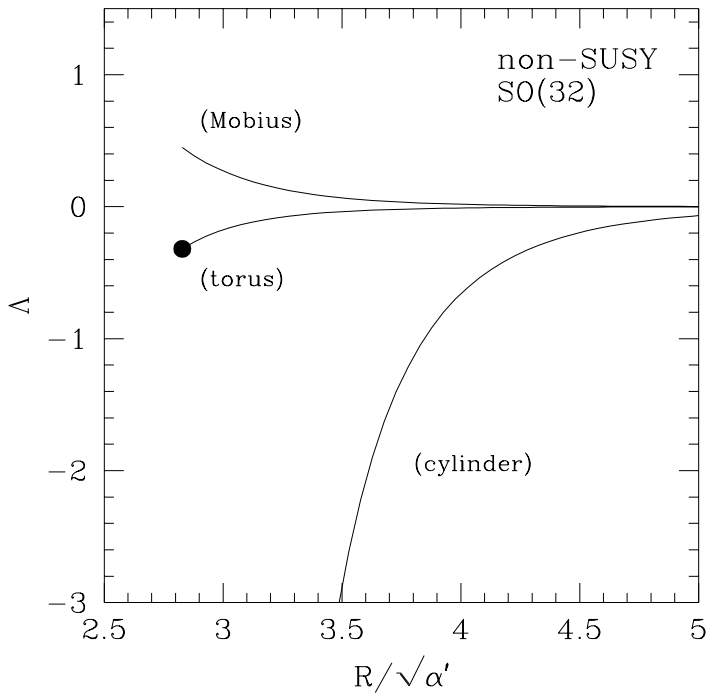}}
\caption{
  The non-vanishing contributions to the one-loop cosmological constant
    $\tilde \Lambda$
   for the non-supersymmetric $SO(32)$ Type~I interpolating model,
    plotted in units of $\half \calM^{10}$,
   as a function of the radius $R$ of the compactified dimension.
   This Type~I model reproduces the supersymmetric $SO(32)$ Type~I string
    as $R\to \infty$, and is non-supersymmetric with gauge group $SO(32)$ for
$R<\infty$.
    The torus amplitude develops a divergence below
$R^\ast/\sqrt{\alpha'}\equiv 2\sqrt{2}\approx 2.83$,
    which reflects the appearance of a tachyon in the torus amplitude below
this radius.
    The M\"obius contribution is tachyon-free for all radii, while
    the cylinder contribution diverges asymptotically as $R$ approaches
$R^\ast$ from above.
    The total cosmological constant is essentially equal to the cylinder
    contribution, and is not shown.}
\label{interplambdaA}
\end{figure}

The evaluation of this integral involves the same subtleties as our previous
calculation of the contribution from the M\"obius strip, and can be handled
similarly.  As before, we rewrite (\ref{expressionloop}) in the form
\beqn
 \Lambda_C^{(9)} &\equiv&
     -{\calM^9\over 128\sqrt{2}}\,\left( {\rm Tr}\,\gamma_g\right)^2\,
     \int_1^\infty  {dt \over t^{11/2}}~
           {f_2^8 (q^{1/2})\over f_1^8 (q^{1/2}) } \,
         \sum_{m=-\infty}^{\infty} \, (-1)^m \,q^{m^2 a^2} \nonumber\\
     && ~-~ {\calM^9\over 128\sqrt{2}}\,\left( {\rm Tr}\,\gamma_g\right)^2\,
     {1\over \sqrt{2}\,a}\,
     \int_1^\infty  dt^\prime  ~
           {f_4^8 (\tilde q^{1/2})\over f_1^8 (\tilde q^{1/2}) } \,
      \sum_{m=-\infty}^{\infty} \, \tilde q^{(m+1/2)^2/4a^2} ~
\eeqn
where $\tilde q\equiv e^{-2\pi t'}$, and then evaluate the first and
second lines separately.
Unfortunately, unlike the previous case, we see that the second
line still has a divergence from the $t^\prime\to\infty$ range of the
integration
if $a^2\geq 8$ (or if $R\leq R^\ast$).
At $R=R^\ast$, this is a divergence due to a massless state in the tree channel
({\it i.e.}\/, a tadpole divergence), and for $R<R^\ast$, this is a divergence
due
to a tachyon in the tree channel.
Thus, the cylinder amplitude is finite only for $R>R^\ast$.
It is important to note that unlike the previous divergences at $R^\ast$
(such as that for the torus amplitude), this is {\it not}\/ a discontinuous
divergence;
rather, as $R$ approaches $R^\ast$ from above, the magnitude of the cylinder
amplitude
grows without bound, with $R=R^\ast$ serving as the asymptote.
These results are plotted in Fig.~\ref{interplambdaA}.

It is clear from Fig.~\ref{interplambdaA}
that although the torus and M\"obius contributions
are not affected by the change in $\gamma_g$ from $SO(16)\times SO(16)$ to
$SO(32)$,
this change induces a very large negative cylinder contribution which is
proportional
to $({\rm Tr}\, \gamma_g)^2$ and which dominates over all other contributions.
This then significantly changes the behavior of the total cosmological constant
of the model.
Without this contribution, as plotted in Fig.~\ref{interplambdaI}
for the \sosixteen\ case, the total
cosmological constant is finite and positive for all $R\geq R^\ast$.
Indeed, although the total cosmological constant has a negative divergence for
$R<R^\ast$,
for all radii $R\geq R^\ast$ it is finite and positive, and induces a flow
towards the
direction of increasing radius.  In the present non-supersymmetric
$SO(32)$ case, by contrast, the behavior
is entirely
different:  the divergence is continuous rather than discontinuous, and the
total cosmological
constant is strictly negative.  Thus the direction of the flow is towards
smaller radii,
where tachyons and their associated divergences develop.

Despite these observations, we stress that in the range $R\geq R^\ast$, this
Type~I model
is nevertheless a consistent (if unstable) theory.  It has no tadpole
divergences, is tachyon-free,
and can be considered to be the strong-coupling dual of the heterotic
interpolating Model~A.
Moreover, it is easy to see that the massless states of these two models agree
exactly.
Indeed, the instability of this Type~I model is completely analogous to the
instability
of Model~A.

Let us now consider the D1-brane soliton of this theory.
As with the \sosixteen\ theory, we begin by compactifying
the $x_1$ direction
of the supersymmetric $SO(32)$ soliton
on a circle of radius $R$.
However, unlike the \sosixteen\ case, we now must project by the element
\beq
         \calY' ~\equiv~ \calT \,(-1)^F\, \gamma'_g~.
\eeq
Here $\calT$ and $(-1)^F$ are the same operators as we had in the \sosixteen\
case, and $\gamma'_g={\bf 1}_{32}$.
As before, we can use the orbifold notation of
Sect.~3.2 to represent the action of $(-1)^F \gamma'_g$
on the heterotic soliton as $\tilde R_{SC} R^{(1)}_{SC}$.  As indicated
in Fig.~\ref{orbrelations},
is precisely the ten-dimensional orbifold that yields the non-supersymmetric
heterotic $SO(32)$ theory from the supersymmetric heterotic $SO(32)$ theory.

The remaining calculation proceeds exactly as in the \sosixteen\ case.
Specifically,
the right-moving Class~A and Class~B sectors are identical to those in the
\sosixteen\ case, and their eigenvalues under $\calY'$ in the present case
are the same as their eigenvalues under $\calY$ in the \sosixteen\ case.
Likewise, for the left-movers,
there are only minor differences in the allowed sectors and eigenvalues.
Specifically, for the Class~A sectors we find the following results:
\beq
\begin{tabular}{||c|c||c|c||c|c||}
\hline
  \multicolumn{2}{||c||}{right-movers} & \multicolumn{4}{c||}{left-movers}
\\
\hline
\hline
  \multicolumn{2}{||c||}{~} & \multicolumn{2}{c||}{Ramond} &
\multicolumn{2}{c||}{NS} \\
\hline
   $\calY'$ & sector & $\calY'$ & sector &  $\calY'$ & sector \\
\hline
 $+1$ & $V_8$ & $-1$ & $S_{16}^{(1)}  S_{16}^{(2)}$
    &  $+1$ & $I_{16}^{(1)}  I_{16}^{(2)}$ \\
 $-1$ & $S_8$ & $-1$ & $C_{16}^{(1)}  C_{16}^{(2)}$
    & $+1$ & $V_{16}^{(1)}  V_{16}^{(2)}$ \\
\hline
\end{tabular}
\label{classAtablenew}
\eeq
while for the Class~B sectors, we find:
\beq
\begin{tabular}{||c|c||c|c||c|c||}
\hline
  \multicolumn{2}{||c||}{right-movers} & \multicolumn{4}{c||}{left-movers}
\\
\hline
\hline
  \multicolumn{2}{||c||}{~} & \multicolumn{2}{c||}{Ramond} &
\multicolumn{2}{c||}{NS} \\
\hline
   $\calY'$ & sector & $\calY'$ & sector &  $\calY'$ & sector \\
\hline
 $+1$ & $I_8$ & $+1$ & $I_{16}^{(1)}  V_{16}^{(2)}$
    &  $+1$ & $S_{16}^{(1)}  C_{16}^{(2)}$ \\
 $+1$ & $C_8$ & $+1$ & $V_{16}^{(1)}  I_{16}^{(2)}$
    & $+1$ & $C_{16}^{(1)}  S_{16}^{(2)}$ \\
\hline
\end{tabular}
\label{classBtablenew}
\eeq
Note that the Class~B table is substantially simplified relative
to (\ref{classBtable}).  This simplification occurs because the requirement
of $SO(32)$ symmetry only permits $\psi^{A_1}_{-n} \psi^{A_2}_{-n}$ or
$\psi^{A_1}_{-r} \psi^{A_2}_{-r}$ moding patterns.

Following our previous rules for combining left- and right-moving sectors,
we then obtain the following soliton sectors:
\beqn
   {\rm Class~A:}&&~~~
     V_8 \,  I_{16}^{(1)} I_{16}^{(2)} ~,~~
     V_8 \, V_{16}^{(1)} V_{16}^{(2)} ~,~~
     S_8 \, S_{16}^{(1)} S_{16}^{(2)} ~,~~
     S_8 \, C_{16}^{(1)} C_{16}^{(2)}  \nonumber\\
   {\rm Class~B:}&&~~~
     I_8 \,  I_{16}^{(1)} V_{16}^{(2)} ~,~~
     I_8 \,  V_{16}^{(1)} I_{16}^{(2)} ~,~~
     C_8 \,  S_{16}^{(1)} C_{16}^{(2)} ~,~~
     C_8 \,  C_{16}^{(1)} S_{16}^{(2)} ~.
\label{resultingsectorsnew}
\eeqn
Once again, upon comparison with (\ref{nonsusyso32}),
we see that these are precisely the set of sectors that comprise
the heterotic non-supersymmetric $SO(32)$ theory.
Thus, we have succeeded
in realizing the non-supersymmetric $SO(32)$ theory as a soliton
of its Type~I dual.

\subsection{The non-supersymmetric $SO(8)\times SO(24)$ theory}

Until now, we have not discussed the tachyonic, non-supersymmetric
heterotic $SO(8)\times SO(24)$ theory.  However, this theory is
on the same footing as the other tachyonic non-supersymmetric heterotic models
(such as the non-supersymmetric $SO(32)$ model discussed above),
and can be analyzed similarly.

For this model, we shall use a notation in which $\overline{\chi}$, $\chi$, and
$\tilde \chi$ respectively represent the
characters of the right-moving $SO(8)$ Lorentz group, the left-moving internal
$SO(8)$
gauge group, and the left-moving internal $SO(24)$ gauge group.
In terms of these characters,
the partition function of the heterotic $SO(8)\times SO(24)$ model can
then be written as:
\beqn
     Z &=&  \chibar_I \, (\chi_I \tilde \chi_S + \chi_S \tilde \chi_I )
       ~+~  \chibar_V \, (\chi_I \tilde \chi_I + \chi_S \tilde \chi_S )
\nonumber\\
     &&  ~-~  \chibar_S \, (\chi_V \tilde \chi_V + \chi_C \tilde \chi_C )
         ~-~  \chibar_C \, (\chi_V \tilde \chi_C + \chi_C \tilde \chi_V )~.
\eeqn
In addition to the gravity multiplet, the complete massless spectrum
of this model consists of the following representations of $SO(8)\times
SO(24)$:
\beqn
          {\rm vectors}:&~~& ({\bf 28}, {\bf 1}) \oplus ({\bf 1},{\bf
276})\nonumber\\
          {\rm spinors}:&~~& ({\bf 8_V}, {\bf 24})_+ \oplus
              ({\bf 8_C}, {\bf 24})_- ~.
\eeqn
Once again, cancellation of the irreducible gravitational
anomaly is manifest, even without
supersymmetry.
In addition, this string model contains
bosonic tachyons with left- and right-moving
masses $M_R^2=M_L^2= -1/2$ transforming
in the $({\bf 8_S},{\bf 1})$ representation of $SO(8)\times SO(24)$.

This model can be realized as the $\tilde R_{SC} R^{(24)}_{VC}$ orbifold
of the supersymmetric $SO(32)$ model, where $\tilde R$ indicates the
action on the right-movers (so that $\tilde R_{SC}=(-1)^F$) and where
$R^{(24)}$
indicates the action on the left-moving $SO(8)$ factors.
Since this is a $\IZ_2$ orbifold, by our previous arguments there must
exist a nine-dimensional heterotic model that interpolates between the
supersymmetric $SO(32)$ model and the non-supersymmetric $SO(8)\times SO(24)$
model.
We shall refer to this as Model~F.
Following the procedure presented in Sect.~3.4, we find that this
interpolating model has the partition function
\beqn
    Z_F ~=~  Z^{(7)}_{\rm boson} \,\times \,\bigl\lbrace ~
    \phantom{+}&\calE_0 &  \lbrack
          \chibar_V \,(\chi_I\tilde \chi_I + \chi_S\tilde\chi_S)  ~-~
          \chibar_S \,(\chi_V\tilde \chi_V + \chi_C\tilde\chi_C)
          \rbrack\nonumber\\
   +&\calE_{1/2}  & \lbrack
          \chibar_I \,(\chi_I\tilde \chi_S + \chi_S\tilde\chi_I)  ~-~
          \chibar_C \,(\chi_V\tilde \chi_C + \chi_C\tilde\chi_V)
          \rbrack\nonumber\\
   +&\calO_0 & \lbrack
          \chibar_V \,(\chi_V\tilde \chi_V + \chi_C\tilde\chi_C)  ~-~
          \chibar_S \,(\chi_I\tilde \chi_I + \chi_S\tilde\chi_S)
          \rbrack\nonumber\\
   +&\calO_{1/2} & \lbrack
          \chibar_I \,(\chi_V\tilde \chi_C + \chi_C\tilde\chi_V)  ~-~
          \chibar_C \,(\chi_I\tilde \chi_S + \chi_S\tilde\chi_I)
          \rbrack ~~\bigr\rbrace~.\nonumber\\
     ~&& ~
\eeqn
It is clear from this partition function that this interpolating
model is tachyon-free for all radii $R\geq R^\ast$ where
again the critical radius is $R^\ast =2\sqrt{2\alpha^\prime}$.
The cosmological constant $\tilde \Lambda(R)$ for this interpolating
model is plotted in Fig.~\ref{otherhetplots}.

\begin{figure}[thb]
\centerline{\epsfxsize 4.0 truein \epsfbox {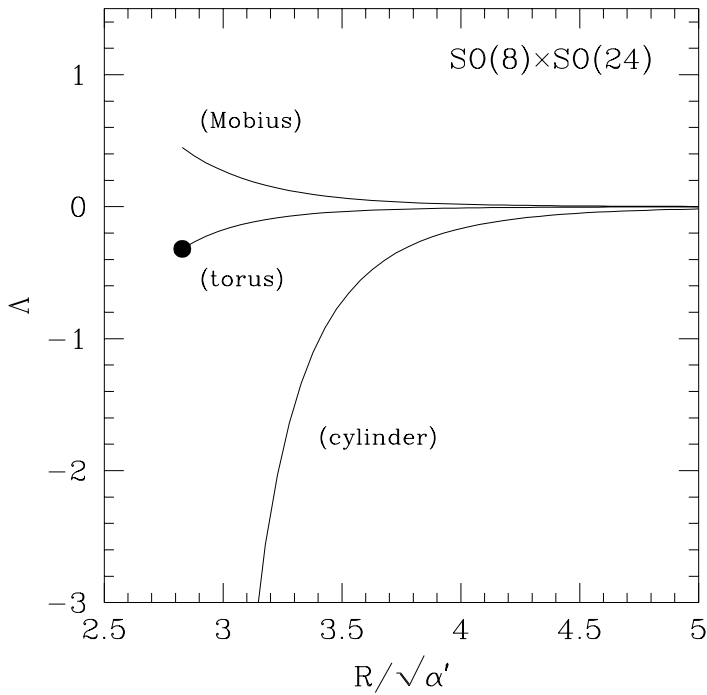}}
\caption{
  The non-vanishing contributions to the one-loop cosmological constant
    $\tilde \Lambda$
   for the non-supersymmetric $SO(8)\times SO(24)$ Type~I interpolating model,
    plotted in units of $\half \calM^{10}$,
   as a function of the radius $R$ of the compactified dimension.
   This Type~I model reproduces the supersymmetric $SO(32)$ Type~I string
    as $R\to \infty$, and is non-supersymmetric with gauge group $SO(8)\times
SO(24)$
    for $R<\infty$.
    The torus amplitude develops a divergence below
$R^\ast/\sqrt{\alpha'}\equiv 2\sqrt{2}\approx 2.83$,
    which reflects the appearance of a tachyon in the torus amplitude below
this radius.
    The M\"obius contribution is tachyon-free for all radii, while
    the cylinder contribution diverges asymptotically as $R$ approaches
$R^\ast$ from above.
    The total cosmological constant is essentially equal to the cylinder
    contribution, and is not shown.}
\label{interplambdaF}
\end{figure}

As was true for the above non-supersymmetric $SO(32)$ case,
there is only one interpolating Type~II model that connects
to the Type~IIB model and that breaks supersymmetry.  Therefore,
the strong-coupling dual of the $SO(8)\times SO(24)$ heterotic interpolating
model must again be realized through the same orientifold as the
\sosixteen\ model, and can be realized by instead choosing the Wilson line
\beq
          \gamma_g^{\prime\prime} ~=~ \pmatrix{ {\bf 1}_{8} & 0 \cr
           0 & -{\bf 1}_{24} \cr}~.
\label{gammaXnew}
\eeq
The rest of our analysis of this Type~I model then proceeds as before:
for $R\geq R^\ast$ all tadpole divergences are cancelled, and the model
is tachyon-free.  As a function of the nine-dimensional radius, its total
cosmological constant receives the same contributions as for the
non-supersymmetric
$SO(32)$ case:  the torus and M\"obius contributions are unchanged, and the
cylinder contribution is reduced by a factor of four due to the change in
the value of $({\rm Tr}\, \gamma_g)^2$ from $(32)^2$ to $(16)^2$.
The resulting contributions are plotted in Fig.~\ref{interplambdaF}.

The analysis of the soliton also proceeds as in the \sosixteen\ case.
We now project the supersymmetric $SO(32)$ soliton by the element
\beq
         \calY'' ~\equiv~ \calT \,(-1)^F\, \gamma''_g~,
\eeq
which can be identified with the element
$\tilde R_{SC} R^{(24)}_{SC}  R^{(24)}_{VS}= \tilde R_{SC} R^{(24)}_{VC}$,
just as on the heterotic side.
Performing the projection by $\calY''$ then yields Class~A and Class~B
sectors for the soliton, just as for the \sosixteen\ case.
We find the following results.  For the Class~A sectors, we have
\beq
\begin{tabular}{||c|c||c|c||c|c||}
\hline
  \multicolumn{2}{||c||}{right-movers} & \multicolumn{4}{c||}{left-movers}
\\
\hline
\hline
  \multicolumn{2}{||c||}{~} & \multicolumn{2}{c||}{Ramond} &
\multicolumn{2}{c||}{NS} \\
\hline
   $\calY''$ & sector & $\calY''$ & sector &  $\calY''$ & sector \\
\hline
 $+1$ & $V_8$ & $+1$ & $S_{8}  S_{24}$
    &  $+1$ & $I_{8}  I_{24}$ \\
 $-1$ & $S_8$ & $-1$ & $C_{8}  C_{24}$
    & $-1$ & $V_{8}  V_{24}$ \\
\hline
\end{tabular}
\label{classAtablenewnew}
\eeq
while for the Class~B sectors, we now have
\beq
\begin{tabular}{||c|c||c|c||c|c||}
\hline
  \multicolumn{2}{||c||}{right-movers} & \multicolumn{4}{c||}{left-movers}
\\
\hline
\hline
  \multicolumn{2}{||c||}{~} &
  \multicolumn{2}{c||}{Ramond} &
  \multicolumn{2}{c||}{NS}  \\
\hline
   $\calY''$ & sector & $\calY''$ & sector & $\calY''$ & sector \\
\hline
 $+1$ & $I_8$
      & $-1$ & $I_{8}  V_{24}$
           & $+1$ & $S_{8}  C_{24}$ \\
 $+1$ & $C_8$
      & $+1$ & $V_{8}  I_{24}$
           & $-1$ & $C_{8}  S_{24}$ \\
    ~ & ~
      & $+1$ & $S_{8}  I_{24}$
           & $+1$ & $I_{8}  S_{24}$ \\
    ~ & ~
      & $+1$ & $C_{8}  V_{24}$
           & $+1$ & $V_{8}  C_{24}$ \\
    ~ & ~
      & $+1$ & $I_{8}  C_{24}$
           & $-1$ & $C_{8}  I_{24}$ \\
    ~ & ~
      & $+1$ & $V_{8}  S_{24}$
           & $-1$ & $S_{8}  V_{24}$ \\
\hline
\end{tabular}
\label{classBtablenewnew}
\eeq
Joining these sectors together in accordance with the assumptions
made for the \sosixteen\ case (and respecting the symmetry under exchange
of the $SO(8)$ and $SO(24)$ gauge factors), this then yields the following
sectors:
\beqn
   {\rm Class~A:}&&~~~
     \overline{V}_8 \,  I_{8} I_{24} ~,~~
     \overline{V}_8 \, S_{8} S_{24} ~,~~
     \overline{S}_8 \, V_{8} V_{24} ~,~~
     \overline{S}_8 \, C_{8} C_{24}  \nonumber\\
   {\rm Class~B:}&&~~~
     \overline{I}_8 \,  I_{8} S_{24} ~,~~
     \overline{I}_8 \,  S_{8} I_{24} ~,~~
     \overline{C}_8 \,  V_{8} C_{24} ~,~~
     \overline{C}_8 \,  C_{8} V_{24} ~.
\label{resultingsectorsnewnew}
\eeqn
As expected, this is precisely the set of sectors that comprise the
non-supersymmetric $SO(8)\times SO(24)$ heterotic string, with the Class~A
sectors playing the role of the untwisted sectors, and the Class~B sectors
playing the role of the twisted sectors required by modular invariance.
Once again, we see that modular invariance has ``magically'' been restored 
on the soliton
through the action of the half-rotation operator $\calT$.

We thus conclude that this Type~I theory is the dual of the non-supersymmetric
$SO(8)\times SO(24)$ interpolating
theory.  As in the $SO(32)$ case, their massless states
agree exactly, and the D1-brane soliton of the Type~I theory 
has the correct behavior.
Furthermore, the instability of the Type~I theory exactly mirrors
the instability of the heterotic model.

\subsection{The non-supersymmetric $U(16)$ theory}

As indicated in Fig.~\ref{wheels}, there is one other non-supersymmetric
heterotic string model which can be realized as a $\IZ_2$ orbifold of the
supersymmetric $SO(32)$ heterotic string model --- this is the
$U(16)$ model.
Strictly speaking, at this point we should distinguish between the gauge
group $SO(32)$ and ${\rm Spin}(32)/\IZ_2$, for
the $\IZ_2$ that generates this orbifold is a $\IZ_2$ in
${\rm Spin}(32)/\IZ_2$ but not in $SO(32)$.

Given this $\IZ_2$ orbifold relation, it is clear from the discussion
in Sect.~3 that a nine-dimensional heterotic string model exists which
interpolates between the supersymmetric $SO(32)$ string and the
non-supersymmetric
$U(16)$ string, and it can be constructed in precisely the same manner as
the previous interpolating models.  Furthermore, since Model~B$^\prime$ is the
unique nine-dimensional non-supersymmetric Type~II model that connects to
the Type~IIB model as $R\to \infty$, we once again find that our previous
orientifold must again provide the strong-coupling dual for this heterotic
interpolating
model.  Indeed, the only subtlety here is in the choice of the Wilson line:
rather than choose any of the {\it symmetric}\/ Wilson lines $\gamma_g$, we
instead choose the anti-symmetric Wilson line
\beq
         \gamma'''_g ~=~ \pmatrix { 0 &   - {\bf 1}_{16} \cr
          {\bf 1}_{16} & 0 \cr}~.
\eeq
This then generates a $U(16)$ gauge symmetry, and the rest of the analysis
follows as before.
The only difference in the calculation of the one-loop cosmological constant
is that the cylinder contribution cancels as for
the \sosixteen\ case and that there are no open-string tachyonic divergences
for $R>0$.  However, the M\"obius strip contribution now has the opposite
sign relative to the \sosixteen\ case, and therefore once again the theory
is unstable.

\subsection{The other non-supersymmetric ten-dimensional theories}

Finally, we consider the remaining non-supersymmetric ten-dimensional theories
shown in Fig.~\ref{wheels}, namely the $SO(16)\times E_8$, $(E_7\times
SU(2))^2$,
and $E_8$ theories.

As indicated in Fig.~\ref{wheels}, these theories can all be realized as
$\IZ_2$ orbifolds of the supersymmetric $E_8\times E_8$ theory.
Unfortunately, they cannot be realized as $\IZ_2$ orbifolds of the
supersymmetric
$SO(32)$ theory, and consequently there do not exist nine-dimensional
interpolating
models of the sort we have been considering which interpolate between the
supersymmetric
$SO(32)$ model and any of these non-supersymmetric models.  Indeed, these
models
are connected to the $SO(32)$ model only through $\IZ_2\times \IZ_2$ orbifolds,
but such orbifolds cannot be ``interpolated'' in the manner we have discussed
here.  Our reasoning is as follows.
In order to achieve the $SO(32)$ theory at $R=\infty$, all $\IZ_2$ elements
must be adjoined to $\IZ_n$ elements for some $n> 1$, where these $\IZ_n$
elements are translations by $2\pi k R/n$, where $k\not=0$ (mod $n$).
It is easy to see that this is impossible, and thus no interpolations can
be constructed.
Consequently, since we cannot connect these models smoothly to
the supersymmetric $SO(32)$ model, we do not expect them to
have Type~I strong-coupling duals.

Of course, this result is precisely what we expected,
for we already know that it is impossible to realize
exceptional gauge groups perturbatively
via Type~I Chan-Paton factors on nine-branes.
Indeed, all of these remaining non-supersymmetric string models have
gauge groups with exceptional factors.
Thus, our inability to connect these heterotic models to the supersymmetric
$SO(32)$ heterotic model via a smooth interpolating model can in fact be taken
as further evidence for the legitimacy of our entire approach based
on such interpolations.

It is interesting, however, that these models all have $\IZ_2$ interpolations
with the $E_8\times E_8$ model.
Indeed, as shown in Fig.~\ref{interpfig},
Model~D interpolates between the $SO(16)\times E_8$ model and the $E_8\times
E_8$
model.
This suggests that strong-coupling duals for these remaining non-supersymmetric
heterotic theories might be realized via deformations of M-theory.

We also note from Fig.~\ref{interpfig} that the $SO(16)\times SO(16)$ theory
is the only theory that has $\IZ_2$ interpolations to {\it both}\/ the
supersymmetric $SO(32)$ theory and the $E_8\times E_8$ theory.
This suggests that even though we have not succeeded in finding a dual
for the $SO(16)\times SO(16)$ theory itself (and have instead only found
a dual for its nine-dimensional interpolating Model~B),
there might be a deformation of M-theory in which the analogue of
the $R\to 0$ limit can be taken.  This would result in an $M$-theoretic
dual for the full ten-dimensional $SO(16)\times SO(16)$ theory.

Finally, we make a remark concerning the supersymmetric $E_8\times E_8$ theory
itself.
Although this theory has a known strong-coupling dual, namely M-theory,
it is evident from Figs.~\ref{orbrelations} and \ref{interpfig} that this
theory can itself be realized as a $\IZ_2$ orbifold of the $SO(32)$ string.
Consequently there exists a nine-dimensional interpolating model
(specifically Model~A in Fig.~\ref{interpfig}) which connects the
supersymmetric $SO(32)$ theory  to the supersymmetric $E_8\times E_8$ theory.
This connection is not new, and is essentially equivalent to the
connection discussed in Ref.~\cite{Ginsparg} which made use of
background field variations.
What this connection implies, however, is that it is possible to realize
the strong-coupling dual of the $E_8\times E_8$ theory at $R=0$ as a {\it
Type~I}\/ theory
(or equivalently, the ten-dimensional $E_8\times E_8$ theory as a
Type~I$^\prime$ theory).
More precisely, just as for the \sosixteen\ case,
it will be possible to realize a Type~I dual for the heterotic interpolating
Model~A.
Even more interestingly, we will also be able to realize the $E_8\times E_8$
heterotic theory as a Type~I soliton.
This will be discussed in Ref.~\cite{e8paper}.

\section{Interpretation and Stability Analysis}
\setcounter{footnote}{0}

Given the results of the prior sections, we are now in a position
to address some long-standing questions pertaining to the perturbative
and non-perturbative stability of these non-supersymmetric strings.
Note that all of the duality relations we have found apply to tachyon-free
interpolating theories.  This is important, since these
are the only theories for which a stability analysis is meaningful.
We shall therefore focus on our $SO(16)\times SO(16)$ interpolating model.
The analysis of the previous section makes it abundantly clear that the
qualitative behavior of this case will be quite different from the
behavior of the other cases.

\subsection{Basic relations}

We begin by recalling some basic facts that will be important for our analysis.
First, we define our variables:  we shall denote by $\lambda_H^\ten$ the
heterotic
ten-dimensional dimensionless coupling, $R_H$ the radius of the heterotic
compactification,
$\lambda_I^\ten$ the ten-dimensional Type~I dimensionless coupling,
and $R_I$ the radius of the Type~I compactification.
The corresponding nine-dimensional couplings are then given by
\beq
  \lambda_I^\nine ~=~ {\lambda_I^\ten \over \sqrt{R_I}} ~,~~~~~~
  \lambda_H^\nine ~=~ {\lambda_H^\ten \over \sqrt{R_H}} ~.
\label{ninetencouplings}
\eeq
Note that with these conventions, the nine-dimensional couplings are
dimensionful.

Let us also recall how the cosmological constants $\Lambda$ are to be
interpreted.
In general, these cosmological constants are simply one-loop vacuum energies or
zero-point functions, and from conformal invariance it is known that the
one-loop dilaton
one-point function (or dilaton tadpole diagram) is always proportional to
$\Lambda$.
Thus in non-supersymmetric theories there will generally be a non-zero dilaton
potential
whose slope is given by $\Lambda$.  The existence of this dilaton potential
then pushes
the dilaton $\phi$, and with it the corresponding coupling $\lambda\sim
e^\phi$,
in a direction determined by the sign of $\Lambda$.  With our present sign
conventions
in which fermionic string states contribute positively to $\Lambda$, we have
that
\beqn
   \Lambda > 0 ~~~&\Longrightarrow&~~~ \hbox{flow to weak coupling}~\lambda
 {}~\nonumber\\
   \Lambda < 0 ~~~&\Longrightarrow&~~~ \hbox{flow to strong coupling}~\lambda
 {}~.
\label{lambdaflow}
\eeqn
This qualitative result holds both on the Type~I and heterotic sides,
and well as for cosmological constants and couplings in arbitrary dimensions
$D$.
These results are consistent with the interpretations in Ref.~\cite{scales}.
Likewise, we can also easily interpret the function $\Lambda(R)$ and its effect
on the radius $R$,
for in this case $\Lambda(R)$ itself serves as the potential that pushes the
radius $R$
towards different values.
Thus, in analogy with (\ref{lambdaflow}), we find that
\beqn
   \partial \Lambda/\partial R > 0 ~~~&\Longrightarrow&~~~ \hbox{flow to small
radius}~R ~\nonumber\\
   \partial \Lambda/\partial R < 0 ~~~&\Longrightarrow&~~~ \hbox{flow to large
radius}~R~.
\label{radiusflow}
\eeqn

Next, let us recall the standard
predictions of strong/weak coupling duality in the supersymmetric
$SO(32)$ case.
In the uncompactified limit, we have the pure ten-dimensional heterotic/Type~I
relation
\beq
             R_I=R_H=\infty:~~~~~~~  \lambda_H^\ten ~=~ {1\over
\lambda_I^\ten}~.
\label{dualten}
\eeq
If we compactify both sides on circles of radii $R_H$ and $R_I$ respectively,
this then
implies the relations
\beq
                  R_H~=~ {R_I\over \sqrt{\lambda_I^\ten}} ~,~~~~~~
                  R_I~=~ {R_H\over \sqrt{\lambda_H^\ten}} ~.
\label{dualnineone}
\eeq
Using (\ref{ninetencouplings}) to express these results in terms of
the nine-dimensional couplings $\lambda_{I,H}^\nine$, we then find
\beq
        \lambda_I^\nine ~=~ {1\over [\lambda_H^\ten]^{3/4} R_H^{1/2}}~,~~~~~
        \lambda_H^\nine ~=~ {1\over [\lambda_I^\ten]^{3/4} R_I^{1/2}}~.
\label{dualninetwo}
\eeq
Finally, we also recall that $T$-duality takes $R\to R'=1/R$ while preserving
$\lambda^\nine$.
This implies that $(\lambda^\ten)^\prime = \sqrt{\alpha'} \lambda^\ten/R$ or
equivalently
$\lambda^\ten=\sqrt{\alpha'}(\lambda^\ten)^\prime/R'$.

Given these relations, our first task is to consider what sorts of relations we
expect to have for our {\it non}\/-supersymmetric \sosixteen\ case
in which the supersymmetric ten-dimensional heterotic and Type~I $SO(32)$
theories are
compactified on a circle with a twist in such a way that supersymmetry is
broken.
It is here that the power of our {\it continuous interpolations}\/ becomes
evident.
In particular, note that as long as $R_I> R^\ast \equiv 2\sqrt{2\alpha'}$,
there are
 {\it no discontinuities}\/ in passing from our nine-dimensional
non-supersymmetric
theories to the limiting supersymmetric  ten-dimensional theories.
Indeed, supersymmetry is restored in a smooth fashion as a function of the
radius $R$,
with no discrete changes or phase transitions.  This then implies that
(\ref{dualten})
should hold as $R\to\infty$, even in our generally non-supersymmetric case, and
this in turn
implies that
the nine-dimensional relations (\ref{dualnineone}) and (\ref{dualninetwo})
should hold as well.
We emphasize again that it is only the power of our approach based on smooth
interpolations
that enables us to show that such relations continue to be valid.

Given these results, we can then examine the physics in
different situations by appropriately  choosing the initial values of a primary
set of
chosen parameters, such as $\lambda_H^\ten$ and $R_H$.
As is evident from the above discussion, we shall also need to distinguish two
phases
of our Type~I theory depending on whether $R_I> R^\ast$ or $R_I<R^\ast$.
Note that it is the Type~I radius that is crucial since it is this radius that
governs whether tachyons develop and/or extra massless states appear.
We shall denote the $R_I> R^\ast$ case as ``Phase~I'', and denote the case
with $R_I<R^\ast$ as ``Phase II''.  Thus Phase~I is tachyon-free, Phase~II is
generally
tachyonic, and extra massless states appear on the boundary between the two
where the phase transition occurs.

Note that in terms of our fundamental heterotic parameters
$(\lambda_H^\ten,R_H)$,
the condition $R_I>R^\ast$ is equivalent to
\beq
 {\rm Phase~I:}~~~~~~~~   \lambda_H^\ten ~<~ {R_H^2\over (R^\ast)^2} ~,~~~
   \lambda_H^\nine~<~ {R_H^{3/2}\over (R^\ast)^2} ~.
\label{PhaseIlimits}
\eeq
The fact that $\lambda_H^\ten$ and $\lambda_H^\nine$ are bounded only from
above implies
that our tachyon-free Phase~I is compatible with weak heterotic coupling.
Likewise, using heterotic/Type~I duality, we find that
\beq
   \lambda_I^\nine~>~ {(R^\ast)^2 \over R_H^{5/2}}~.
\label{otherPhaseIlimit}
\eeq

\vfill\eject
\subsection{Analysis for Phase~I:~~  $R_I>R^\ast$}

\medskip
\noindent  {\sl \underbar{Case~I}:~~  Perturbative on heterotic side}
\bigskip

Let us begin our analysis of the \sosixteen\ string by focusing on the case
in which the heterotic side is perturbative.
This can be arranged by taking $\lambda_H^\ten$ sufficiently small relative to
$R_H$, so that $\lambda_H^\nine$ is perturbative.
In such situations, we can trust our perturbative calculation of the one-loop
cosmological constant $\Lambda_H^\nine(R)$, a plot of which
is shown in Fig.~\ref{interplambda}.
  From this result, it is immediately apparent that for all $R_H>0$, the theory
flows to weak coupling and large radius.
We thus conclude that in this case,
the theory flows to the supersymmetric $SO(32)$ heterotic string.

There are a number of subtleties which should be discussed in deriving this
result.
First, the radius $R_H$ is not the only modulus on the heterotic side, and one
might question whether there exist other moduli whose flows might alter this
result.
Indeed, in nine dimensions there are a total of 17 moduli:  one is the radius
$R_H$ whose variations we have been discussing, and the other 16 are the Narain
compactification moduli which can be interpreted as the expectation
values of the 16 $U(1)$ gauge symmetries of the original $SO(32)$.
The crucial point, however,
is that flows along any of these other 16 directions
will break our \sosixteen\ gauge group.  Since we know that \sosixteen\
is the unique gauge group for which a consistent tachyon-free theory exists,
any flows in these other 16 directions will necessarily introduce tachyons and
completely destabilize the theory.  At the very least, this would lead to a
phase transition and drastically change the nature of the theory.  Of course,
such a possibility is always a generic risk when dealing with arbitrary flows
in the moduli space of non-supersymmetric string models.  Thus, in our
subsequent
analysis we shall not address this possibility, but we must be aware that it
exists.

The second subtlety concerns the actual $R=\infty$ limit.
As discussed in Refs.~\cite{Nair,GV},
at this limiting point one actually must perform a volume-dependent
rescaling of the spacetime metric in order to ensure that the
gravitational term remains in canonical form.
This has the potential to alter the direction of flow at this limiting point,
so that at $R=\infty$ the attractive force pulling the theory to $R=\infty$
might discontinuously become a repulsive force pushing the theory away from
$R=\infty$.
The general result of such an analysis is as follows \cite{Nair}.
First, we define the quantity $\tilde \Lambda(R)$
as in (\ref{tildelambdadefI}), so that
$\tilde\Lambda$ has no radius dependence
in the $R\to \infty$ limit and approaches a constant.
We then define
\beq
  W~\equiv~ \tilde \Lambda \, R^{-2/(D-2)}
         =~ \tilde \Lambda \, R^{-2/7}~.
\eeq
The general result is then that
\beqn
      \lim_{R\to\infty} \, \left({\partial W\over \partial R}\right) <0
 {}~~~~&\Longrightarrow&~~~~
        \hbox{attracts to}~ R=\infty \nonumber\\
      \lim_{R\to\infty} \, \left({\partial W\over \partial R}\right) >0
 {}~~~~&\Longrightarrow&~~~~
        \hbox{repels from}~ R=\infty~.
\eeqn
In our nine-dimensional case, we see that $W\sim R^{-2/7}$.  Thus the flow
towards $R_H=\infty$
is preserved all the way up to {\it and including}\/ the $R_H=\infty$ limit.

Finally, we may ask what happens at $R_H=0$.  Na\"\i vely,
the plot in in Fig.~\ref{interplambda} would seem to suggest that at
this point the theory flows to weak coupling, but is otherwise stable at
$R_H=0$.
In actuality, however, this limit is a good deal more subtle, and will be
discussed below.

Note that, as expected,  none of the results in this case depended in any way
on the existence of
our heterotic/Type~I duality.  However, they depend crucially on the existence
of our
nine-dimensional interpolating models.  Such interpolations  enable these flows
to be discussed in terms of
of a single parameter $R$ whose variations
neither alter the gauge group (except at $R=\infty$) nor introduce tachyons,
but nevertheless break supersymmetry.

\bigskip
\noindent  {\sl \underbar{Case~II}:~~  Perturbative on Type~I side}
\bigskip

We now address the opposite extreme of Phase~I, namely the case when our theory
is perturbative on the Type~I side.  This situation can be realized in
terms of the heterotic parameters $(\lambda_H^\ten, R_H)$
as follows.
First, we choose an initial, large, fixed value of $\lambda_H^\ten$ which,
because we are in Phase~I, must be consistent with (\ref{PhaseIlimits}).
This can always be arranged.
Then, we simply choose $R_H$ sufficiently large that the lower bound in
(\ref{otherPhaseIlimit})
is sufficiently small.  Thus $\lambda_I^\nine$ can be perturbative, even within
Phase~I.

The perturbativity of the Type~I theory implies that our calculation of the
Type~I cosmological constant can be trusted.
The results of this calculation are shown in Fig.~\ref{interplambdaI};  note
that because we are in Phase~I, with $R_I>R^\ast$, the total cosmological
constant
is finite and positive.
Our analysis is then precisely as for the above perturbative heterotic case,
and we conclude that our \sosixteen\ Type~I theory flows to the supersymmetric
$SO(32)$ Type~I theory.

We also remark that the Type~I soliton plays little role in this analysis.
This is, of course, to be expected.  Recall that the mass of the soliton
goes as
\beq
        M_{\rm soliton} ~\sim~ {T_F\, R_I\over \lambda_I^\ten}~~~~ {\rm
where}~~~
    T_F~\equiv~ {1\over 2\pi\alpha'}~.
\eeq
Thus, since $\lambda_I^\ten$ is small in Case~II of Phase~I, and since in this
limit
we find that $R_I\to\infty$, the soliton is always heavy and in fact becomes
unstable.
Thus we do not expect this soliton to have any effect on the analysis.

Thus, we conclude that within Phase~I, we have good evidence that
our non-supersymmetric interpolating heterotic and Type~I
models are dual to each other.
Indeed, even though we have only considered the cases in which either the
heterotic
or Type~I theories are perturbative, the absence of phase transitions or
discontinuities
in Phase~I suggests that this result persists even to other intermediate ranges
for
the heterotic and Type~I couplings.
We therefore conclude that throughout Phase~I, our heterotic \sosixteen\
interpolating model
flows to the ten-dimensional supersymmetric $SO(32)$ theory.

\subsection{Conjecture for Phase~II:~~  $R_I<R^\ast$}

Let us now turn to Phase~II.  In terms of our fundamental heterotic
parameters $(\lambda_H^\ten, R_H)$, this is the region for which
\beq
 {\rm Phase~II:}~~~~~~~~   \lambda_H^\ten ~>~ {R_H^2\over (R^\ast)^2} ~,~~~
   \lambda_H^\nine~>~ {R_H^{3/2}\over (R^\ast)^2} ~.
\label{PhaseIIlimits}
\eeq
Unlike the analogous Phase~I constraints (\ref{PhaseIIlimits}),
these constraints show that Phase~II is generally not compatible with
weak heterotic coupling except at very small radii $R_H$.

The fact that this phase corresponds on the Type~I
side to the tachyonic range $R_I<R^\ast$ suggests that the dual
theory in this range, if one exists, is not likely to be the Type~I theory
we have constructed.  This is not entirely unexpected, since we have already
seen that there is a phase transition in our Type~I theory at $R_I=R^\ast$.
This phase transition is evidenced by
the fact that the Type~I one-loop cosmological constant has a discontinuity at
this radius due to the sudden appearance of tachyons.

The question that we must address, then, is the nature of a possible dual
theory
in this range.  One natural possibility, of course, is M-theory.
Recall that M-theory may be defined as the eleven-dimensional theory
whose compactification on a line segment of length $\rho$ gives the
$E_8\times E_8$ string.  If we further compactify this string on a circle
of radius $R_H$, then $\rho$ in M-theory units is given as
\beq
           \rho ~=~ \left( {\lambda_H^\ten\over R_H} \right)^{2/3}~.
\eeq
Using the Phase~II constraints (\ref{PhaseIIlimits}), we then find
the bound
\beq
         \rho ~>~ { [\lambda_H^{(10)}]^{1/3}\over (R^{\ast})^{2/3}}~.
\eeq
Thus we see that $\rho$ is bounded from below, and can easily be large.
Phase~II is thus the region in which we expect an M-theory description
to be valid.

If we expect to realize a non-supersymmetric dual via M-theory,
the next step is to find solutions of M-theory in nine dimensions for
which supersymmetry is broken.
This issue has been studied in Ref.~\cite{horava},
and the low-energy analysis seems to indicate that there are no stable
solutions of  M-theory on a line segment which break spacetime supersymmetry
in nine dimensions.

We therefore face two possibilities.
The first possibility is that enough additional spacetime dimensions of the
theory are
compactified so that a stable non-supersymmetric solution of M-theory on a line
segment is achieved.  This would seem to require at least three additional
compactified
dimensions, leading to a theory in $D\leq 6$.
This is an attractive possibility, and might be especially relevant for
lower-dimensional
non-supersymmetric theories and their strong-coupling duals.

The second possibility would be to remain in $D=9$ and $D=10$.  However,
from the above arguments, we know that the theory will be forced to
flow to a {\it supersymmetric}\/ point.
There are, of course, only two possible supersymmetric points to which the
theory in Phase~II could flow:  the heterotic $SO(32)$ theory, or the
$E_8\times E_8$
theory.
We believe that $E_8\times E_8$ is the more natural candidate, and can offer
several arguments in its favor.

First, we know that the Phase~I theory already flows to the $SO(32)$ theory.
It seems implausible that {\it despite a discontinuous phase transition}\/,
both phases of the theory could behave similarly.
We therefore believe that the behavior in Phase~II
should be fundamentally different from that in Phase~I.

Second, we know that it is sensible, at least perturbatively,
that the $SO(16)\times SO(16)$ theory could flow to the
supersymmetric $E_8\times E_8$ theory.  Indeed, this behavior is
already exhibited in the heterotic interpolating Model~C
(as sketched in Fig.~\ref{interpfig}).  It is not unreasonable
to expect that this perturbative behavior might have a non-perturbative
realization as well.

Our third argument, however, is perhaps the most compelling.
Let us consider our \sosixteen\ soliton at weak coupling $\lambda_I^\ten$.
(This region is roughly equivalent to large $\lambda_H^\ten$, even though
we do not expect our duality mappings to hold here.)
As we shall now argue, in this region
the soliton has the massless worldsheet fields
of the supersymmetric $E_8\times E_8$ heterotic string.
Thus, we expect the fundamental \sosixteen\ string at strong coupling
to behave as a fundamental $E_8\times E_8$ heterotic string.

In order to see this behavior of the D-string soliton,
let us first recall how our soliton was derived in Sect.~5.
In Sect.~5,
we found that our soliton was comprised of a number of right-moving
and left-moving sectors which in turn
arose in two categories that we called Class~A and Class~B.
These sectors were listed in (\ref{classAtable}) and (\ref{classBtable}).
In order to construct the full soliton theory,
our procedure for joining these separate sectors
involved a set of assumptions which were discussed
in Sect.~5.  Among these were two critical assumptions:  we did not
permit Class~A and Class~B sectors to mix, and we permitted left- and
right-moving
sectors to combine and survive the $\calY$ projection as long as they {\it
together}\/
had a combined $\calY$ eigenvalue $+1$.  The first of these assumptions
is essentially a selection rule, and the second was implemented to reflect
the {\it interactions}\/ that we expected to occur between left- and
right-moving
sectors at strong coupling.
At weak (or zero) coupling, however, both of these should assumptions no longer
apply.
Specifically, we should no longer impose a selection rule between Class~A and
Class~B sectors, and likewise we should impose a more stringent
$\calY$-projection that forces each sector, whether left-moving or
right-moving,
to individually have the $\calY$-eigenvalue $+1$.
If we make these new weak-coupling assumptions,
then our Class~A and Class~B tables collapse into the single table
\beq
\begin{tabular}{||c|c||c|c||c|c||}
\hline
  \multicolumn{2}{||c||}{right-movers} & \multicolumn{4}{c||}{left-movers}
\\
\hline
\hline
  \multicolumn{2}{||c||}{~} & \multicolumn{2}{c||}{Ramond} &
\multicolumn{2}{c||}{NS} \\
\hline
   $\calY$ & sector & $\calY$ & sector &  $\calY$ & sector \\
\hline
 $+1$ & $V_8$ & $+1$ & $S_{16}^{(1)}  S_{16}^{(2)}$
    &  $+1$ & $I_{16}^{(1)}  I_{16}^{(2)}$ \\
 $+1$ & $I_8$
      & $+1$ & $V_{16}^{(1)}  I_{16}^{(2)}$
           & $+1$ & $S_{16}^{(1)}  C_{16}^{(2)}$ \\
 $+1$ & $C_8$
      & $+1$ & $I_{16}^{(1)}  S_{16}^{(2)}$
           & $+1$ & $S_{16}^{(1)}  I_{16}^{(2)}$ \\
    ~ & ~
      & $+1$ & $V_{16}^{(1)}  C_{16}^{(2)}$
           & $+1$ & $C_{16}^{(1)}  V_{16}^{(2)}$ \\
    ~ & ~
      & ~  & ~
           & $+1$ & $I_{16}^{(1)}  C_{16}^{(2)}$ \\
    ~ & ~
      & ~  & ~
           & $+1$ & $V_{16}^{(1)}  S_{16}^{(2)}$ \\
\hline
\end{tabular}
\label{finaltable}
\eeq
and we find (after imposing our remaining constraints)
that the only surviving sector combinations are:
\beqn
   &&~~~
     V_8 \,  I_{16}^{(1)} I_{16}^{(2)} ~,~~
     V_8 \, S_{16}^{(1)} S_{16}^{(2)} ~,~~
     V_8 \,  S_{16}^{(1)} I_{16}^{(2)} ~,~~
     V_8 \, I_{16}^{(1)} S_{16}^{(2)} ~,~~\nonumber\\
   &&~~~
     C_8 \,  I_{16}^{(1)} I_{16}^{(2)} ~,~~
     C_8 \, S_{16}^{(1)} S_{16}^{(2)} ~,~~
     C_8 \,  S_{16}^{(1)} I_{16}^{(2)} ~,~~
     C_8 \, I_{16}^{(1)} S_{16}^{(2)} ~,~~\nonumber\\
   &&~~~
     I_8 \,  V_{16}^{(1)} C_{16}^{(2)} ~,~~
     I_8 \,  C_{16}^{(1)} V_{16}^{(2)} ~.~~
\eeqn
Remarkably, the first two lines are precisely the set of sectors that comprise
the
supersymmetric $E_8\times E_8$ heterotic string.
Moreover, the extra sectors in the last line are purely massive, and since
they have left- and right-moving momentum modings $m'=m/2=1/2$,  we see from
(\ref{momentump1}) that their total momenta grow as $1/(2R_I)$.
Thus as $R_I\to 0$, these sectors become infinitely massive and completely
decouple,
leaving us with a soliton whose massless states are those of
heterotic $E_8\times E_8$ string.
Indeed, the \sosixteen\ theory is the only non-supersymmetric theory whose
soliton behaves this way when these assumptions are so modified.

Putting all of this together,
we therefore make the conjecture that in Phase~II, our \sosixteen\
theory flows to the strongly coupled $E_8\times E_8$ theory.

\subsection{Conjecture for the boundary between Phase~I and Phase~II:~~
$R_I=R^\ast$}

Finally, we make two brief comments concerning the boundary between Phase~I and
Phase~II.
In terms of our heterotic variables, this boundary line corresponds to
\beq
             {\rm boundary:}~~~~~~~
                     \lambda_H^\ten ~=~ {R_H^2 \over (R^\ast)^2}~.
\label{boundary}
\eeq

In order to analyze this boundary line ``perturbatively'', we choose to examine
the region on the boundary where $\lambda_H^\ten$ and $R_H$ are both small.
This means that we should really be using the $T$-dual coupling
$\lambda_H^\prime\equiv \sqrt{\alpha'} \lambda_H^\ten/R_H$, in terms of
which (\ref{boundary}) becomes
\beq
                     \lambda_H^\prime  ~=~ {\sqrt{\alpha'}\, R_H \over
(R^\ast)^2}~.
\eeq
Thus, if we now examine the $R_H=0$ limit (which corresponds to the
actual ten-dimensional \sosixteen\ string), we learn that {\it any}\/ non-zero
coupling $\lambda_H^\prime$ will push the theory into Phase~II.
Thus, our first conclusion is that the ten-dimensional \sosixteen\ string
is likely to behave as an $E_8\times E_8$ string for any non-zero coupling.

Our second comment concerns the stability of the boundary between the two
phases.
At $R_H=0$ (or sufficiently close to $R_H=0$), the analysis of Ref.~\cite{Nair}
can be used to show that there is a repulsive ``force'' which prevents falling
into Phase~I.  Although this result is derived only for the boundary line
sufficiently near $R_H=0$, we expect that this result can be extended by
continuity
arguments to cover the entire boundary line.
We therefore expect that the entire boundary line is stable against
passing into Phase~I.


\section{Conclusions and Discussion}
\setcounter{footnote}{0}

In this paper, we have undertaken an analysis
of the extent to which strong/weak coupling duality can be extended
to non-supersymmetric strings.  We focused primarily on the tachyon-free
\sosixteen\ string, but also considered a variety of other non-supersymmetric
tachyonic heterotic strings in ten dimensions.  Using an approach involving
interpolating models, we were able to continuously connect these
heterotic non-supersymmetric models to the heterotic supersymmetric $SO(32)$
model
for which a strong coupling dual is known.  In this way we were then able to
generate a set of non-supersymmetric Type~I models which are dual to the
heterotic interpolating models.
Specifically, in each case (and in the tachyon-free range 
$R>R^\ast\equiv 2\sqrt{2\alpha'}$), we found
that the massless
spectra agreed exactly, and that the D1-brane soliton of the Type~I theory
could be understood as the corresponding heterotic theory.
This latter observation followed as a result of a novel method that we
developed
for analyzing the solitons of non-supersymmetric Type~I theories.
As far as we are aware, our results imply the first known duality relations
between
non-supersymmetric tachyon-free theories.

\begin{figure}[thb]
\centerline{
         \epsfxsize 4.0 truein \epsfbox {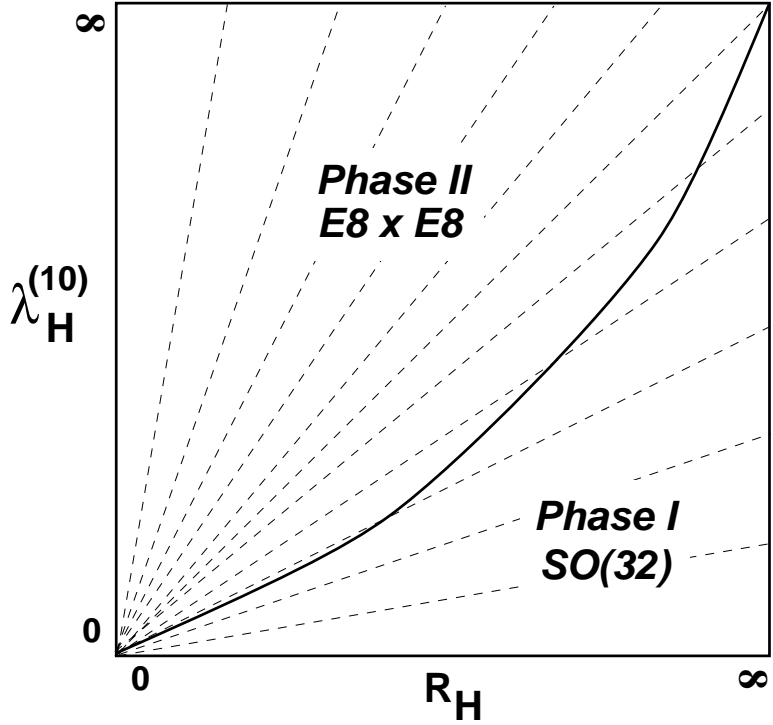}
  }
\caption{
     The proposed phase diagram for the \protect\sosixteen\ interpolating
     model.  For $\lambda_H^\ten < R_H^2 / (R^\ast)^2$, the theory
     is in Phase~I and flows to the ten-dimensional weakly coupled heterotic
     supersymmetric $SO(32)$ theory.
     For $\lambda_H^\ten > R_H^2 / (R^\ast)^2$, by contrast, the theory
     is expected to flow to the ten-dimensional strongly coupled supersymmetric
     $E_8\times E_8$ string.  On the boundary, the theory is stable against
     passing into Phase~I.  The dotted lines are contours of constant
     $T$-dual coupling $\lambda_H^\prime \equiv
\sqrt{\alpha'}\lambda_H^\ten/R_H$.
    }
\label{phases}
\end{figure}

The existence of these non-supersymmetric duality relations then enabled us to
examine the perturbative and non-perturbative stability of the
non-supersymmetric
\sosixteen\ string, and our results can be summarized in Fig.~\ref{phases}.
We found that this theory naturally has two distinct phases
depending on the relative value of its ten-dimensional coupling
$\lambda_H^\ten$
and its radius $R_H$ of compactification.
If $\lambda_H^\ten < R_H^2 / (R^\ast)^2$, our theory is completely
tachyon-free,
and is expected to flow to the ten-dimensional weakly coupled $SO(32)$ theory.
By contrast, if $\lambda_H^\ten > R_H^2 / (R^\ast)^2$,
the theory is expected to flow to the ten-dimensional strongly coupled
$E_8\times E_8$ string.  This includes the $R_H=0$ limiting point which
should correspond to the pure, ten-dimensional \sosixteen\ string.

Our results raise a number of interesting questions which will hopefully
be addressed in the future.

Perhaps the most obvious question concerns the situation in lower dimensions.
This question clearly has important theoretical and phenomenological
ramifications.  It may seem somewhat disappointing that the ten- and
nine-dimensional \sosixteen\ theories apparently flow to new theories for
which supersymmetry is restored.
Might there exist {\it stable}\/ non-supersymmetric theories in lower 
dimensions?
Our methods can undoubtedly be generalized to lower dimensions, and one would
expect the phase structure to be much richer.
Indeed, there are certainly many more non-supersymmetric tachyon-free 
theories in lower dimensions than there are in ten 
dimensions \cite{lowerdim}, and hence there are many more candidates.

\begin{figure}[tb]
\centerline{\epsfxsize 4.0 truein \epsfbox {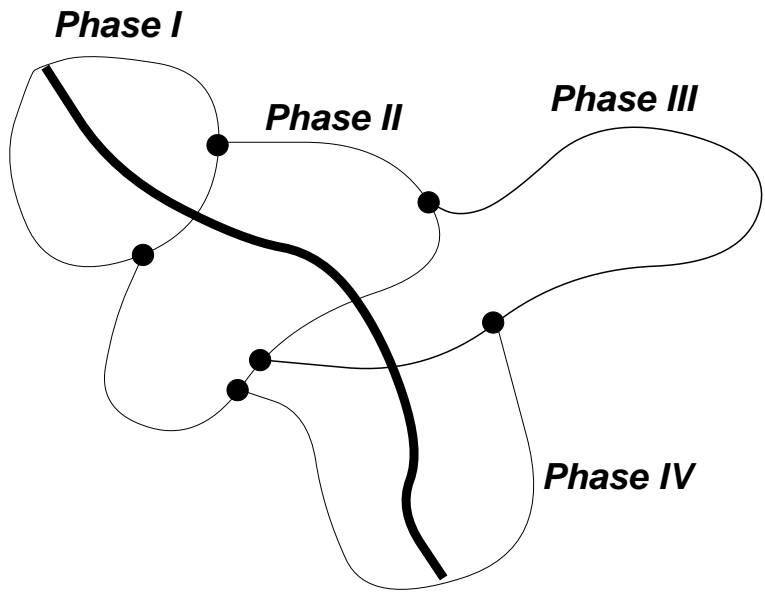}}
\caption{A proposed generalization of Fig.~\protect\ref{phases} for
    string theories in lower dimensions.
    We show four distinct phases, with the dark
    line indicating a submanifold
    of supersymmetric theories.  This supersymmetric submanifold
    runs through the larger manifold
    like a spine.
    Within each phase, we conjecture that strong/weak coupling duality
    can be extended
    from the supersymmetric spine throughout that phase.
    The dots indicate the intersections of boundaries between phases,
    and are conjectured to correspond to stable non-supersymmetric
    string theories.}
\label{amoeba}
\end{figure}

In fact, one might imagine the following situation.
If we assume that the space of supersymmetric string theories in a given
dimension forms a closed submanifold within the larger manifold of
all self-consistent theories (both supersymmetric and non-supersymmetric),
then perhaps this larger manifold consists of subregions with boundaries
such that each subregion has an intersection with the submanifold of
supersymmetric
theories.
We can then imagine that whatever strong/weak coupling duality
relation is valid on the manifold of supersymmetric theories within a given
subregion
can be extended throughout the entire subregion.
This situation is sketched in Fig.~\ref{amoeba}.

In many ways, Fig.~\ref{amoeba} can be viewed as a natural generalization of
Fig.~\ref{phases} in that it generalizes
the situation that we found in the case of our nine-dimensional interpolating
models.
Fig.~\ref{phases} shows, in some sense, a section of
the large manifold of self-consistent nine-dimensional theories,
and the submanifold of supersymmetric theories corresponds to the top and right
axes of Fig.~\ref{phases}.  The right axis, clearly, corresponds to the
supersymmetric
$SO(32)$ theory, and the top axis, because of the infinite coupling and
our above conjectures, essentially corresponds to the supersymmetric $E_8\times
E_8$
theory.  Fig.~\ref{phases} indicates how the
supersymmetric strong/weak coupling duality relations are extended
to the interior of the diagram:  points in Phase~I flow to the
right axis, and points in Phase~II flow to the top axis.

Note that there is a critical point at the lower left corner of
Fig.~\ref{phases},
corresponding to the ten-dimensional \sosixteen\ theory at zero coupling.  This
theory is, of course, completely stable, for it has no interactions and hence
no
dynamics.  In some sense, this stability arises because this point sits on the
intersection of boundaries between the two subregions of the theory.
It is then natural to imagine that even in lower dimensions, as sketched in
Fig.~\ref{amoeba},
there will be similar manifolds of {\it stable}\/ non-supersymmetric
tachyon-free
theories which are on the {\it intersections}\/ of boundaries of the different
subregions.
Note that while some subregions are expected to have string-theoretic duals,
other subregions will have duals that are better described via M-theory.

Another natural question raised by our results concerns
whether additional evidence for these non-supersymmetric
dualities can be found.
In supersymmetric theories, BPS states can be used to
provide evidence for duality.
Do (stable) tachyon-free string theories contain analogues of BPS states which
might be used for the same purpose?
In this connection, the notion of ``misaligned supersymmetry''
\cite{misaligned}
might play a crucial role.  Indeed, even though these string theories are
non-supersymmetric,
we have seen in (\ref{supertracerelations}) that many remarkable constraints
continue to govern the resulting string spectrum.
It is natural to conjecture that
misaligned supersymmetry might be sufficiently powerful to guarantee
the BPS-like stability of certain states in the spectra of
stable, non-supersymmetric tachyon-free strings.

But perhaps most of all, our results demonstrate that the entire
notion of duality may not be as directly
tied to supersymmetry as has been thought.
This suggests that a reformulation of the ideas and methodologies
of strong/weak coupling duality in string theory might be in order.

\bigskip
\medskip
\leftline{\large\bf Acknowledgments}
\medskip

We are happy to thank K.S. Babu, K. Intriligator, J. March-Russell,
R. Myers, S. Sethi, F. Wilczek, E. Witten, and especially A. Sagnotti
for discussions.  This work was supported in part by
NSF Grant No.\  PHY-95-13835
and DOE Grant No.\ DE-FG-0290ER40542.

\setcounter{section}{0}   
\Appendix{Free-Fermionic Realizations of Models}

In this Appendix, we give the explicit free-fermionic realizations
of the models we have considered in this paper.

\subsection{Ten-Dimensional Models}

In ten dimensions, are concerned with five different heterotic models:
the supersymmetric $SO(32)$ and $E_8\times E_8$ models, the
non-supersymmetric tachyon-free $SO(16)\times SO(16)$ model,
and the non-supersymmetric tachyonic $SO(32)$ and $SO(16)\times E_8$ models.
Using the complex-fermion notation of Ref.~\cite{KLST}, these models can be
realized
as follows.
If we begin with the
single fermionic boundary condition vector
\beq
      \V{0} ~=~ \lbrack \, (\half)^4 \,|\, (\half)^{16} \,\rbrack~,
\eeq
we obtain the {\it non}\/-supersymmetric $SO(32)$ string model.
If we now introduce the additional boundary condition vector
\beq
      \V{1} ~=~ \lbrack \, (0)^4 \,|\, (\half)^{16} \,\rbrack~,
\eeq
we obtain the {\it supersymmetric}\/ $SO(32)$ model.
In order to produce the remaining models, we introduce a twist
corresponding to the additional boundary condition vector
\beq
      \V{2} ~=~ \lbrack \, (0)^4 \,|\, (\half)^{8}\,(0)^8 \,\rbrack ~.
\eeq
It then turns out that we can obtain the $E_8\times E_8$ and $SO(16)\times
SO(16)$ models
depending on our choice of the corresponding GSO phase.
In the free-fermionic notation, such phase choices are given by specifying
the values of the independent parameters $k_{20}$ and $k_{21}$.
These two parameters  reflect
the chosen phases of the sectors corresponding to $\V{2}$ relative
to those corresponding to $\V{0}$ and $\V{1}$ respectively.
If $k_{20}=k_{21}$, the resulting model preserves supersymmetry;
if $k_{20}\not= k_{21}$, supersymmetry is broken.
We then find that
\beqn
     k_{20}=k_{21}
    ~&\Longrightarrow&~ {\rm produces}~ E_8\times E_8~ {\rm model} \nonumber\\
     k_{20}\not=k_{21}
    ~&\Longrightarrow&~ {\rm produces}~ SO(16)\times SO(16)~ {\rm model} ~.
\eeqn
The fact that these two models differ merely by a phase in a GSO projection
means that the states which are projected out of the spectrum in order to
produce
the $E_8\times E_8$ model are kept in the $SO(16)\times SO(16)$ model, and vice
versa.
Finally, the non-supersymmetric $SO(16)\times E_8$ model can be obtained
by deleting $\V{1}$ from the set of boundary condition vectors.
It should be noted that other non-supersymmetric tachyonic heterotic string
models can be obtained by adding additional boundary condition vectors to
the above list, but these
will not be necessary for our purposes.

In general, any two models related through a $\IZ_2$ orbifold
can be realized in the free-fermionic construction
as above through the relative addition or subtraction of a single
boundary-condition vector.
For example, since deleting the vector $\V{1}$ from
the supersymmetric $SO(32)$ model
produces the non-supersymmetric $SO(32)$ model,
these two models are related by a $\IZ_2$ orbifold.
Thus, simply by examining the above realizations of our ten-dimensional
models, we can immediately deduce the pattern
of $\IZ_2$ orbifold relations indicated in Fig.~\ref{orbrelations}.
There is, however, one subtlety, for it would not be immediately apparent from
the above fermionic realizations that the $SO(16)\times SO(16)$ string can
be realized as an orbifold of the $E_8\times E_8$ string.  However, this
orbifold relation does in fact exist, and  follows from
an alternative fermionic realization of the non-supersymmetric
\sosixteen\ model which starts from the above realization
of the $E_8\times E_8$ model, but which then adds the boundary-condition vector
\beq
     \V{3} ~=~ \lbrack \, (0)^4 \,|\, (\half)^{4}\,(0)^4\,(\half)^4\,(0)^4
\,\rbrack ~
\eeq
with corresponding phases $k_{30}\not= k_{31}$.

We now give the free-fermionic realizations of the ten-dimensional Type~II
models.
These models can all be realized in the free-fermionic construction
through the set of basis vectors:
\beqn
      \V{0} &=& \lbrack \, (\half)^4 \,|\, (\half)^{4} \,\rbrack~\nonumber\\
      \V{1} &=& \lbrack \, (\half)^4 \,|\, (0)^{4} \,\rbrack~.
\eeqn
If we take $k_{00}={1/2}$, we obtain the Type~IIA model,
whereas if we take $k_{00}=0$, we obtain the Type~IIB model.
By contrast, if we delete the vector $\V{1}$, we then
obtain the non-supersymmetric models:
with $k_{00}={1/2}$ we obtain the Type~0A model,
and with $k_{00}={0}$ we obtain the Type~0B model.

\subsection{Nine-Dimensional Models}

We now give the
explicit free-fermionic construction of the nine-dimensional
interpolating models presented in Sect.~3.5.
As we discussed in Sect.~3.1, the free-fermionic construction
will yield nine-dimensional models formulated at the specific
radius $R=\sqrt{2\alpha'}$.  However, using (\ref{fflimit}) in
reverse, it is then a simple matter to extrapolate these models
to arbitrary radius.  Such radius extrapolations do not affect the
self-consistency of these models.
It will therefore be sufficient to construct these models
at the specific radius $R=\sqrt{2\alpha'}$.

To do this, we
employ the free-fermionic construction as follows.
First, we introduce the following generic set of fermionic
boundary condition vectors:
\beqn
      \V{0} &=& \lbrack \, (\half)^4 \,(\half)\,|\, (\half)^{16}
\,(\half)\,\rbrack~, \nonumber\\
      \V{1} &=& \lbrack \, (0)^4 \,(\half)\, |\, (\half)^{16}
\,(\half)\,\rbrack~, \nonumber\\
      \V{2} &=& \lbrack \, (0)^4 \,(0)\, |\, (\half)^{16} \,(0)\,\rbrack~ ~.
\eeqn
Corresponding to these vectors,  there then exist three independent relative
GSO projection phases which may be chosen.
In the free-fermionic notation of Ref.~\cite{KLST}, these three phases are
given by specifying
the values of the three parameters
$\lbrace k_{10}, k_{20}, k_{21}\rbrace$.
Since each of these parameters can take two possible values
(either 0 or $\half$, corresponding to plus or minus signs in
the corresponding GSO projections), this {\it a priori}\/ leads
to eight possible models.
However, there is a great redundancy, and indeed we find that there
are only two distinct models (modulo the duality under which the limiting $a\to
\infty$
and $a\to 0$ endpoints are exchanged).
In general, supersymmetry is preserved if
$k_{20}=k_{21}$,
and broken otherwise.
For $k_{20}=k_{21}$ we obtain the trivial {\it untwisted}\/ compactification
of the $SO(32)$ string model, while for $k_{20}\not= k_{21}$ we obtain Model~A
which interpolates between the ten-dimensional supersymmetric and
non-supersymmetric
$SO(32)$ string models.

In order to obtain the additional interpolating models, we now introduce
the additional boundary-condition vector
\beq
      \V{3} ~=~ \lbrack \, (0)^4 \,(0)\,|\, (\half)^{8}\,(0)^8\, (0) \,\rbrack
 {}~.
\eeq
Taken together with the previous vectors, this now leaves us
with six independent relative
GSO projection phases
$\lbrace k_{10}, k_{20}, k_{21}, k_{30}, k_{31}, k_{32}\rbrace$
which may be chosen.
In general, supersymmetry is preserved if
$k_{20}=k_{21}$ and $k_{30}=k_{31}$,
and broken if either of these two conditions is not met.
After taking into account various redundancies,
this leaves us with six distinct models.
The following restrictions on the allowed values of these
six parameters will then suffice to specify our
string models:
\beqn
    \lbrace k_{20}=k_{21},\, k_{30}\not=k_{31},\, k_{30}=k_{32} \rbrace
       &~\Longrightarrow~& {\rm Model~B}~\nonumber\\
    \lbrace k_{20}\not=k_{21},\, k_{30}=k_{32} \rbrace
       &~\Longrightarrow~& {\rm Model~D}~\nonumber\\
    \lbrace k_{20}=k_{21},\, k_{30}=k_{31},\, k_{30}\not=k_{32} \rbrace
       &~\Longrightarrow~& {\rm Model~E}~\nonumber\\
    \lbrace k_{20}\not=k_{21},\, k_{30}\not=k_{32} \rbrace
       &~\Longrightarrow~& {\rm Model~G}~\nonumber\\
    \lbrace k_{20}=k_{21},\, k_{30}=k_{31},\, k_{30}=k_{32} \rbrace
       &~\Longrightarrow~& {\rm Model~X}~\nonumber\\
    \lbrace k_{20}=k_{21},\, k_{30}\not=k_{31},\, k_{30}\not=k_{32} \rbrace
       &~\Longrightarrow~& {\rm Model~Y}~.
\label{kvalues}
\eeqn
Model~B is the nine-dimensional model of Sect.~3.5 which interpolates
between the supersymmetric $SO(32)$ model and the
non-supersymmetric \sosixteen\ model, while Model~D likewise interpolates
between the $E_8\times E_8$ model and the
$SO(16)\times E_8$ model.
These models therefore succeed in interpolating
between supersymmetric and non-supersymmetric
ten-dimensional models.
By contrast,
Model~E interpolates between the ten-dimensional
supersymmetric $SO(32)$ and $E_8\times E_8$ models, while Model~G
interpolates between the non-supersymmetric \sosixteen\
and $SO(16)\times E_8$ models.
Similarly,
Models~X and Y are trivial {\it untwisted}\/ compactifications
of the $E_8\times E_8$ and $SO(16)\times SO(16)$ strings respectively.

Note that for each of these models,
there exists a ``dual'' model which exchanges the role of radius with inverse
radius.
Thus, in the dual model, the limiting ten-dimensional models are exchanged.
Within the free-fermionic construction, there are two ways of obtaining the
dual of a given model.  The first is merely to flip the value of $k_{10}$ from
0 to $\half$
or vice versa.  The second is to simultaneously flip the values of all
of the other independent parameters
$\lbrace k_{20}, k_{21}\rbrace$
or
$\lbrace k_{20}, k_{21}, k_{30}, k_{31}, k_{32}\rbrace$.
This has the effect of exchanging
$a\leftrightarrow 1/(2a)$, which is equivalent to
exchanging $\calE_{1/2}\leftrightarrow\calO_0$.

Finally, we give the free-fermionic construction of the remaining
Model~C which interpolates between the supersymmetric
$E_8\times E_8$ model and the non-supersymmetric \sosixteen\ model.
This model can be realized
through the introduction of the additional boundary-condition vector
\beq
     \V{4} ~=~ \lbrack \, (0)^4 \,(0)\, |\,
     (\half)^{4}\,(0)^4\,(\half)^4\,(0)^4 \,(0)\,\rbrack ~.
\eeq
We then take the GSO projection phase $k_{31}=1/2$, setting
$k_{ij}=0$ for all other cases with $i>j$.

We now present our interpolating Type~II models.
We follow the same procedure as for the heterotic models, and compactify
these ten-dimensional Type~II models on circles of arbitrary radii.
At the fermionic radius $a=1/\sqrt{2}$, the resulting models can be realized
via the nine-dimension free-fermionic construction as follows.  We begin
with the set of basis vectors:
\beqn
      \V{0} &=& \lbrack \, (\half)^4 \,(\half)\, |\, (\half)^{4}
\,(\half)\,\rbrack~\nonumber\\
      \V{1} &=& \lbrack \, (\half)^4 \,(\half)\, |\, (0)^{4}
\,(\half)\,\rbrack~\nonumber\\
      \V{2} &=& \lbrack \, (\half)^4 \,(0)\, |\, (0)^{4} \,(0)\,\rbrack~.
\eeqn
We then have four independent GSO-projection phases
$\lbrace k_{00},k_{10},k_{20},k_{21}\rbrace$ that must be specified.
Generally, $N=2$ supersymmetry is preserved if $k_{20}=k_{21}$
and completely broken to $N=0$ otherwise.
We then find that there exist four physically distinct nine-dimensional
Type~II models, as follows:
\beqn
    \lbrace k_{20}\not=k_{21},\, k_{00}+k_{10}+k_{20}\not\in\IZ \rbrace
       &~\Longrightarrow~& {\rm Model~A'}~\nonumber\\
    \lbrace k_{20}\not=k_{21},\, k_{00}+k_{10}+k_{20}\in\IZ  \rbrace
       &~\Longrightarrow~& {\rm Model~B'}~\nonumber\\
      \lbrace k_{20}=k_{21},\, k_{00}+k_{10}+k_{20}\in\IZ \rbrace
         &~\Longrightarrow~& {\rm Model~C'}~\nonumber\\
      \lbrace k_{20}=k_{21},\, k_{00}+k_{10}+k_{20}\not\in\IZ \rbrace
         &~\Longrightarrow~& {\rm Model~D'}~.
\eeqn
It turns out that Models~${\rm C}^\prime$ and D$^\prime$
are trivial untwisted compactifications
of the Type~IIB and Type~IIA models respectively.
They therefore interpolate between the Type~IIA and IIB theories in opposite
directions ({\it i.e.}\/, Models~C$^\prime$ and D$^\prime$ are $T$-duals of
each other, with both collected together as the single Model~C$^\prime$ in
Fig.~\ref{interpfig2}).
By contrast, Models~${\rm A}^\prime$ and ${\rm B}^\prime$ are
non-supersymmetric for general
values of the radius, and interpolate between the four ten-dimensional models
as indicated in Fig.~\ref{interpfig2}.
In each case, the $a\leftrightarrow 1/(2a)$ dual of a
given model can be realized by flipping the values of $k_{00}$ and $k_{10}$
from 0 to 1/2 and vice versa.

Finally, we conclude with a brief comment concerning
our heterotic Model~C.  In Ref.~\cite{IT},
a heterotic nine-dimensional string model
was constructed which is claimed to interpolate between
the ten-dimensional
$E_8\times E_8$
and $SO(16)\times SO(16)$ heterotic models.
The authors provide the following partition function\footnote{
  In writing this partition function, we do not distinguish
  between $\chibar_S$ and $\chibar_C$,
  nor between $\chi_S$ and $\chi_C$, since the original authors
  of Ref.~\cite{IT} have not done so.
  We also have corrected a typographical error on the
  final line of Eq.~(13) of Ref.~\cite{IT}:  as is necessary for
   modular invariance, the first term on this line should
   be $\vartheta_4^8$ rather than $\vartheta_2^8$.}
as corresponding to their model:
\beqn
    Z ~=~  Z^{(7)}_{\rm boson} \,\times \,\bigl\lbrace ~
    \phantom{+}&\calE_0 &  \lbrack
    \chibar_S \,(\chi_I^2 + \chi_S^2) ~-~ \chibar_V \,(2\, \chi_I \chi_S)
\rbrack\nonumber\\
   +&\calE_{1/2}  & \lbrack
     \chibar_I \,(2\,\chi_V \chi_S) ~-~ \chibar_S \,(\chi_V^2 + \chi_S^2)
\rbrack\nonumber\\
   +&\calO_0 & \lbrack
    \chibar_S \,(2\,\chi_I \chi_S) ~-~ \chibar_V \,(\chi_I^2 + \chi_S^2)
\rbrack\nonumber\\
   +&\calO_{1/2} & \lbrack
     \chibar_I \,(\chi_V^2 + \chi_S^2) ~-~ \chibar_S \,(2\,\chi_V \chi_S)
\rbrack ~~\bigr\rbrace~.
\label{ZIT}
\eeqn
This expression is indeed modular-invariant,
and in the limit $a\to 0$, this does reproduce the
$E_8 \times E_8$ partition function.
However, in the limit $a\to\infty$,
this does not yield the partition function of the
$SO(16)\times SO(16)$ model.
Indeed, this limit results in an expression which is
not a valid partition function.

It is not difficult to show that the function (\ref{ZIT})
cannot correspond to a valid, self-consistent string model.
First, we observe that it violates
spin-statistics:  terms proportional to $\chibar_S$ (which come from
spacetime fermionic sectors) should always appear
in the partition function with minus signs, and terms proportional
to $\chibar_I$ and $\chibar_V$ (spacetime bosonic sectors) should always appear
with plus signs.  This is not the case for (\ref{ZIT}), in which mixed signs
appear.
Second, we also observe that this expression is inconsistent with the presence
of
gravitons.  We know that the graviton state, with structure $\psi_{-1/2}^\mu
|0\rangle_R\otimes
X_{-1}^\nu |0\rangle_L$, must always appear in a sector corresponding to a term
of
the form $\chibar_V \chi_I^2$.  However, for general radius, such a term
can contain massless states only if it arises in the $\calE_0$ sector.
Unfortunately, no such term $\calE_0 \chibar_V \chi_I^2$ arises in (\ref{ZIT}),
which signals an inconsistency in this expression.
For these two reasons, we doubt the validity of the expression (\ref{ZIT})
as the partition function of a self-consistent string model.

Although the expression (\ref{ZIT}) is clearly invalid as a partition function,
this does not necessarily imply that the {\it model}\/ constructed
in Ref.~\cite{IT} is invalid.  It may simply be that
the expression (\ref{ZIT}) does not, in fact, correspond to the
model claimed.  In any case, our orbifold and free-fermionic constructions
of Model~C, along with the correct partition function given in (\ref{9dparts}),
explicitly demonstrate the existence of a nine-dimensional string model with
the desired interpolating properties.
Moreover, our general construction procedure demonstrates that in fact
a variety of such models exist, as shown in Fig.~\ref{interpfig},
illustrating that such interpolating models are a completely general feature.

\bigskip
\bigskip

\vfill\eject

\bibliographystyle{unsrt}

\begin{thebibliography}{99}


\bibitem{W}  E. Witten, \NPB{443}{95}{85}.
\bibitem{PW}  J. Polchinski and E. Witten, \NPB{460}{96}{525}.
\bibitem{HW}  P. Ho\v{r}ava and E. Witten, \NPB{460}{96}{506}.
\bibitem{scherkschwarz}   J. Scherk and J.H. Schwarz, \PLB{82}{79}{60}.
\bibitem{misaligned}  K.R. Dienes, \NPB{429}{94}{533};
        hep-th/9409114;  hep-th/9505194.
\bibitem{supertraces}
       K.R. Dienes, M. Moshe, and R.C. Myers,
     {\it Phys.\ Rev.\ Lett.}\/ {\bf 74} (1995) 4767;
     hep-th/9506001.
\bibitem{review}  K.R. Dienes, {\it Phys.\ Rep.}\/ {\bf 287} (1997) 447
[hep-th/9602045].
\bibitem{antoniadis}
    I. Antoniadis, C. Bachas, D.C. Lewellen, and T. Tomaras,
\PLB{207}{88}{441};\\
    I. Antoniadis, \PLB{246}{90}{377};\\
    I. Antoniadis, C. Mu\~{n}oz, and M. Quir\'{o}s, \NPB{397}{93}{515};\\
    I. Antoniadis and K. Benakli, \PLB{326}{94}{69}.
\bibitem{others}
         S. Ferrara, C. Kounnas, and M. Porrati,
             \NPB{197}{87}{135};  \PLB{206}{88}{25};
              \NPB{304}{88}{500};\\
         S. Ferrara, C. Kounnas, M. Porrati, and F. Zwirner, \NPB{318}{89}{75}.
\bibitem{others2}
         C. Bachas, hep-th/9503030;\\
         J.G. Russo and A.A. Tseytlin, \NPB{461}{96}{131};\\
         A.A. Tseytlin, hep-th/9510041;\\
         M. Spalinski and H.P. Nilles, \PLB{392}{97}{67};\\
         I. Shah and S. Thomas, hep-th/9705182.
\bibitem{KLTclassification}    H. Kawai, D.C. Lewellen, and S.-H.H. Tye,
          \PRD{34}{86}{3794}.
\bibitem{AGMV} L. Alvarez-Gaum\'e, P. Ginsparg, G. Moore, and C. Vafa,
          \PLB{171}{86}{155}.
\bibitem{DH}  L.J. Dixon and J.A. Harvey, \NPB{274}{86}{93}.
\bibitem{SW}  N. Seiberg and E. Witten, \NPB{276}{86}{272}.
\bibitem{sen}  A. Sen, {\it Mod.\ Phys.\ Lett.}\/ {\bf A11} (1996) 1339;
       \NPB{474}{96}{361}.
\bibitem{Rohm}  R. Rohm, \NPB{237}{84}{553}.
\bibitem{IT}  H. Itoyama and T.R. Taylor, \PLB{186}{87}{129}.
\bibitem{GV}  P. Ginsparg and C. Vafa, \NPB{289}{87}{414}.
\bibitem{Sagnotti}  M. Bianchi and A. Sagnotti, \PLB{247}{90}{517};\\
      A. Sagnotti, hep-th/9509080.
\bibitem{BG}  O. Bergman and M.R. Gaberdiel, hep-th/9701137.
\bibitem{Gates}  S.J. Gates, Jr.\ and V.G.J. Rodgers, hep-th/9704101.
\bibitem{DabHull}  A. Dabholkar, \PLB{357}{95}{307};\\
           C.M. Hull, \PLB{357}{95}{545}.
\bibitem{short}  J.D. Blum and K.R. Dienes, hep-th/9707148.
\bibitem{KLT}  H. Kawai, D.C. Lewellen, and S.-H.H. Tye,
                {\it Nucl.\ Phys.}\/ {\bf B288} (1987) 1.
\bibitem{ABK} I. Antoniadis, C. Bachas, and C. Kounnas,
                {\it Nucl.\ Phys.}\/ {\bf B289} (1987) 87.
\bibitem{KLST}      H. Kawai, D.C. Lewellen, J.A. Schwartz, and S.-H.H. Tye,
                 \NPB{299}{88}{431}.
\bibitem{DHS}
      M. Dine, P. Huet, and N. Seiberg, \NPB{322}{89}{301}.
\bibitem{DLP}
      J. Dai, R.G. Leigh, and J. Polchinski,
      {\it Mod.\ Phys.\ Lett.}\/ {\bf A4} (1989) 2073.
\bibitem{orientifolds}
     See, {\it e.g.}\/:\\
   A. Sagnotti, in Proceedings of {\it Cargese 1987:  Non-Perturbative
   Quantum Field Theory}\/, eds.\ G. Mack \etal\  (Plenum, 1988), p.\  521;\\
   P. Ho\v{r}ava, \NPB{327}{89}{461};  \PLB{231}{89}{251};\\
   J. Dai, R.G. Leigh, and J. Polchinski,
    {\it Mod.\ Phys.\ Lett.}\/ {\bf A4} (1989) 2073;\\
   G. Pradisi and A. Sagnotti, \PLB{216}{89}{59};\\
   M. Bianchi and A. Sagnotti,
       \NPB{361}{91}{519};\\
   E.G. Gimon and J. Polchinski, \PRD{54}{96}{1667}.
\bibitem{PC}  J. Polchinski and Y. Cai, \NPB{296}{88}{91}.
\bibitem{cancellambda}
   See, {\it e.g.}\/:\\
      G. Moore, \NPB{293}{87}{139};
        Erratum: {\it ibid.}\/ {\bf B299} (1988) 847;\\
      T.R. Taylor, \NPB{303}{88}{543};\\
     J. Balog and M.P. Tuite, \NPB{319}{89}{387};\\
      K.R. Dienes, \PRD{42}{90}{2004};\\
     T. Gannon and C.S. Lam, \PRD{46}{92}{1710}.
\bibitem{lowerdim}
     K.R. Dienes, \PRL{65}{90}{1979}.
\bibitem{Ginsparg}
     P. Ginsparg, \PRD{35}{87}{648}.
\bibitem{e8paper}
     J.D. Blum and K.R. Dienes, to appear.
\bibitem{scales}
     M. Dine and N. Seiberg, \PRL{55}{85}{366}; \PLB{162}{85}{299}.
\bibitem{Nair}
     V.P. Nair, A. Shapere, A. Strominger, and F. Wilczek,
      \NPB{287}{87}{402}.
\bibitem{horava}  P.~Ho\v{r}ava, \PRD{54}{96}{7561}.

\end{thebibliography}

\end{document}